\documentclass[final,5p,times,twocolumn,authoryear]{elsarticle}

\usepackage{amsmath}
\usepackage{amssymb}
\usepackage{lipsum}

\usepackage{lineno}
\usepackage{booktabs}
\usepackage{threeparttable}

\def\tsc#1{\csdef{#1}{\textsc{\lowercase{#1}}\xspace}}
\tsc{WGM}
\tsc{QE}
\newcommand{\rhoh}{\hat{\rho}}
\newcommand{\Hh}{\hat{H}}
\newcommand{\Sh}{\hat{S}}
\newcommand{\Fh}{\hat{F}}

\newcommand{\Bh}{\hat{B}}
\newcommand{\Qh}{\hat{Q}}
\newcommand{\Heff}{\hat{H}_{\mathrm{eff}}}
\newcommand{\rhoc}{\rhoh_{\mathrm{c}}}
\newcommand{\SEAQT}{\mathrm{SEAQT}}
\newcommand{\GKSL}{\mathrm{GKSL}}

\journal{Annals of Physics}

\begin{document}

\begin{frontmatter}



\title{A Description of the Quantum Mpemba Effect using the Steepest-Entropy-Ascent Quantum Thermodynamics Framework}


\author[1]{L. E. Rocha-Soto}
\affiliation[1]{organization={Universidad de Guanajuato},
            addressline={Carretera Salamanca - Valle de Santiago km 3.5 + 1.8}, 
            city={Salamanca},
            postcode={36885}, 
            state={Guanajuato},
            country={Mexico}}
\author[1]{C. E. Damian-Ascencio}

\author[2]{A. Salda\~na-Robles}
\affiliation[2]{organization={Universidad de Guanajuato},
            addressline={Carretera Irapuato-Silao km 9 }, 
            city={Irapuato},
            postcode={36500}, 
            state={Guanajuato},
            country={Mexico}}
            
\author[1]{S. Cano-Andrade}

\begin{abstract}
The quantum Mpemba effect describes the unusual relaxation of a quantum system where if starting from a state far from equilibrium reaches equilibrium faster than starting from a closer state. In this work, the steepest-entropy ascent quantum thermodynamics framework is used to model this effect in a three-level ion system coupled to a short-life level which acts as an environment. The four-level Hilbert space is reduced to an effective three-dimensional description via the Feshbach projection, where the resulting model parameters are determined by a differential evolution algorithm. Predictions of both the steepest-entropy-ascent and Lindblad frameworks agree well with experimental data for all three initial conditions considered. In addition it is shown that, independently of the Mpemba condition, the
relaxation parameter $\tau_D$ is an emergent thermodynamic quantity whose step-function time dependence arises from a near-indetermination at the metastable state, resolved by the two-timescale structure of the dissipative dynamics, with a step height $\tau_D^{+}/\tau_D^{-} (|\lambda_f|/|\lambda_s|)\,\mathfrak{r}$ set by the ratio of the linearised relaxation rates at the metastable state times a geometric factor $\mathfrak{r}$ fixed by the curvature of the entropy and
free-energy-variance surfaces there. Furthermore, it is also established the correspondence between the Lindblad and steepest-entropy-ascent order parameters as spectral and geometric suppression of the same slow dissipative mode, and it is shown that for isolated systems with a negative nonequilibrium inverse temperature, as realized in the case studied here, the genuine Mpemba free-energy ordering is equivalent to an entropy ordering.
\end{abstract}




\end{frontmatter}




\section{Introduction}\label{sec:Int}

In classical thermodynamics there exists an unusual effect in which a system initially at a high temperature undergoes a dynamical crossover to a colder state faster than if starting at a lower temperature. This phenomenon is known as Mpemba effect. One of the earliest attempts to explain this effect dates back to Aristotle, who attributed it to the intensification of a property due to the presence of its opposite in the surrounding environment \cite{aristotle_meteorologica_1952}. In this sense, heat would be compressed by cold, leading water to freeze more rapidly when it had been previously warmed. Several years later, the same effect was rediscovered by Erasto Mpemba, who observed that a mixture of hot milk used for ice cream froze faster than another mixture initially cooler. This observation motivated a series of controlled experiments conducted by Mpemba and Osborne, in which water samples prepared at different initial temperatures were cooled under similar conditions, leading to the same conclusion \cite{Mpemba_1969,Kell_1969}.

Although the Mpemba effect was originally regarded as a purely thermal phenomenon, it has been shown that it is instead a manifestation of a more general anomalous relaxation behavior that can arise in a wide variety of systems. Different physical mechanisms have been proposed to account for this behavior, depending on the nature of the system and its relevant parameters. As a consequence, no single, universal theoretical framework has yet been established to fully describe the effect \cite{Kumar_2023}. Nowadays, these relaxation effects have recently been observed experimentally in single-particle and many-particle open quantum systems \cite{chatterjee2025direct, schnepper2025experimental, zhang2025observation, aharony2024inverse, joshi2024observing, ares2023entanglement, yu2025quantum, liu2024symmetry, liu2025symmetry}.

Several theoretical frameworks have been developed to explain this phenomenon, considering both Markovian \cite{caceffo2024entangled,Wang_2024,ivander2023hyperacceleration,Manikandan_2021,zhou2023accelerating,Carollo_Lesa_2022,first_Qmpemba,Multiple_mpemba_2024,nava2019lindblad,qian2025intrinsic, chattopadhyay2026anomaly, das2025role} and non-Markovian dynamics \cite{PhysRevLett.134.220403,5xrr-x2rm,PhysRevE.101.052106, sapui2026ergotropic}, where the speed of relaxation from one state to another is quantified using information-theoretic or geometric measures, such as the Kullback-Leibler divergence or the Hilbert-Schmidt distance. Moreover, most of these studies focus on anomalous relaxation processes leading to general fixed points rather than true thermalization, which would instead yield a stable equilibrium state \cite{Moroder_2024}. Within Markovian dynamics, and specifically in the Gorini-Kossakowski-Sudarshan-Lindblad framework described by the Lindblad master equation \cite{manzano2020short}, Davies maps emerges as an approach which ensures that the steady state corresponds to a stable equilibrium state \cite{alicki1976detailed,Dann2021,ROGA2010311,DAVIES1979421} and allows the use of the nonequilibrium free energy as a metric to characterize the so-called genuine quantum Mpemba effect \cite{chatterjee2025direct,schnepper2025experimental}.

In this context, the steepest-entropy-ascent quantum thermodynamics (SEAQT) framework is presented as an alternative approach to describe these relaxation processes from a thermodynamic perspective \cite{beretta1984quantum, beretta1985quantum}. Within this framework, the state of a quantum system evolves in time along the direction of maximum entropy production subject to the constraints imposed by conserved quantities, until a stable equilibrium state is reached. The resulting dynamics is described by the SEAQT equation of motion, which is constructed using geometrical principles by defining a set of generators of the motion that span the state space where the system evolves. Each generator is associated with a conserved quantity that must remain constant during the evolution, while the entropy increases \cite{beretta2009nonlinear, beretta2010maximum, beretta2014steepest}. For an isolated system, the identity operator and the Hamiltonian are chosen as generators of the motion, and therefore the resulting equation preserves the trace of the density operator and keeps the energy constant. Consequently, the dynamics is consistent with the conservation of probability and satisfies both the first and second laws of thermodynamics. Moreover, this choice of generators leads to a stable equilibrium state corresponding to the canonical ensemble. If the particle number operator is also included among the generators of the motion, the stable equilibrium state corresponds to the grand canonical ensemble \cite{saldana2025model}.

In this work, both the SEAQT framework and the Gorini-Kossakowski-Sudarshan-Lindblad (GKSL) framework are employed to predict the relaxation observed in the experiment reported by Zhang \emph{et al.} \cite{zhang2025observation}. The remainder of the paper is organized as follows: section \ref{sec:MatMod} presents the mathematical models used for describing the Mpemba effect; section \ref{sec:Res} presents the results and a discussion of the findings; and finally section \ref{sec:Conc} concludes the paper.

\section{\label{sec:MatMod}Mathematical Model}

\subsection{\label{sec:Case}Case of study}
	
\begin{figure}
    \centering
    \includegraphics[width=0.7\linewidth]{ 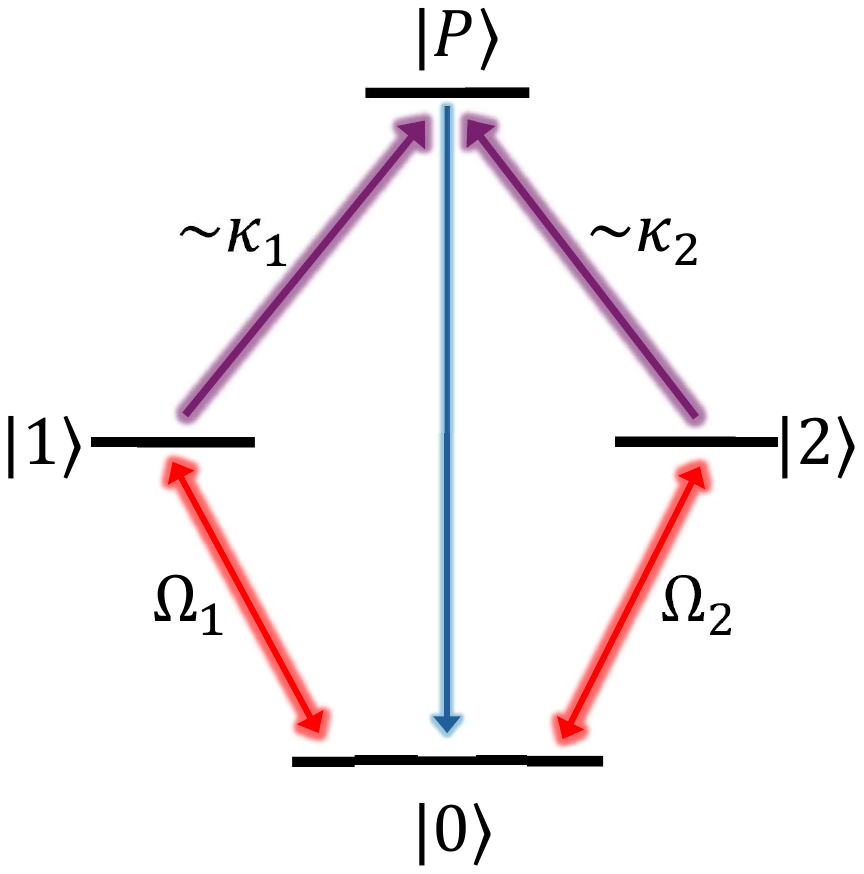}
    \caption{Schematic representation of the Mpemba effect experiment developed by Zhang \emph{et al.} \cite{zhang2025observation}.}
    \label{fig: experimental}
\end{figure}

The experiment by \cite{zhang2025observation} considers an ion whose internal states are resonantly coupled by laser fields, as shown in Figure \ref{fig: experimental}. The ground state $|0\rangle$ is resonantly coupled to the excited states $|1\rangle$ and $|2\rangle$. The states $|1\rangle$ and $|2\rangle$ are coupled to a short-lived auxiliary state $|P\rangle$, which rapidly decays to the state $|0\rangle$. The coupling between the states $|P\rangle$ and $|2\rangle$ is very weak, and the state $|1\rangle$ exhibits a faster decay due to its stronger coupling with $|P\rangle$. Due to the fast decay of the state $|P\rangle$, this dissipation channel can be effectively interpreted as an environmental interaction that induces a loss of population of $|1\rangle$ and $|2\rangle$ at rates $\kappa_1$ and $\kappa_2$, respectively, with $\kappa_2 \ll \kappa_1$. This allows the system to be modeled and described as an open quantum system within the Lindblad framework. Similar considerations have been adopted in other related works \cite{zhang2020single,zhang2022dynamical,Reiter_2012}.

Three different initial conditions are prepared to study the state dynamics of the system, where it is shown that an initial state minimizing the overlap with the slowest decay mode exhibits an accelerated relaxation towards a stable equilibrium state.

For modeling the experiment, a Hilbert space 
\begin{equation}\label{eq: hilbert}
	\mathcal{H}=\mathrm{span}\{|0\rangle, |1\rangle, |2\rangle\}
\end{equation}
is considered, where the Hamiltonian is
\begin{equation} \label{eq: H_lindblad}
 \hat{H} = \sum_{i=1}^2 \frac{\Omega_i}{2} \left( |0\rangle\langle i| + |i\rangle\langle 0|\right) =
 \frac{1}{2}
\begin{bmatrix}
0 & \Omega_1 & \Omega_2 \\
\Omega_1 & 0 & 0 \\
\Omega_2 & 0 & 0
\end{bmatrix}
\end{equation}
and the coupling to the $|P\rangle$ state is given by the jump operators, such as
\begin{equation}
    J_1 = \sqrt{\kappa_1}\,|0\rangle\langle1|, \quad J_2 = \sqrt{\kappa_2}\,|0\rangle\langle2|
\end{equation}
It should be noted that the GKSL description uses the bare Hamiltonian of
Eq.~(\ref{eq: H_lindblad}), since the auxiliary level $|P\rangle$ is already
accounted for by the jump operators; the SEAQT description of
Sec.~\ref{subsec:feshbach} instead uses the Feshbach-projected Hamiltonian
$\Heff$, which carries the level shifts induced by $|P\rangle$ but no
anti-Hermitian part. Using $\Heff$ in Eq.~(\ref{eq:lindb}) would double count
the auxiliary level.
\subsection{\label{subsec:lindblad-fw}Lindblad framework}

The evolution of an open quantum system that interacts weakly with its environment, while neglecting system-environment correlations, is modeled by the quantum master equation
\begin{equation} \label{eq:lindb}
    \frac{d\hat \rho}{dt}=-\frac{i}{\hbar}[\hat{H},\hat{\rho}] + \sum_i\left(J_i \hat{\rho} J_i^\dagger-\frac{1}{2}\{J_i^\dagger J_i, \hat{\rho} \}\right)
\end{equation}
which is known as Lindblad equation, where $\hbar$ is the reduced Planck constant, $\hat{H}$ is the Hamiltonian operator, $\hat{\rho}$ is the reduced density operator of the system, $J_i$ are the Lindblad operators which represent the coupling of the system with the environment. This equation provides a general framework for Markovian dynamics \cite{lindblad1976generators, gorini1976completely, manzano2020short}. However, it is not valid for modeling systems with strong couplings to the environment, because it is based on the Born–Markov approximation \cite{nakatani2010quantum}. The Lindblad equation is linear in $\hat{\rho}$; thus, it can be written in terms of the Liouvillian superoperator $\mathcal{L}$, such as \cite{rivas2012open}
\begin{equation}
    \frac{d\hat\rho}{dt}=\mathcal{L}\hat\rho(t)
\end{equation}

If the density matrix is vectorized, then the matrix representation of the Liouvillian is given as \cite{minganti2018spectral}
\begin{align}
\mathcal{L} = & -i(\hat{H}\otimes \mathbb{I} - \mathbb{I}\otimes \hat{H}^\mathrm{T}) \nonumber\\
& + \sum_i \frac{1}{2} \left( 2J_i \otimes J_i^* - J_i^\dagger J_i \otimes \mathbb{I} - \mathbb{I}\otimes J_i^\mathrm{T} J_i^* \right)
\end{align}
and the right, $\hat{R}$, and left, $\hat{L}$, eigenmatrices of the Liouvillian can be obtained by solving
\begin{equation}
    \mathcal{L}\hat{R}_i = \lambda_i \hat{R}_i \quad \text{and} \quad \hat{L}_i\mathcal{L}=\lambda_i^* \hat{L}_i
\end{equation}
where $\lambda_i$ are the corresponding eigenvalues sorted in descending order and represent the relaxation rates of the eigenmode $\hat{R}_i$. Thus, the solution of the Lindblad equation can be expressed as
\begin{equation} \label{eq: solution}
    \hat{\rho}(t) = e^{\mathcal{L}t}\hat{\rho}_{in}=\sum_{i=0}^{d^2-1} \left(\mathrm{Tr}(\hat{L}_i\hat{\rho}_{in})e^{\lambda_i t}\hat{R}_i \right)
\end{equation}
where $d$ is the dimension of the Hilbert space. The eigenvalue $\lambda_0=0$ corresponds to a mode that does not evolve in time, since its exponential factor $\exp(\lambda_0t)=1$. Its associated right eigenmatrix, $\hat{R}_0$, represents the stationary state, $\hat{\rho}_{ss}$ \cite{minganti2018spectral,zhang2025observation,Carollo_Lesa_2022,zhou2023accelerating}. Thus, Equation~(\ref{eq: solution}) can be written as
\begin{equation} \label{eq: solution_Lindblad}
    \hat{\rho}(t) = \hat{\rho}_{ss} + \sum_{i=1}^{d^2-1} \left(\mathrm{Tr}(\hat{L}_i\hat{\rho}_{in})e^{\lambda_i t}\hat{R}_i \right)
\end{equation}

If the initial state, $\hat{\rho}_{in}$, is chosen such that $\mathrm{Tr}(\hat{L}_1\hat{\rho}_{in})=0$, the relaxation mode corresponding to $\lambda_1$ vanishes, leading to an exponential acceleration of the relaxation process. This is known as the strong quantum Mpemba effect \cite{first_Qmpemba}.

\subsection{\label{subsec:seaqt}SEAQT framework}
	
The SEAQT equation of motion for a quantum system is given as
\begin{equation} \label{eq:gen_seaqt}
\frac{d\hat{\rho}}{dt}
= -\frac{i}{\hbar}[\hat{H},\hat{\rho}]
- \frac{1}{\tau_D}\,\hat{D}
\end{equation}
where $\hat{H}$ and $\hat{\rho}$ are the Hamiltonian and the density operator, respectively, $[\hat{H},\hat{\rho}] = \hat{H}\hat{\rho} - \hat{\rho}\hat{H}$, $\hbar$ is Planck's modified constant, $\tau_D$ is a characteristic relaxation time that can be a positive constant or a functional of the density operator, and $\hat{D}$ is an operator constructed such that the dynamics of the system is driven irreversibly along the path of maximum entropy production, and is given as

\begin{equation}\label{eq: D_general}
    \hat D = \frac{1}{2}\left[\sqrt{\hat \rho}\,\tilde{D}+(\sqrt{\hat \rho}\,\tilde{D})^\dagger \right]
\end{equation}
where the operator $\tilde{D}$ is 
\begin{equation} \label{eq:tildeD}
	 \tilde{D}=\frac{\left| \begin{matrix}
	 \sqrt{\hat\rho} \ln\hat{\rho} &  \sqrt{\hat \rho}\hat I  & \sqrt{\hat\rho}\hat H\\ 
	 (\hat I, \ln\hat{\rho}) & (\hat I, \hat I) & (\hat I, \hat H)\\
	 (\hat H, \ln\hat{\rho}) & (\hat H, \hat I) & (\hat H,\hat H)
	 \end{matrix}\right|}{\left| \begin{matrix}
	 (\hat I, \hat I) & (\hat I, \hat H)\\
	 (\hat H, \hat I) & (\hat H, \hat H)
	 \end{matrix} \right|}
\end{equation}
Here, $(\cdot,\cdot)$ is the Hilbert-Schmidt inner product between any two Hermitian operators $\hat{X}$ and $\hat{Y}$ acting on $\mathcal{H}$ and is given as
\begin{equation}
    (\hat{X}, \hat{Y})=\langle \hat{X} \hat{Y} \rangle = \frac{1}{2} \mathrm{Tr}\!\left( |\hat\rho| \{ \hat{X}, \hat{Y} \} \right)
\end{equation}
where $\{\hat{X},\hat{Y} \} = \hat{X} \hat{Y} + \hat{Y} \hat{X}$, and $|\hat\rho| = \sqrt{\hat\rho^\dag \hat\rho}$.

If each generator of the motion is expressed in terms of the deviation operator
\begin{equation}
\Delta \hat{X} = \hat{X} - \langle \hat{X} \rangle \hat I
\end{equation}
where $\langle \hat{X} \rangle = \mathrm{Tr} (\hat{\rho} \hat{X})$, then Equation~(\ref{eq:tildeD}) can be written in the equivalent form
\begin{equation} \label{eq: D_modified}
    \tilde D = \frac{
\left|
\begin{matrix}
    -\sqrt{\hat \rho} \Delta \hat S & 0 & \sqrt{\hat \rho} \Delta \hat H \\
    -\langle \hat S \rangle & 1 & \langle \hat H \rangle \\
    -\langle \hat H \hat S \rangle + \langle \hat H \rangle \langle \hat S \rangle & 0 & \langle \hat H^2 \rangle - \langle \hat H \rangle^2
\end{matrix}
\right|
}{\langle \hat H^2 \rangle - \langle \hat H \rangle^2} 
\end{equation}
where the entropy operator $\hat{S}$ is given as
\begin{equation} \label{eq: entropy}
	\hat{S} = -k_B \hat{B} \ln \hat{\rho} 
\end{equation}
and $\hat{B}$ is the projector onto the range of $\hat{\rho}$; that is, the subspace spanned by the eigenvectors of $\hat{\rho}$ with nonzero eigenvalues, so that $\hat B \ln \hat \rho$ is always well defined, such as \cite{beretta2009nonlinear}
\begin{equation} \label{eq: B_projector}
    \hat{B} = \hat{P}_{\mathrm{Ran}(\hat{\rho})} = \hat{I} - \hat{P}_{\mathrm{Ker}(\hat{\rho})}
\end{equation}
Accordingly, the entropy operator may be written as
\begin{equation}
    \hat{S}
    = -k_B\hat{P}_{\mathrm{Ran}(\hat{\rho})}\ln\hat{\rho}
    = - k_B\ln\!\left(\hat{\rho} + \hat{P}_{\mathrm{Ker}(\hat{\rho})}\right)
\end{equation}
thus, Equation~(\ref{eq: D_modified}) can be written as
\begin{equation} \label{eq: D_simplified}
	\tilde{D} = \sqrt{\hat{\rho}} \left( \beta\,\Delta \hat{H} - \Delta \hat{S}\right)
\end{equation}
Here
\begin{equation} \label{eq: beta}
   k_B \beta
    = \frac{\langle \hat{H} \hat{S} \rangle - \langle \hat{H} \rangle \langle \hat{S} \rangle}
           {\langle \hat{H}^2 \rangle - \langle \hat{H} \rangle^2}
    = \frac{\sigma_{\hat{H}\hat{S}}}{\sigma_{\hat{H}\hat{H}}}
\end{equation}
where, $\sigma_{\hat{H}\hat{S}}=\mathrm{Cov}(\hat{H},\hat{S})$ and  $\sigma_{\hat{H}\hat{H}}=\mathrm{Var}(\hat{H})$ represent the covariance between the Hamiltonian and the entropy operators, and the variance of the Hamiltonian, respectively. In what follows $k_B = 1$ is considered.

At stable equilibrium, $\beta$, as given by Equation~(\ref{eq: beta}), converges to the thermodynamic inverse temperature. This follows from the equilibrium fluctuation-response relations for equilibrium states, which assert that covariances of observables are given by second derivatives of the Massieu potential, $\ln Z$, with respect to their conjugate variables.  In particular, the entropy operator is affine in the Hamiltonian in the canonical ensemble, leading to $\sigma_{\hat{H}\hat{S}}= k_B \, \beta\, \sigma_{\hat{H}\hat{H}}$. Here, the parameter $\beta$ is interpreted as a nonequilibrium inverse temperature.

A nonequilibrium free-energy operator can be written as
\begin{equation}\label{eq: free energy}
	\hat{F} = \hat{H} - \beta^{-1} \hat{S}
\end{equation}
so that Equation~(\ref{eq: D_simplified}) becomes as
\begin{equation}
	\tilde{D} = \beta \sqrt{\hat{\rho}}\, \Delta \hat{F}
\end{equation}
Accordingly, Equation~(\ref{eq: D_general}) can be written as
\begin{equation} \label{eq: D_final}
    \hat{D} = \frac{\beta}{2} \left\{ \Delta \hat{F}, \hat{\rho} \right\}
\end{equation}
Consequently, the SEAQT equation of motion, Equation~(\ref{eq:gen_seaqt}), takes the form
\begin{equation} \label{eq: seaqt_eq}
	\frac{d \hat{\rho}}{dt}
	= - \frac{i}{\hbar} \left[ \hat{H}, \hat{\rho} \right]
	- \frac{\beta}{2 \tau_D} \left\{ \Delta \hat{F}, \hat{\rho} \right\}
\end{equation}
Two remarks on Eq.~(\ref{eq: seaqt_eq}) are in order. First, $\beta$ enters only
through the product
\begin{equation}
\frac{\beta}{2}\bigl\{\Delta\Fh,\rhoh\bigr\}
=\frac{1}{2}\bigl\{\beta\,\Delta\Hh-\Delta\Sh,\,\rhoh\bigr\} ,
\label{eq:nonsingular}
\end{equation}
which is Eq.~(\ref{eq: D_simplified}) rewritten. The right-hand side is regular
as $\beta\to 0$, whereas the form involving $\Fh=\Hh-\beta^{-1}\Sh$ is not; the
two are algebraically identical wherever $\beta\neq 0$, and all results reported
below were obtained by integrating Eq.~(\ref{eq:nonsingular}).
 
Second, and consequently, an exactly pure state is a stationary point of the
dissipative dynamics. Because the entropy operator is defined with the range
projector $\Bh$ of Eq.~(\ref{eq: entropy}), a pure state has $\Sh=0$,
$\Delta\Sh=0$ and, by Eq.~(\ref{eq: beta}), $\beta=0$; the generator
$\beta\Delta\Hh-\Delta\Sh$ then vanishes identically, and since the symplectic
term preserves purity, the state does not relax. The initial conditions of
Sec.~\ref{subsec: initial conditions} are therefore taken as mixed states,
consistent with the finite entropy of the experimentally prepared states.

The generators of the motion adopted here, $\hat{I}$ and $\hat{H}$, correspond to
an isolated system. When the system is instead coupled to a reservoir, the
reservoir enters the equation of motion through an additional generator whose
effect, in the hypoequilibrium description \cite{beretta2026evolution}, is an additive constant contribution
fixed by the reservoir inverse temperature rather than a modification of the
dissipative geometry~\cite{damian2026hagedorn}. Since the case studied here
is isolated, that contribution is absent and $\langle\hat{H}\rangle$ is conserved
along the entire evolution. Finally, noting that
\begin{equation}
\frac{1}{2}\left\{ \Delta \hat{F}, \hat{\rho} \right\}
= \frac{1}{2}\left\{ \hat{F}, \hat{\rho} \right\} - \hat{\rho}\langle \hat{F} \rangle
\end{equation}
the SEAQT state dynamics can be interpreted as the combined evolution of the density operator under a symplectic contribution, compatible with Schr\"odinger dynamics, and a dissipative contribution driven by free-energy fluctuations. This equation of motion preserves the unitary trace and nonnegativity of the density operator \cite{beretta2009nonlinear,montanez2022decoherence}.

In the long-time limit, $t \to \infty$, the system relaxes towards a stable equilibrium state. At stable equilibrium, the dynamics becomes stationary, implying that
\begin{equation}
\frac{d \hat{\rho}}{dt} = [\hat{H}, \hat{\rho}] =  0
\end{equation}
and also,  the SEAQT dynamics coincides with a Gibbs state of the form
\begin{equation}
\hat{\rho}_{\mathrm{eq}}
= \frac{e^{-\beta_{\mathrm{eq}} \hat{H}}}{Z(\beta_{\mathrm{eq}})}
\end{equation}
where $Z(\beta_\mathrm{eq}) = \mathrm{Tr} (e^{-\beta\mathrm{eq} \hat H} )$, and $\beta_{\mathrm{eq}}$ is the inverse temperature characterizing the stable equilibrium state. As a consequence, at stable equilibrium the entropy variance is given as
\begin{equation} \label{eq: sigmaSS_eq}
\sigma_{\hat S \hat S}^{\mathrm{eq}}
= \beta_{\mathrm{eq}}^2 \sigma_{\hat H \hat H}^{\mathrm{eq}}
= \beta_{\mathrm{eq}}^2 \frac{\partial^2 \ln Z}{\partial \beta^2}
\end{equation}
In addition, the covariance between energy and entropy satisfies the relation
\begin{equation}
\left(\sigma_{\hat H \hat S}^{\mathrm{eq}}\right)^2
= \sigma_{\hat H \hat H}^{\mathrm{eq}}
  \sigma_{\hat S \hat S}^{\mathrm{eq}}
\end{equation}
reflecting the fact that entropy fluctuations at stable equilibrium are entirely determined by energy fluctuations. This relation allows the equilibrium inverse temperature parameter in the SEAQT formalism to be expressed as~\cite{beretta2009nonlinear}
\begin{equation}\label{eq: beta_eq}
\beta_{\rm eq}^{\rm SEAQT}
  =\operatorname{sgn}\!\left(\sigma^{\rm eq}_{\hat{H}\hat{S}}\right)
   \left(\frac{\sigma^{\rm eq}_{\hat{S}\hat{S}}}
              {\sigma^{\rm eq}_{\hat{H}\hat{H}}}\right)^{1/2} \,.
\end{equation}

It is observed that the SEAQT dynamics describe a trajectory towards a maximum entropy production path connecting an initial nonequilibrium state to a final stable equilibrium state. For isolated systems, this trajectory preserves the mean energy, whereas for systems coupled to a thermal reservoir, the dynamics drive the system towards a thermal equilibrium characterized by a fixed temperature of the reservoir.

The heat capacity can also be defined within the SEAQT framework by analogy with its equilibrium statistical-mechanical expression~\cite{tong2011statistical}
\begin{equation} \label{eq: Cv}
C = \beta^2 \sigma_{\hat H \hat H} = \frac{\sigma_{\hat H \hat S}^2}{\sigma_{\hat H \hat H}}
\end{equation}
which, at equilibrium, reduces to
\begin{equation}
C_{\mathrm{eq}}
= \sigma_{\hat S \hat S}^{\mathrm{eq}}
= \beta_{\mathrm{eq}}^2
  \frac{\partial^2 \ln Z}{\partial \beta^2}
\end{equation}
ensuring full consistency with the classical thermodynamic definition of heat capacity.

\subsection{Dynamical Order Parameter}\label{subsec:dinamparam}
In Markovian open quantum systems governed by the Gorini–Kossakowski–Sudarshan–Lindblad (GKSL) master equation, the quantum Mpemba effect can be characterized in terms of the spectral properties of the Liouvillian superoperator. In particular, the accelerated relaxation associated with this effect occurs when the initial state has a vanishing overlap with the slowest decaying mode. Denoting the left eigenoperator associated with this mode by \(\hat{L}_1\), a Mpemba state \(\hat{\rho}_{\mathrm{M}}\) satisfies
\begin{equation}
    \mathrm{Tr}\!\left(\hat{L}_1\hat{\rho}_{\mathrm{M}}\right)=0.
\end{equation}
This condition eliminates the contribution of the slowest relaxation mode in Eq.~\ref{eq: solution_Lindblad}. To distinguish genuinely anomalous relaxation from the trivial case of a state that is initially closer to equilibrium, the Mpemba state must also be compared with a reference initial state \(\hat{\rho}_{\mathrm{in}}\).

Within this framework, we define the scalar functional
\begin{equation} \label{eq: phi_gksl}
    \Phi_{\mathrm{GKSL}}(\hat{\rho})
    =
    \mathrm{Tr}\!\left(\hat{L}_1\hat{\rho}\right).
\end{equation}
A state \(\hat{\rho}_{\mathrm{M}}\) is identified as a Mpemba state when
\begin{equation}
    \Phi_{\mathrm{GKSL}}(\hat{\rho}_{\mathrm{M}})=0
\end{equation}
and
\begin{equation}
    D(\hat{\rho}_{\mathrm{M}},\hat{\rho}_{\mathrm{eq}})
    >
    D(\hat{\rho}_{\mathrm{in}},\hat{\rho}_{\mathrm{eq}}),
\end{equation}
where \(D\) denotes the Hilbert–Schmidt distance, \(\hat{\rho}_{\mathrm{eq}}\) is the equilibrium state, and \(\hat{\rho}_{\mathrm{in}}\) is a reference initial state. The quantity \(\Phi_{\mathrm{GKSL}}\) therefore acts as an order parameter that distinguishes two dynamical regimes: a normal relaxation regime, in which the dynamics are governed by the slowest decay rate, and an accelerated relaxation regime, in which the contribution of the slowest mode vanishes.

Unlike the GKSL master equation, the steepest-entropy-ascent quantum thermodynamics (SEAQT) equation is nonlinear in \(\hat{\rho}\) and, in general, does not admit a state-independent modal decomposition analogous to the Liouvillian spectrum. We therefore seek a coordinate in state space that identifies the sector associated with slow relaxation.

\ref{app:slow_population_order_parameter} presents an analysis of 100 randomly generated initial states that satisfy \(\Phi_{\mathrm{GKSL}}\approx 0\). In particular, the initial population associated with the state \(\lvert 2\rangle\) exhibits clustering around its corresponding equilibrium value. This behavior suggests that the relevant quantity is an effective slow coordinate dominated by the population of the weakly relaxing state \(\lvert 2\rangle\), with additional contributions from the coherent directions dynamically coupled to this population through the effective Hamiltonian.

This observation motivates the introduction of a Hermitian, Hamiltonian-dressed slow-coordinate operator \(\hat{Q}_2\) and the corresponding SEAQT order parameter
\begin{equation}
    \Phi_{\mathrm{SEAQT}}(\hat{\rho})
    =
    \left|
    \langle\hat{Q}_2\rangle_{\hat{\rho}}
    -
    p_2^{\mathrm{eq}}
    \right|,
\end{equation}
where
\begin{align}
    \hat{Q}_2(\boldsymbol{\alpha})
    &=
    \hat{\Pi}_2
    +
    \sum_{n=1}^{N}
    \alpha_n\mathcal{D}_H^{\,n}[\hat{\Pi}_2],
    \\
    \hat{\Pi}_2
    &=
    \lvert 2\rangle\langle 2\rvert,
    \\
    \mathcal{D}_H[\hat{A}]
    &=
    \frac{i}{\Omega_1}
    [\hat{H}_{\mathrm{eff}},\hat{A}],
    \\
    p_2^{\mathrm{eq}}
    &=
    \mathrm{Tr}\!\left(
    \hat{\rho}_{\mathrm{eq}}\hat{\Pi}_2
    \right),
\end{align}
and
\begin{equation}
    \langle\hat{Q}_2\rangle_{\hat{\rho}}
    =
    \mathrm{Tr}\!\left(\hat{\rho}\hat{Q}_2\right).
\end{equation}
Here, the coefficients \(\alpha_n\) are assumed to be real so that \(\hat{Q}_2\) remains Hermitian.

The SEAQT Mpemba condition can then be expressed as
\begin{equation}
    \Phi_{\mathrm{SEAQT}}(\hat{\rho}_{\mathrm{M}})
    \approx 0.
\end{equation}
This condition indicates that the effective slow coordinate is already close to its equilibrium value, even though the full state remains far from equilibrium. Consequently, the slow dynamical sector is effectively bypassed, whereas the remaining directions in state space may continue to relax through the fast dynamical channel.

A detailed derivation and justification of this order parameter are provided in \ref{app:slow_population_order_parameter}. It is also shown that
\begin{equation}
\Phi_{\GKSL}(\rhoh_{\mathrm{in}})\simeq 0
\;\Longleftrightarrow\;
\Phi_{\SEAQT}(\rhoh_{\mathrm{in}})\simeq 0.
\label{eq:equiv}
\end{equation}
thus, the proposed order parameter provides a means of identifying whether the relaxation governed by \(\tau_D\) can be described by a single regime or whether two distinct relaxation regimes must be considered, as discussed below.

\subsection{Entropy production and the thermodynamic definition of $\tau_D$}
\label{sec:tauD-def}
The quantity $\tau_D$ in Eq.~(\ref{eq: seaqt_eq}) is the internal relaxation
time, i.e.\ the time scale on which the system undergoes dissipative dynamics.
It has been treated either as a positive constant
\cite{beretta2009nonlinear,beretta2010maximum,younis2022predicting,yamada2019low,Cano_2015,McDonald_2023,saldana2025steepest}
or as a functional of $\hat{\rho}$
\cite{beretta2009nonlinear,beretta2010maximum,smith2016comparing,montanez2020loss,montanez2022decoherence}.
In the present work, rather than postulating a functional form for $\tau_D$, we extract it from the
entropy production relation implied by the equation of motion itself.
Taking the time derivative of $\langle\Sh\rangle=\mathrm{Tr}(\rhoh\Sh)$ with
$\rhoh$ evolving under Eq.~(\ref{eq: seaqt_eq}), the symplectic term does not contribute,
since $\Sh=-k_B\Bh\ln\rhoh$ commutes with $\rhoh$ and
$\mathrm{Tr}\!\left(\ln\rhoh\,[\Hh,\rhoh]\right)=0$. Only the dissipative
term survives, giving
\begin{equation}
\frac{d\langle \Sh\rangle}{dt}
= -\frac{\beta}{\tau_D}\,\sigma_{\Fh\Sh},
\qquad
\frac{d\langle \Hh\rangle}{dt}
= -\frac{\beta}{\tau_D}\,\sigma_{\Fh\Hh},
\label{eq:rates}
\end{equation}
where $\sigma_{\hat{X}\hat{Y}}=\tfrac{1}{2}\mathrm{Tr}\big(\rhoh\{\Delta\hat X,\Delta\hat Y\}\big)$.
The definition of $\beta$ in Eq.~(\ref{eq: beta}), is precisely the statement
$\sigma_{\Fh\Hh}=\sigma_{\Hh\Hh}-\beta^{-1}\sigma_{\Hh\Sh}=0$, so energy
conservation for the isolated system is automatic and not an additional
assumption. Using $\Fh=\Hh-\beta^{-1}\Sh$ together with $\sigma_{\Fh\Hh}=0$
yields $\sigma_{\Fh\Sh}=-\beta\,\sigma_{\Fh\Fh}$, and Eq.~\eqref{eq:rates}
collapses to the compact entropy production relation
\begin{equation}
\frac{d\langle \Sh\rangle}{dt}
=\frac{\beta^{2}\,\sigma_{\Fh\Fh}}{\tau_D}\;
\label{eq:entprod}
\end{equation}
which is manifestly non-negative, since $\sigma_{\Fh\Fh}\ge 0$ is a variance
and $\tau_D>0$; the second law is therefore satisfied identically, for any
$\tau_D$. Equivalently, substituting Eq.~(\ref{eq: beta}),
\begin{equation}
\beta^{2}\sigma_{\Fh\Fh}
=\sigma_{\Sh\Sh}-\beta^{2}\sigma_{\Hh\Hh}
=\sigma_{\Sh\Sh}-C ,
\label{eq:sigmaFF-closed}
\end{equation}
with $C$ the nonequilibrium heat capacity of Eq.~(\ref{eq: Cv}). Equation~(\ref{eq: sigmaSS_eq}), which
states $\sigma^{\mathrm{eq}}_{\Sh\Sh}=\beta_{\mathrm{eq}}^{2}\sigma^{\mathrm{eq}}_{\Hh\Hh}$,
is thus the statement that $\sigma_{\Fh\Fh}$ vanishes at stable equilibrium.
 
Rearranging Eq.~\eqref{eq:entprod} identifies $\tau_D$ not as a fitting
parameter but as the ratio of the thermodynamic driving force to the actual
dissipation rate,
\begin{equation}
\;\tau_D(t)=\frac{\beta^{2}(t)\,\sigma_{\Fh\Fh}(t)}{\;d\langle \Sh\rangle/dt\;}
=\frac{\sigma_{\Sh\Sh}(t)-C(t)}{\;d\langle \Sh\rangle/dt\;}
\label{eq:tauD-exact}
\end{equation}
Every quantity on the right-hand side is a state functional already computed
along the trajectories of Section ~\ref{sec:Res}, so Eq.~\eqref{eq:tauD-exact} converts the
optimization-based parametrization into a measurable object.
 
\subsection{The $0/0$ indetermination at the metastable state}
\label{sec:tauD-indet}
 
For the initial condition $\rhoh_{\mathrm{in}}=|\mathrm{sME}\rangle\langle
\mathrm{sME}|$ the entropy develops the plateau visible in Fig.~\ref{fig: E-S dynamics tau(t)}c,
corresponding to a partially canonical metastable state $\rhoc$ in which the
population of the weakly coupled state $|2\rangle$ is still far from its
equilibrium value. Because the entropy operator is defined with the
range projector $\Bh$ of Eq.~(\ref{eq: B_projector}), $\rhoc$ is canonical \emph{within the
subspace it occupies}, so that
\begin{equation}
\Delta\Fh\big|_{\rhoc}=0
\quad\Longrightarrow\quad
\sigma_{\Fh\Fh}\big|_{\rhoc}=0
\quad\text{and}\quad
\frac{d\langle \Sh\rangle}{dt}\bigg|_{\rhoc}=0 .
\label{eq:indet}
\end{equation}
Both the numerator and the denominator of Eq.~\eqref{eq:tauD-exact} therefore
become small at $\rhoc$. The cancellation is exact only when $\Heff$ is block
diagonal with respect to the partition
$\mathrm{span}\{|0\rangle,|1\rangle\}\oplus\mathrm{span}\{|2\rangle\}$; for the
effective Hamiltonian obtained here the residual couplings
$H^{\mathrm{eff}}_{02}=0.030\,\Omega_1$ and
$H^{\mathrm{eff}}_{12}=-0.106\,\Omega_1$ leave a remainder of order
$|H^{\mathrm{eff}}_{12}|^{2}$, so that $\tau_D$ at the metastable state is large
but finite rather than strictly indeterminate. Its value there is fixed not by
the state alone but by the direction along which the trajectory enters and
leaves it. This is what regularizes the step over a finite crossover width and
justifies the logistic parametrization of Sec.~\ref{sec:tauD-logistic} on
physical rather than numerical grounds; the idealized $0/0$ limit is recovered
as $H^{\mathrm{eff}}_{12}\to 0$.

Let $t_{\mathrm{c}}$ denote the time at which the trajectory passes through
$\rhoc$ and write $\rhoh(t)=\rhoc+\delta\rhoh(t)$. Expanding to second order,
\begin{align}
\langle \Sh\rangle &= S_{\mathrm{c}}-\tfrac{1}{2}\,\mathcal{G}(\delta\rhoh,\delta\rhoh)+\mathcal{O}(\delta^{3}),
\label{eq:Gexp}\\
\beta^{2}\sigma_{\Fh\Fh} &= \mathcal{Q}(\delta\rhoh,\delta\rhoh)+\mathcal{O}(\delta^{3}),
\qquad \mathcal{Q}\succeq 0 ,
\label{eq:Qexp}
\end{align}
where the quadratic form $\mathcal{Q}$ is positive semidefinite while
$\mathcal{G}$ is \emph{not} sign definite: $\rhoc$ maximizes the entropy only on
the fast manifold, so $\mathcal{G}>0$ along the fast directions and
$\mathcal{G}<0$ along the slow direction through which the system eventually
escapes. The linearized dissipative dynamics of the three-level system has two
well-separated channels: the approach to $\hat{\rho}^{\rm c}$ proceeds along a
fast direction $\hat{v}_f$ at rate $\kappa_1$, and the departure along the slow
direction $\hat{v}_s$ at rate $\kappa_2\ll\kappa_1$. Writing
$\delta\rhoh(t)=c\,e^{\lambda (t-t_{\mathrm{c}})}\hat{v}$ and substituting
Eqs.~\eqref{eq:Gexp}--\eqref{eq:Qexp} into Eq.~\eqref{eq:tauD-exact}, the
factors $c^{2}e^{2\lambda(t-t_{\mathrm c})}$ cancel identically and
\begin{equation}
\tau_D=-\frac{\mathcal{Q}(\hat{v})}{\lambda\,\mathcal{G}(\hat{v})} .
\label{eq:tauD-limit}
\end{equation}
The two one-sided limits are therefore finite and unequal,
\begin{equation}
\tau^{-}_{D}=\frac{\mathcal{Q}(\hat{v}_{f})}{\kappa_1\,\mathcal{G}(\hat{v}_{f})},
\qquad
\tau^{+}_{D}=\frac{\mathcal{Q}(\hat{v}_{s})}{\kappa_2\,|\mathcal{G}(\hat{v}_{s})|},
\label{eq:tauD-pm}
\end{equation}
so that the indetermination is resolved by a genuine step of height
\begin{equation}
\frac{\tau^{+}_{D}}{\tau^{-}_{D}}
=\frac{|\lambda_f|}{|\lambda_s|}\,\mathfrak{r} ,
\qquad
\mathfrak{r}\equiv
\frac{\mathcal{Q}(\hat{v}_{s})}{|\mathcal{G}(\hat{v}_{s})|}\,
\frac{\mathcal{G}(\hat{v}_{f})}{\mathcal{Q}(\hat{v}_{f})} ,
\label{eq:step}
\end{equation}
where $\lambda_f$ and $\lambda_s$ are the linearized relaxation rates along
$\hat{v}_f$ and $\hat{v}_s$. These are not the bare loss rates: the Liouvillian
of the effective model has no eigenvalue at $-\kappa_1$, and its slowest nonzero
eigenvalue, $\lambda_s=-0.01086\,\Omega_1$, exceeds $\kappa_2$ by almost an
order of magnitude, because the Feshbach-induced coupling
$H^{\mathrm{eff}}_{12}$ feeds the population of $|2\rangle$ into the fast
channel. The two ratios coincide only in the limit of vanishing coherent
coupling between $|2\rangle$ and the remaining levels.
 
The rate ratio $|\lambda_f|/|\lambda_s|$ is a property of the dissipative channels
alone, whereas the geometric factor $\mathfrak{r}$ depends on the curvature of
the entropy and free-energy-variance surfaces along the two directions, and
hence on which fast direction the particular initial condition follows into
$\rhoc$. This is the reason the upper limit $\tau^{+}_{D}$ is essentially
common to all trajectories that visit $\rhoc$, it is controlled by
$\hat{v}_{s}$, a property of $\rhoc$, while the lower limit $\tau^{-}_{D}$
is initial-condition dependent.
 
\subsection{Finite crossover and the logistic parametrization}
\label{sec:tauD-logistic}
 
The discontinuous change characterized by Eq.~\eqref{eq:step} is idealized:
the trajectory does not reach $\hat{\rho}_{\mathrm{c}}$ exactly, and its
passage through the metastable region occurs over a finite characteristic
timescale $\delta t$, set by the competition between the residual fast
relaxation and the onset of the slow channel. Smoothing this transition over
$\delta t$ gives the two-asymptote logistic form
\begin{equation}
\tau_D(t)=\tau_D^{-}
+\frac{\tau_D^{+}-\tau_D^{-}}
{1+\exp\!\left[-(\omega_4+\omega_5 t)\right]},
\label{eq:logistic4}
\end{equation}
with the parameter identifications
\begin{equation}
\omega_0\equiv\tau_D^{-},
\qquad
\omega_3\equiv\tau_D^{+}-\tau_D^{-},
\qquad
\omega_5\equiv\delta t^{-1},
\qquad
\omega_4\equiv-\omega_5 t_{\mathrm{c}}.
\label{eq:ident}
\end{equation}
We fit Eq.~\eqref{eq:logistic4} using the four free parameters
$\omega_0$, $\omega_3$, $\omega_4$, and $\omega_5$. For $\omega_5>0$,
$\omega_0=\tau_D^{-}$ and $\omega_0+\omega_3=\tau_D^{+}$ are the early-
and late-time limits, respectively, while
$\omega_3=\tau_D^{+}-\tau_D^{-}$ is the crossover amplitude. Furthermore,
$t_{\mathrm{c}}=-\omega_4/\omega_5$ is the midpoint of the crossover, at
which
\begin{equation}
\tau_D(t_{\mathrm{c}})
=\frac{\tau_D^{-}+\tau_D^{+}}{2},
\end{equation}
and $\omega_5^{-1}$ is its characteristic timescale.
 
Two consequences follow. First, $\omega_0+\omega_3=\tau^{+}_{D}$ is a property
of $\rhoc$ and of the slow channel, and is therefore expected to be common to
initial conditions that transit the metastable state, whereas $\omega_0$ need
not be. Second, the four coefficients are jointly identifiable only when the
transit occurs well inside the measured time window: if $t_{\mathrm{c}}$ lies
near or beyond the last measured time, the fit constrains only one of the two
one-sided limits and no step height can be extracted. Both situations are
encountered in Sec.~\ref{sec:Res}.
 
It should be emphasized that satisfying the Mpemba condition does not by itself
imply that the metastable state is bypassed. The suppression of the slow mode,
Eqs.~(\ref{eq:equiv}), accelerates the approach to the final state, but the
trajectory may still pass through the metastable region and leave it through the
fast channel, as is found here for $|\mathrm{sME}\rangle$.

\subsection{Feshbach projection}\label{subsec:feshbach}

For describing the Mpemba effect, the SEAQT framework describes the state evolution of a three-state isolated system, yielding results comparable to the experimental data of \cite{zhang2025observation}. To achieve this, the four-state Hilbert space is reduced to an effective three-state Hilbert space. This reduction takes into account the coupling between the $|1\rangle$ and $|2\rangle$ states with the $|P\rangle$ state, without explicitly describing the dynamics of $|P\rangle$ because it represents a short-lived state.

A complete Hamiltonian for the four-level system is written as
\begin{equation}
\hat H = \frac{1}{2}
\begin{bmatrix}
0 & \Omega_1 & \Omega_2 & 0 \\
\Omega_1 & 0 & 0 & \Omega_{1P} \\
\Omega_2 & 0 & 0 & \Omega_{2P} \\
0 & \Omega_{1P} & \Omega_{2P} & \Delta
\end{bmatrix}
\end{equation}
where $\Omega_{iP}$ represent the couplings of $|P\rangle$ with $|1\rangle$ and $|2\rangle$, and $\Delta$ represents the detuning between $|0\rangle$ and $|P\rangle$.

Partitioning the Hilbert space as
\begin{equation}
\mathcal{H} = \operatorname{span}\{|0\rangle,|1\rangle,|2\rangle\}\oplus\operatorname{span}\{|P\rangle\} = \mathcal{H}_S \oplus \mathcal{H}_P
\end{equation}
the Hamiltonian can be written in a block form such as
\begin{equation}
\hat H_\mathrm{block} = 
\begin{bmatrix}
\hat H_S & \hat H_{SP} \\
\hat H_{PS} & \hat H_P
\end{bmatrix}
\end{equation}
where
\begin{align}
\hat H_S &= \frac{1}{2}
\begin{bmatrix}
0 & \Omega_1 & \Omega_2 \\
\Omega_1 & 0 & 0 \\
\Omega_2 & 0 & 0
\end{bmatrix} \\
\hat H_P &= \frac{1}{2}\Delta \\
\hat H_{SP} &= \frac{1}{2}
\begin{bmatrix}
0 \\ \Omega_{1P} \\ \Omega_{2P}
\end{bmatrix} \\
\hat H_{PS} &= \frac{1}{2}
\begin{bmatrix}
0 & \Omega_{1P} & \Omega_{2P}
\end{bmatrix}
\end{align}

If the time-independent Schr\"odinger equation is considered,

\begin{equation}\label{Eq:Shrod}
    \hat{H}|\psi\rangle = E|\psi\rangle
\end{equation}
 where
\begin{align}
    \hat H = \hat{H}_\mathrm{block} &=
    \begin{bmatrix}
       \hat H_S & \hat H_{SP} \\
        \hat H_{PS} & \hat H_P
    \end{bmatrix} \\
    |\psi\rangle &=
    \begin{bmatrix}
        |\psi_S\rangle \\
        |\psi_P\rangle
    \end{bmatrix}
\end{align}
therefore
\begin{align}
    \begin{bmatrix}
       \hat H_S & \hat H_{SP} \\
        \hat H_{PS} & \hat H_P
    \end{bmatrix}
    \begin{bmatrix}
        |\psi_S\rangle \\
        |\psi_P\rangle
    \end{bmatrix} &= 
    E
    \begin{bmatrix}
        |\psi_S\rangle \\
        |\psi_P\rangle
    \end{bmatrix}
\end{align}
which can also be represented as 
\begin{align}
    \hat H_S|\psi_S\rangle + \hat H_{SP}|\psi_P\rangle &= E|\psi_S\rangle \label{eq:psiSt} \\
    \hat H_{PS}|\psi_S\rangle + \hat H_{P}|\psi_P\rangle &= E|\psi_P\rangle \label{eq:psiPt}
\end{align}
Rearranging Equation~(\ref{eq:psiPt}) gives
\begin{equation} \label{eq:time-psiP}
    \left(\hat H_P-E \right) |\psi_P\rangle = - \hat H_{PS}|\psi_S\rangle
\end{equation}

Defining $\epsilon=\hat{H}_P-E$, $|\psi_P\rangle$ becomes
\begin{equation}
    |\psi_P\rangle=-\epsilon^{-1}H_{PS}|\psi_S\rangle
\end{equation}
and substituting into Equation~(\ref{eq:psiSt}) and grouping terms gives
\begin{equation}
    \left(\hat H_S - \hat H_{SP}\ \epsilon^{-1} \hat H_{PS}\right)|\psi_S\rangle = E|\psi_S\rangle
\end{equation}
Thus, it is obtained an effective Hamiltonian of the form
\begin{align}
    \hat H_{\mathrm{eff}} &= \hat H_S - \hat H_{SP}\ \epsilon^{-1}\hat H_{PS} \\
    \hat H_{\mathrm{eff}} &= \frac{1}{2}
    \begin{bmatrix}
        0 & \Omega_1 & \Omega_2 \\
        \Omega_1 & -\dfrac{\Omega_{1P}^2}{2\epsilon} & -\dfrac{\Omega_{1P}\Omega_{2P}}{2\epsilon} \\
        \Omega_2 & -\dfrac{\Omega_{1P}\Omega_{2P}}{2\epsilon} & -\dfrac{\Omega_{2P}^2}{2\epsilon}
    \end{bmatrix}
\end{align}
and substituting into Equation (\ref{Eq:Shrod}),
\begin{equation}
    \hat H_{\mathrm{eff}}|\psi_S\rangle = E |\psi_S \rangle
\end{equation}

At this point three unknown parameters, i.e., $\Omega_{1P}$, $\Omega_{2P}$, and $\epsilon$, are observed. According to the supplementary notes of \cite{zhang2025observation},
\begin{equation}
    \kappa_1 \approx \frac{\Omega_{1P}^2}{\gamma}
\end{equation}
where $\gamma$ is the emission rate for the transition $|P\rangle\rightarrow |0\rangle$. Then, according with the values provided in \cite{zhang2025observation},
\begin{equation}
    \kappa_1 = 2\Omega_1 \approx \frac{\Omega_{1P}^2}{\gamma}
\end{equation}
therefore
\begin{equation} \label{eq: kappa1}
     \quad \Omega_{1P} \approx \sqrt{2\Omega_1\gamma}
\end{equation}
In a similar way it can be observed that
\begin{equation} \label{eq: kappa2}
    \Omega_{2P}\approx \sqrt{0.0015\Omega_1\gamma}
\end{equation}

Now, introducing the proportionality constants $a$ and $b$,
\begin{equation} \label{eq: Omegas}
    \Omega_{1P} = \sqrt{\omega_{1P}\,\Omega_1}, \quad \Omega_{2P} = \sqrt{\omega_{2P}\,\Omega_1}
\end{equation}
where
\begin{equation}
    \omega_{1P} = 2a\gamma, \quad \omega_{2P} = 0.0015\,b\gamma
\end{equation}

Thus, the effective Hamiltonian can be rewritten as
\begin{equation}
    H_{\mathrm{eff}} = \frac{1}{2}
    \begin{bmatrix}
        0 & \Omega_1 & \Omega_2 \\
        \Omega_1 & -\dfrac{\omega_{1P}}{2\epsilon}\,\Omega_{1} & -\dfrac{\sqrt{\omega_{1P}\omega_{2P}}}{2\epsilon}\,\Omega_1 \\
        \Omega_2 & -\dfrac{\sqrt{\omega_{1P}\omega_{2P}}}{2\epsilon}\,\Omega_1 & -\dfrac{\omega_{2P}}{2\epsilon}\,\Omega_{1}
    \end{bmatrix}
\end{equation}

Defining
\begin{align}
    \omega_1 &= \frac{\omega_{1P}}{2\epsilon}, &
    \omega_2 &= \frac{\omega_{2P}}{2\epsilon}, &
    \sqrt{\omega_1\omega_2} &= \frac{\sqrt{\omega_{1P}\omega_{2P}}}{2\epsilon}
\end{align}
the resultant effective Hamiltonian becomes
\begin{equation}
    H_{\mathrm{eff}} = \frac{1}{2}
    \begin{bmatrix}
        0 & \Omega_1 & \Omega_2 \\
        \Omega_1 & -\omega_1\Omega_{1} & -\sqrt{\omega_1\omega_2}\,\Omega_1 \\
        \Omega_2 & -\sqrt{\omega_1\omega_2}\,\Omega_1 & -\omega_2\Omega_1
    \end{bmatrix}
\end{equation}

When the von Neumann, Lindblad or SEAQT equations are considered dimensionless, and with $\Omega_2=0.06\Omega_1$, the effective dimensionless Hamiltonian becomes
\begin{equation}
    H_{\mathrm{eff}} = \frac{1}{2}
    \begin{bmatrix}
        0 & 1 & 0.06 \\
        1 & -\omega_1 & -\sqrt{\omega_1\omega_2} \\
        0.06 & -\sqrt{\omega_1\omega_2} & -\omega_2
    \end{bmatrix}
    \label{eq: H_eff}
\end{equation}
where the unknown parameters $\omega_1$ and $\omega_2$ can be obtained by using an optimization method.


\subsection{Initial conditions}\label{subsec: initial conditions}
The initial density operator is constructed from the three basis states
\begin{equation}
|0\rangle=[1\,0\,0]^{\rm T},\qquad |1\rangle=[0\,1\,0]^{\rm T},\qquad
|2\rangle=[0\,0\,1]^{\rm T},
\end{equation}
as the mixed state
\begin{equation}
\hat{\rho}_{\rm in}=(1-\epsilon)\,|\psi\rangle\langle\psi|
                    +\tfrac{\epsilon}{3}\,\hat{I},
\qquad \epsilon=10^{-3},
\end{equation}
with $|\psi\rangle=p_0|0\rangle+p_1|1\rangle+p_2|2\rangle$. The coefficients $p_0$, $p_1$, and $p_2$ are chosen to match the population data observed experimentally in \cite{zhang2025observation}. Three initial conditions are considered, and are given in Table \ref{tab:initial_conditions}.

\begin{table}[!htbp]
\caption{Initial conditions, specified by the populations measured in
\cite{zhang2025observation} renormalized to unit trace. The states are taken as
mixed, $\hat{\rho}_{\rm in}=(1-\epsilon)\hat{\rho}_{\rm meas}+\epsilon\hat{I}/3$
with $\epsilon=10^{-3}$; $|{\rm sME}\rangle$ additionally carries the coherences
obtained from the optimization described below.}
\label{tab:initial_conditions}
\centering
\begin{tabular}{lccc}
\hline
$\hat{\rho}_{\rm in}$ & $p_0$ & $p_1$ & $p_2$ \\
\hline
$\hat{\rho}^{(0)}_{\rm in}$ & 0.9494 & 0.0127 & 0.0380 \\
$\hat{\rho}^{(2)}_{\rm in}$ & 0.0356 & 0.0032 & 0.9612 \\
$|{\rm sME}\rangle\langle {\rm sME}|$ & 0.6575 & 0.1130 & 0.2295 \\
\hline
\end{tabular}
\end{table}

For the case $\hat{\rho}_{\mathrm{in}} = |\mathrm{sME}\rangle\langle
\mathrm{sME}|$, the coefficients are obtained from an optimization subject to
the constraints: 1) $\||\mathrm{sME}\rangle\|=1$, 2)
$\mathrm{Tr}(\hat{\rho}_{\mathrm{sME}})=1$, and 3)
$\mathrm{Tr}(\hat{L}_1\hat{\rho}_{\rm sME})=0$, where $\hat{L}_1$ is the left
Liouvillian eigenmatrix associated with the relaxation mode $\lambda_1$. The regularizing mixing $\epsilon$
was varied over $10^{-4}$--$10^{-2}$ with no significant change in the results
reported below.

\ref{app:A} provides a more detailed study of the state dynamics using different initial states, $|\mathrm{sME}\rangle$, constructed by using different sets of coefficients $p_i$.


\subsection{Optimization method}\label{subsec:optimization}

The optimization minimizes the reduced $\chi$-square
\begin{equation}
\chi^2_\nu=\frac{1}{N}\sum_{i}\sum_{k=0}^{2}
\frac{\bigl[P^{\mathrm{model}}_{|k\rangle}(t_i)-P^{\mathrm{exp}}_{|k\rangle}(t_i)\bigr]^2}
{\sigma^2_{k}(t_i)} ,
\label{eq:chi2}
\end{equation}
where $\sigma_k(t_i)$ are the experimental uncertainties reported in
\cite{zhang2025observation} and each predicted population is obtained from
Eq.~(\ref{eq: Population}) for every $\hat{\rho}(t)$ given by
Eq.~(\ref{eq:nonsingular}), $P_{|i\rangle}^\mathrm{exp}(t)$ and $P_{|i\rangle}^\mathrm{model}(t;\omega_i)$ represent the experimental measurements and the populations predicted of the state $|i\rangle$ at time $t$, respectively. Each predicted population is calculated as
 \begin{equation}
    P_{|i\rangle}^\mathrm{model}(t;\omega_i)=\mathrm{Tr}(\hat{\rho}(t) |i\rangle\langle i|) \,.
    \label{eq: Population}
 \end{equation}
 
The measured populations are renormalized to unit
trace at each time, since a trace-preserving model cannot reproduce populations
whose sum departs from unity.
 
Because $\omega_1$ and $\omega_2$ parametrize the effective Hamiltonian of
Eq.~(\ref{eq: H_eff}), they cannot depend on the initial state. They are
therefore fitted jointly to all three data sets, while the four coefficients of
$\tau_D$ in Eq.~(\ref{eq:logistic4}) are fitted per initial condition. The
minimization alternates between the two blocks: with $\omega_1,\omega_2$ held
fixed, each $\tau_D$ is optimized independently by differential evolution
followed by a Nelder--Mead refinement in the reparametrization
$(\ln\tau^{-}_{D},\,\ln\tau^{+}_{D},\,t_{\mathrm{c}},\,\ln\omega_5)$, which
removes the strong correlation between $\omega_4$ and $\omega_5$ and the several
decades spanned by the relaxation times; the shared pair is then refitted
against the three updated $\tau_D$. Sweeps are iterated until $\chi^2_\nu$
changes by less than $0.5\%$. Both algorithms are those implemented in the
Wolfram Language, with fixed random seeds; the search bounds were
$\tau^{-}_{D}\in[6\times10^{-6},55]$, $\tau^{+}_{D}\in[0.14,403]$,
$t_{\mathrm{c}}\in[-50,300]$ and $\omega_5\in[0.018,55]$.
 
An unconstrained fit in which $\omega_1$ and $\omega_2$ are also allowed to vary
per initial condition reaches a lower mean $\chi^2_\nu$ of $12$, but returns
values of $\omega_2$ differing by more than $50\%$ between initial conditions.
Since the effective Hamiltonian cannot depend on the preparation of the state,
that fit is not admissible and the shared-parameter results are the ones
reported. Energy conservation provides an independent check on the integration:
along all three trajectories the relative drift of $\langle\Hh\rangle$ over six
decades in time does not exceed $4\times10^{-13}$.
 
A comment on the Hermiticity of $\Heff$ is in order. Because $|P\rangle$ is
short lived, the exact Feshbach denominator is complex,
$\epsilon=\hat H_P-E-i\gamma/2$, and its anti-Hermitian part is precisely what
generates the loss rates $\kappa_1$ and $\kappa_2$ of Eq.~\eqref{eq: kappa1}--\eqref{eq: kappa2}. In the present
treatment $\epsilon$ is taken real, so that $\Heff$ of Eq.~\eqref{eq: H_eff} is Hermitian and
the SEAQT evolution it generates conserves energy and probability. The
irreversibility associated with the auxiliary level is not discarded: it is
carried by the dissipative term of Eq.~\eqref{eq: seaqt_eq}, whose strength is set by
$\tau_D$ and, through Eqs.~\eqref{eq:tauD-pm} and \eqref{eq:step}, by the same
rates $\kappa_1$ and $\kappa_2$. The Feshbach projection is therefore used here
to fix the level shifts induced by $|P\rangle$, the coefficients $\omega_1$
and $\omega_2$, while the population loss it induces is described
thermodynamically rather than by a non-Hermitian generator. This is what allows
the SEAQT model to reproduce the measured populations with a strictly conserved
energy expectation value (Fig.~\ref{fig: E-S dynamics tau(t)}b), in contrast with the GKSL description in
which $\langle\Hh\rangle$ decays (Fig.~\ref{fig: E-S dynamics Lindblad}b).

\section{\label{sec:Res}Results and Discussion}
\subsection{\label{subsec:tauD-t}SEAQT case for $\tau_D = \tau_D(t)$}
Table~\ref{tab:00} gives the optimized coefficients for the three initial
conditions. The two effective-Hamiltonian coefficients, $\omega_1=2.9257$ and
$\omega_2=0.01544$, are common to all three by construction, as required by
Eq.~(\ref{eq: H_eff}); the four coefficients of $\tau_D(t)$ differ per initial
condition. The corresponding reduced chi-squares are $29.20$, $12.89$ and
$10.78$ ($\nu=212$), the residual being dominated by the experimental scatter at
long times and, for $\hat{\rho}^{(0)}_{\rm in}$, by the amplitude of the
coherent oscillations over $1\lesssim t\,\Omega_1\lesssim 30$, where the model
oscillates with roughly three times the measured amplitude about the correct
mean.

 \begin{table*}[t]
\centering
\begin{threeparttable}
\caption{Optimized coefficients of
$\tau_D(t)=\omega_0+\omega_3/
[1+\exp[-(\omega_4+\omega_5 t)]]$, with $\omega_1$ and $\omega_2$
fitted jointly to the three data sets.
The quantity $t_{\mathrm{c}}=-\omega_4/\omega_5$ is the midpoint of the
crossover through the metastable region, and $\chi_\nu^2$ is the reduced
chi-square ($\nu=212$).}
\label{tab:00}

\small
\setlength{\tabcolsep}{5pt}
\begin{tabular*}{\textwidth}{
    @{\extracolsep{\fill}}lcccccccc@{}
}
\toprule
$\hat{\rho}_{\mathrm{in}}$
& $\omega_1$
& $\omega_2$
& $\omega_0$
& $\omega_3$
& $\omega_4$
& $\omega_5$
& $t_{\mathrm{c}}$
& $\chi_\nu^2$ \\
\midrule
$\hat{\rho}^{(0)}_{\mathrm{in}}$
& 2.9257
& 0.01544
& 21.868
& 88.184
& $-121.98$
& 3.4600
& 35.25
& 29.20 \\

$\hat{\rho}^{(2)}_{\mathrm{in}}$
& 2.9257
& 0.01544
& $<10^{-5}$\tnote{a}
& 65.457
& $-1.7326$
& 0.07505
& 23.09
& 12.89 \\

$|\mathrm{sME}\rangle\langle\mathrm{sME}|$
& 2.9257
& 0.01544
& 1.1027
& 394.35\tnote{b}
& $-12.901$
& 2.6869
& 4.801
& 10.78 \\
\bottomrule
\end{tabular*}

\begin{tablenotes}[flushleft]
\footnotesize
\item[a] At the lower edge of the search domain
$\tau_D^{-}\in[6\times10^{-6},55]$; the measurement window does not
resolve the early-time dissipation.
\item[b] At the upper edge of the search domain
$\tau_D^{+}\in[0.14,403]$; the slow channel has not completed its
evolution by the last measured time.
\end{tablenotes}
\end{threeparttable}
\end{table*}

Table~\ref{tab:step} collects the one-sided limits and the geometric factor
$\mathfrak{r}$ of Eq.~(\ref{eq:step}). The step height is reportable only for
$\hat{\rho}^{(0)}_{\rm in}$: for $\hat{\rho}^{(2)}_{\rm in}$ the fit drives
$\tau^{-}_{D}$ to the lower end of the search domain and for
$|\mathrm{sME}\rangle$ it drives $\tau^{+}_{D}$ to the upper end, so in both
cases the data constrain only one of the two limits. For the one resolvable
case, $\mathfrak{r}=0.032$: the curvature of the entropy surface along the slow
direction reduces the step well below the ratio of the relaxation rates.

\begin{table}[!htbp]
\caption{One-sided limits of the relaxation time and geometric factor
$\mathfrak{r}$ of Eq.~(\ref{eq:step}), normalized by the ratio of the linearized
relaxation rates $|\lambda_f|/|\lambda_s|=1.7208/0.010860=158.46$.}
\label{tab:step}
\centering
\begin{tabular}{lccccc}
\hline
$\hat{\rho}_{\rm in}$ & $\tau_D^{-}$ & $\tau_D^{+}$ &
$\tau_D^{+}/\tau_D^{-}$ & $\mathfrak{r}$ \\
\hline
$\hat{\rho}^{(0)}_{\rm in}$  & $21.868$        & $110.05$        & $5.033$ & $0.0318$ \\
$\hat{\rho}^{(2)}_{\rm in}$  & $<10^{-5}$\,$^{a}$ & $65.46$      & ---     & ---      \\
$|{\rm sME}\rangle\langle{\rm sME}|$ & $1.1027$ & $>395$\,$^{b}$ & ---     & ---      \\
\hline
\end{tabular}
\end{table}

\begin{figure}[!htb]
    \centering
    a)\includegraphics[width=0.7\linewidth]{ 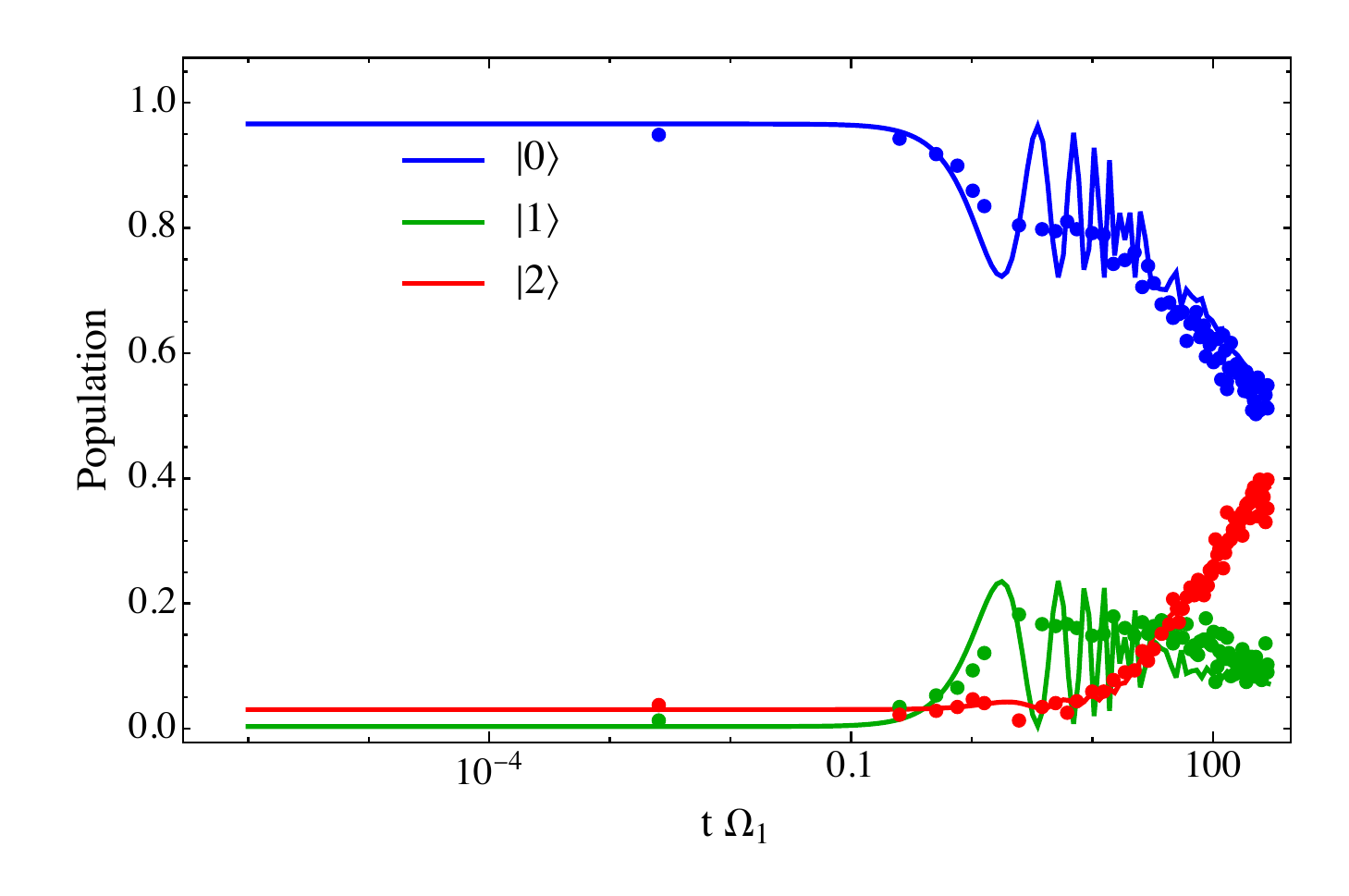}
    b)\includegraphics[width=0.7\linewidth]{ 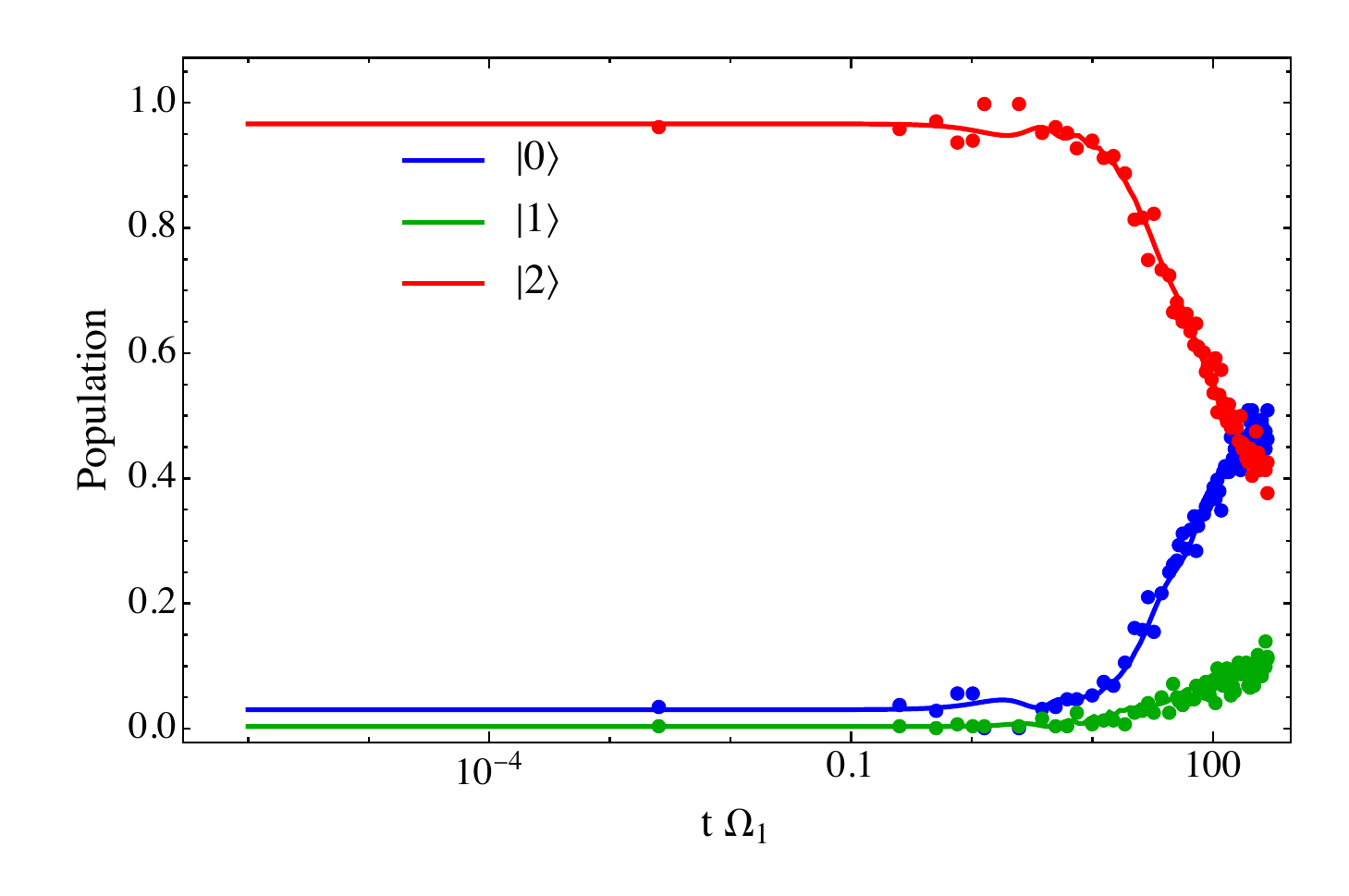}
    c)\includegraphics[width=0.7\linewidth]{ 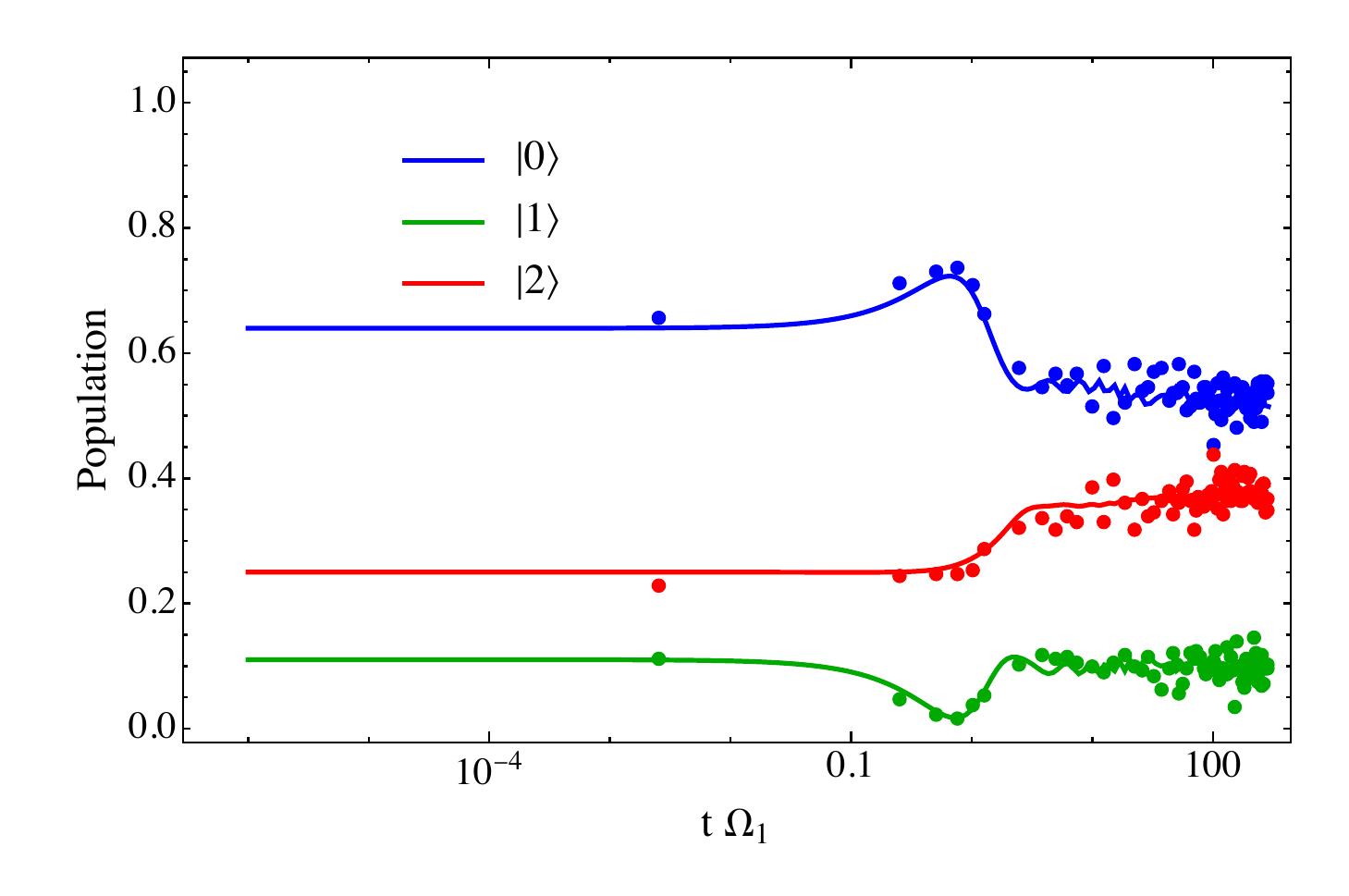}
    \caption{Population dynamics of the SEAQT model with $\tau_D = \tau_D(t)$ for the three initial conditions: a) $|0\rangle\langle0|$, b)$|2\rangle\langle2|$, and c)$|\mathrm{sME}\rangle\langle\mathrm{sME}|$.}
    \label{fig: Population_Dynamics_sq_tau(t)}
\end{figure}

\begin{figure}[!htb]
    \centering
    a)\includegraphics[width=0.7\linewidth]{ 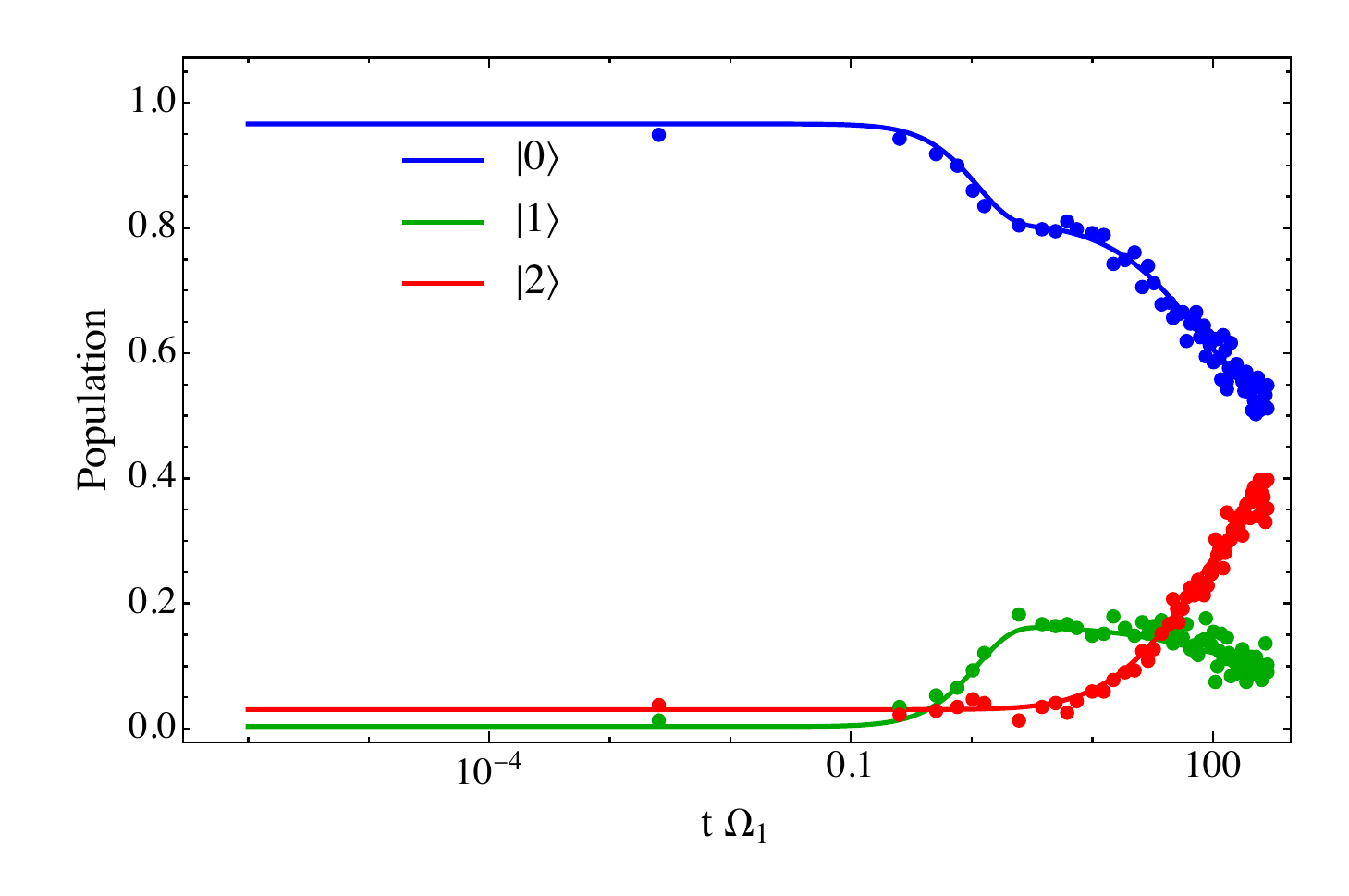}
    b)\includegraphics[width=0.7\linewidth]{ 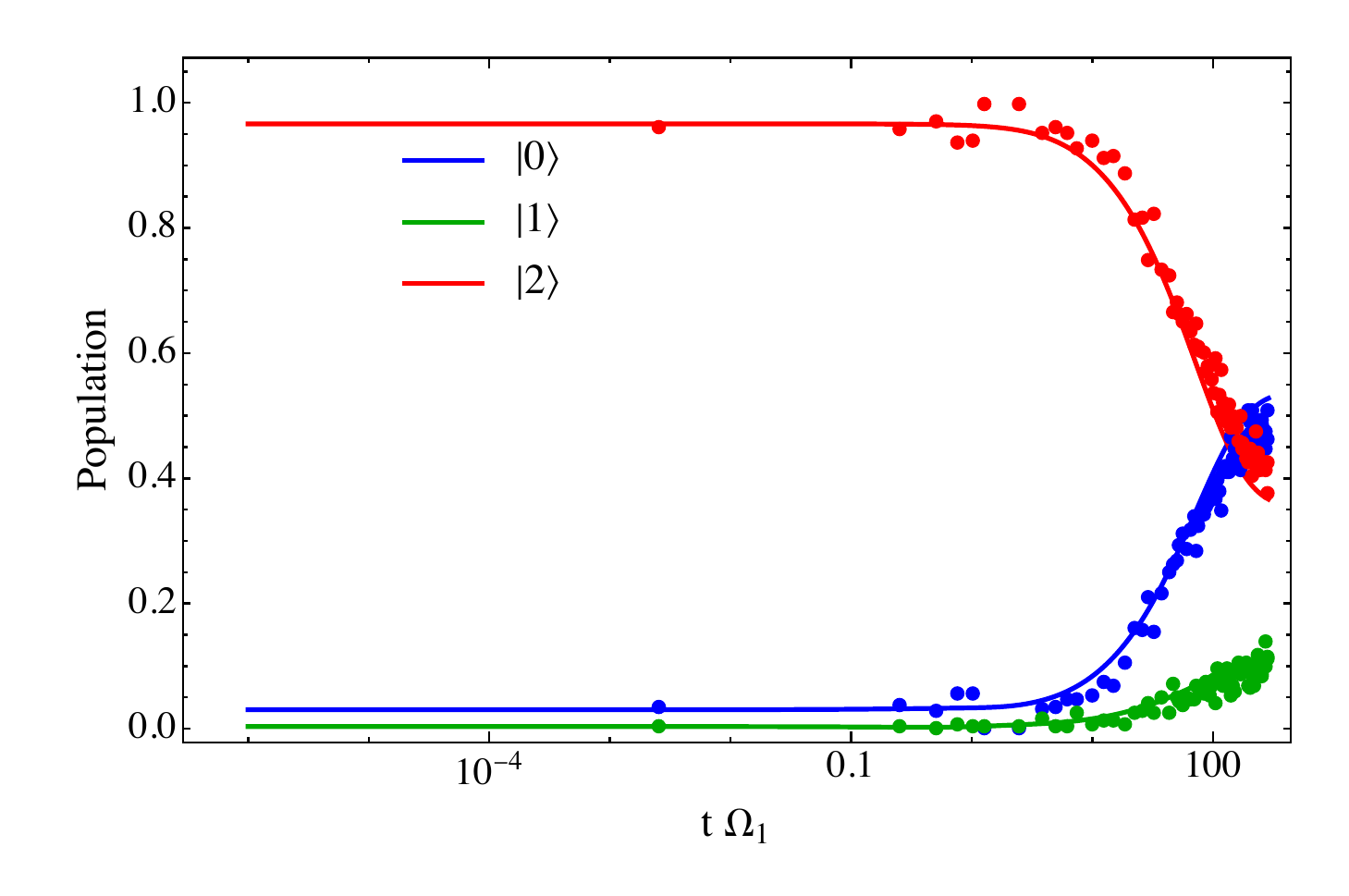}
    c)\includegraphics[width=0.7\linewidth]{ 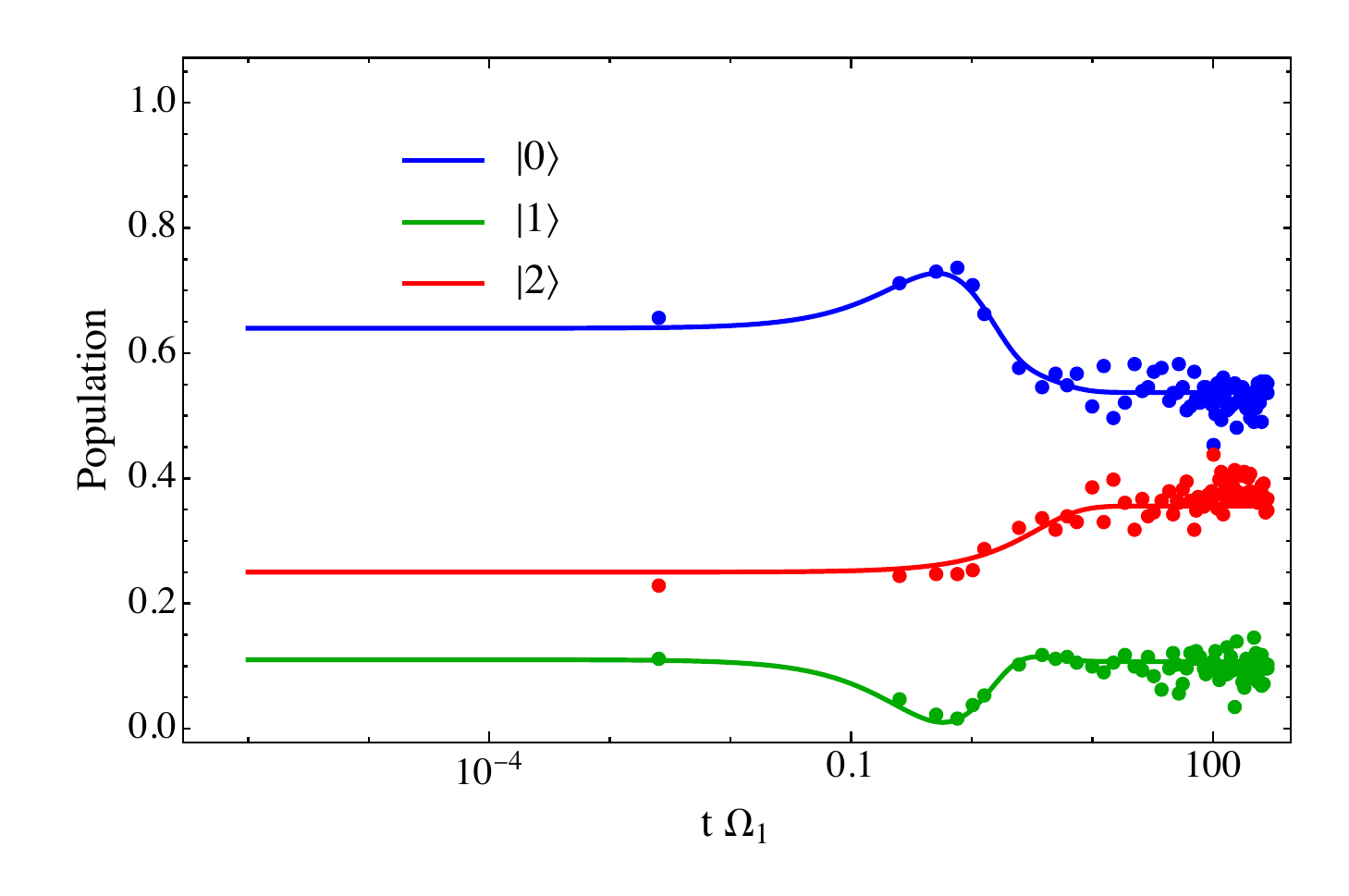}
    \caption{Population dynamics of the Lindblad model for the three initial conditions: a) $|0\rangle\langle0|$, b)$|2\rangle\langle2|$, and c)$|\mathrm{sME}\rangle\langle\mathrm{sME}|$.}
    \label{fig: Population_Dynamics_Lindblad}
\end{figure}

\begin{figure}
    \centering
    \includegraphics[width=0.7\linewidth]{ 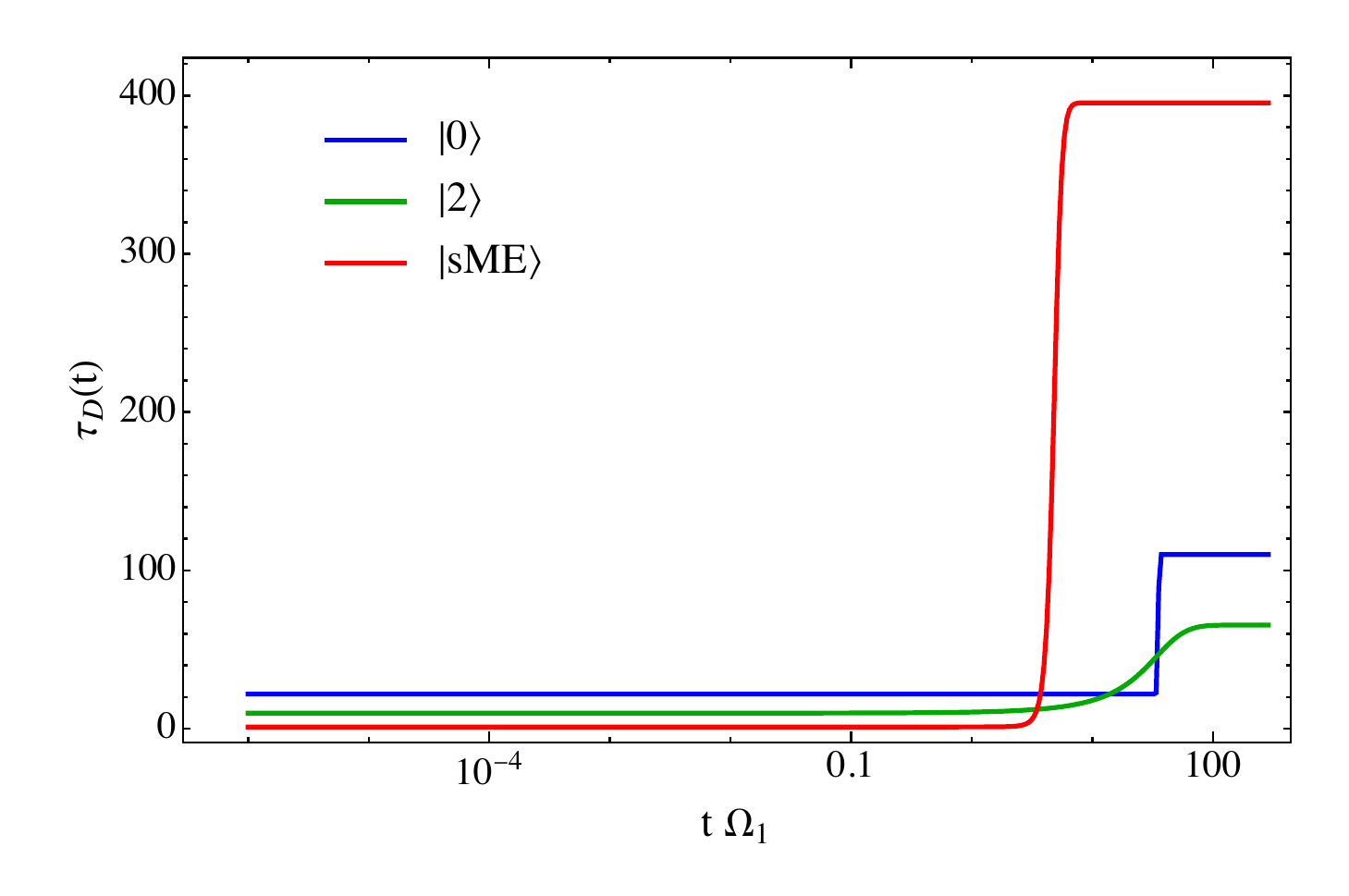}
    \caption{Time evolution of the relaxation parameter, $\tau_D = \tau_D(t)$, for the SEAQT model.}
    \label{fig: tau_time}
\end{figure}

Figure~\ref{fig: Population_Dynamics_sq_tau(t)} and Figure~\ref{fig: Population_Dynamics_Lindblad} show the population dynamics obtained within the SEAQT and the Lindblad frameworks, respectively,  for the three initial conditions and its comparison with experimental data of \cite{zhang2025observation}. The populations are obtained using Equation~(\ref{eq: Population}) for both cases.  It is observed that the predictions obtained with both frameworks fit well the experimental data. Figure~\ref{fig: tau_time} shows the time dependence of $\tau_D(t)$. All three
initial conditions display the two-level structure of
Eq.~(\ref{eq:logistic4}), with a transit time inside the measured window. The
upper limit reached by $|\mathrm{sME}\rangle$ lies at the edge of the search
domain and should be read as a lower bound on $\tau^{+}_{D}$ for that state
rather than as a determination of it.

Table~\ref{tab: table_phi} evaluates both order parameters on the three measured
initial conditions. The two reference states lie on opposite sides of the Mpemba
condition: $\Phi_{\GKSL}$ is negative for $\hat{\rho}^{(0)}_{\rm in}$ and positive
for $\hat{\rho}^{(2)}_{\rm in}$, and correspondingly $\langle\Qh_2\rangle$ falls
below $q_{\star}$ in the first case and above it in the second. This is the
physical content shared by the two criteria: the slow mode is, up to the residual
misalignment of \ref{app:slow_population_order_parameter}, the
displacement of the dressed population of the weakly coupled level $|2\rangle$
from the value it takes on the level set $\mathcal{M}_0$. A state prepared with
too little population in $|2\rangle$ and one prepared with too much both project
onto that mode, with opposite sign, and both must therefore wait for the slow
$\kappa_2$ channel to fill or drain it. The state $|{\rm sME}\rangle$ is prepared
at the crossing, where the projection changes sign, and its order parameter is an
order of magnitude smaller than for either reference state; the slow stage is
suppressed accordingly, which is the acceleration seen in
Fig.~\ref{fig: Population_Dynamics_sq_tau(t)}c. The proportionality
$\Qh^{\perp}_2\propto\hat L^{\perp}_1$ established in \ref{app:slow_population_order_parameter} is visible directly in the
table: the ratio of the signed order parameters is $1.39$, $1.42$ and $1.48$ for
the three states, constant to within a spread of the order of the residual
misalignment $\theta_3=2.36^{\circ}$. Finally, the values quoted are computed from
the populations of Table~\ref{tab:initial_conditions}, whereas the optimization of
Sec.~\ref{subsec: initial conditions} enforces $\mathrm{Tr}(\hat{L}_1\rhoh)=0$ on
the full state; the residual $\Phi_{\GKSL}=0.0652$ is therefore precisely the part
of the overlap carried by the coherences, and it is what the Hamiltonian dressing
of $\Qh_2$ is constructed to absorb.

\begin{table}[!htb]
\centering
\caption{\label{tab: table_phi}Order parameters of Eqs.~(\ref{eq: phi_gksl})
+and~(\ref{eq: phi_seaqt_final}) for the three measured
initial conditions, evaluated with $\boldsymbol{\alpha}$ at $N=3$ and
$q_{\star}=0.1848$.  Both criteria rank the three states identically and
single out $|{\rm sME}\rangle$.}
\begin{tabular}{lccc}
\hline
$\hat{\rho}_{\rm in}$ & $\Phi_{\rm GKSL}$ &
${\rm Tr}(\hat\rho\hat{Q}_2)$ & $\Phi_{\rm SEAQT}$ \\
\hline
$\hat{\rho}^{(0)}_{\rm in}$  & $-0.2024$ & $0.0387$ & $0.1461$ \\
$\hat{\rho}^{(2)}_{\rm in}$  & $ 1.0862$ & $0.9517$ & $0.7669$ \\
$|{\rm sME}\rangle\langle{\rm sME}|$ & $0.0652$ & $0.2289$ & $0.0441$ \\
\hline
\end{tabular}
\end{table}

Figure~\ref{fig: tau_time} shows the time dependence of $\tau_D(t)$ for the SEAQT model. It is observed that dissipation is stronger at early times and progressively weakens as the system approaches stable equilibrium. This behavior is consistent with a dynamics initially dominated by a fast decay channel associated with the state $|1\rangle$, followed by a slower dissipation regime that governs the long-time evolution.

\begin{figure}
    \centering
    a)\includegraphics[width=0.7\linewidth]{ 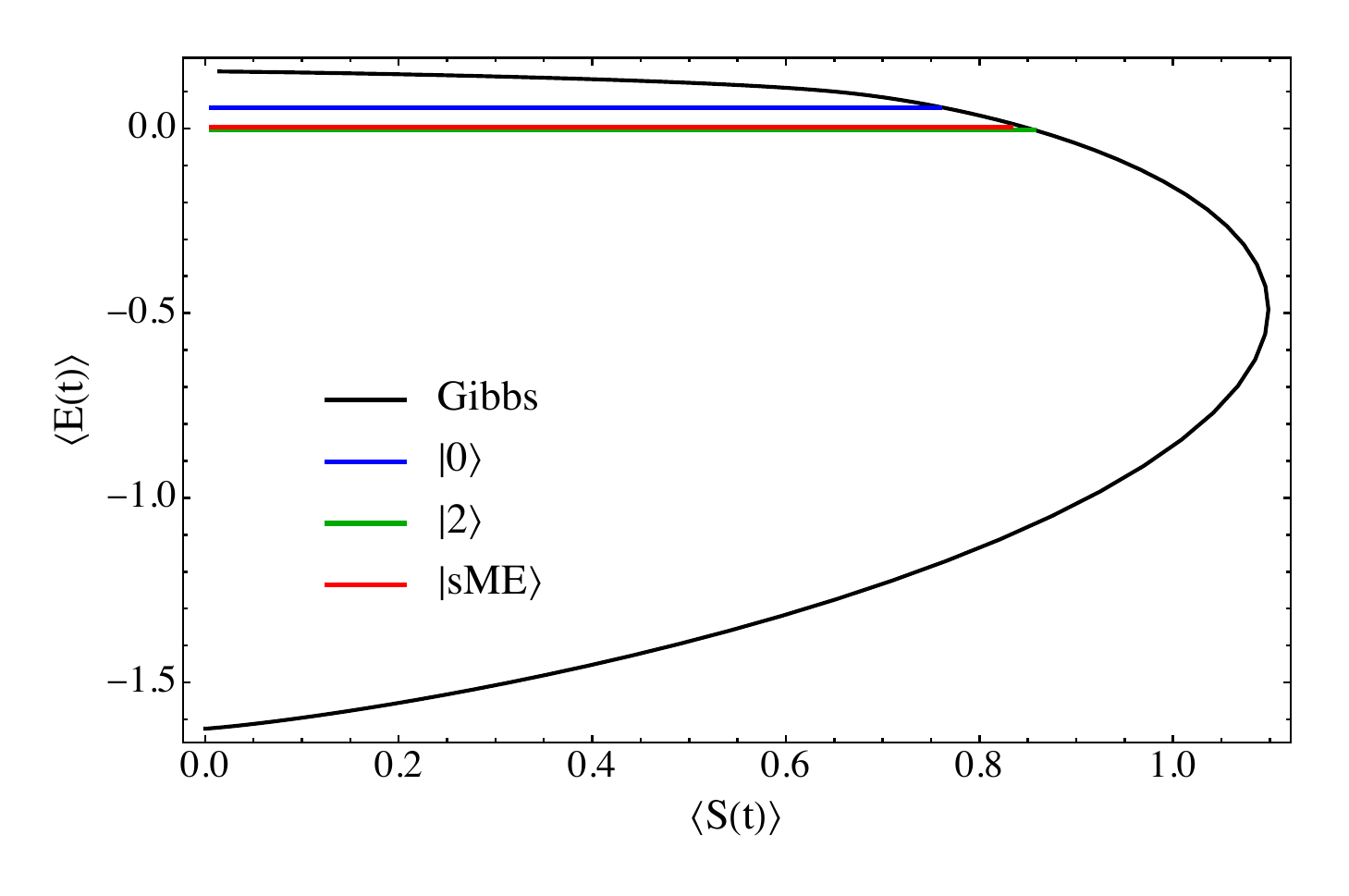}
    b)\includegraphics[width=0.7\linewidth]{ 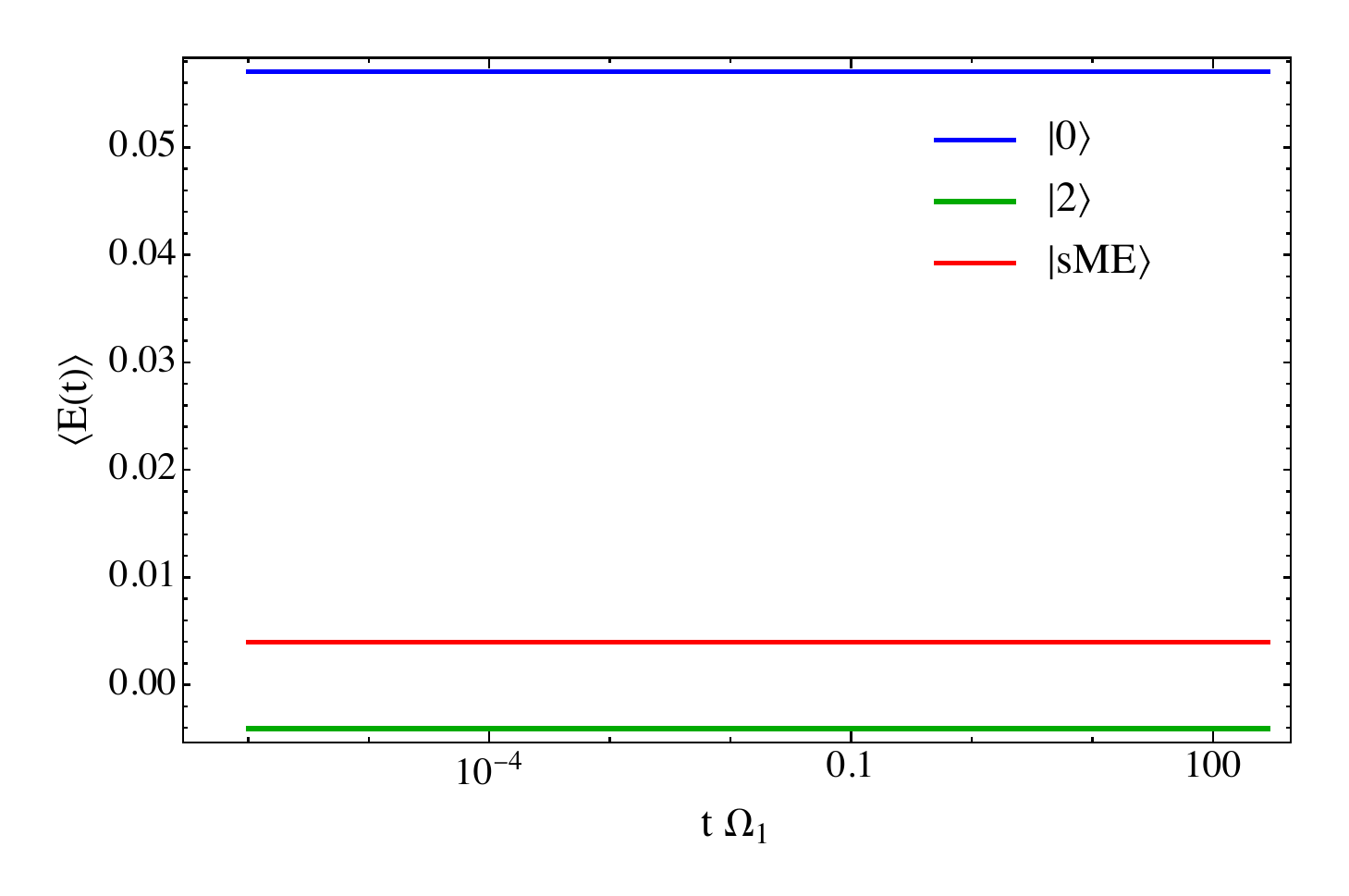}
    c)\includegraphics[width=0.7\linewidth]{ 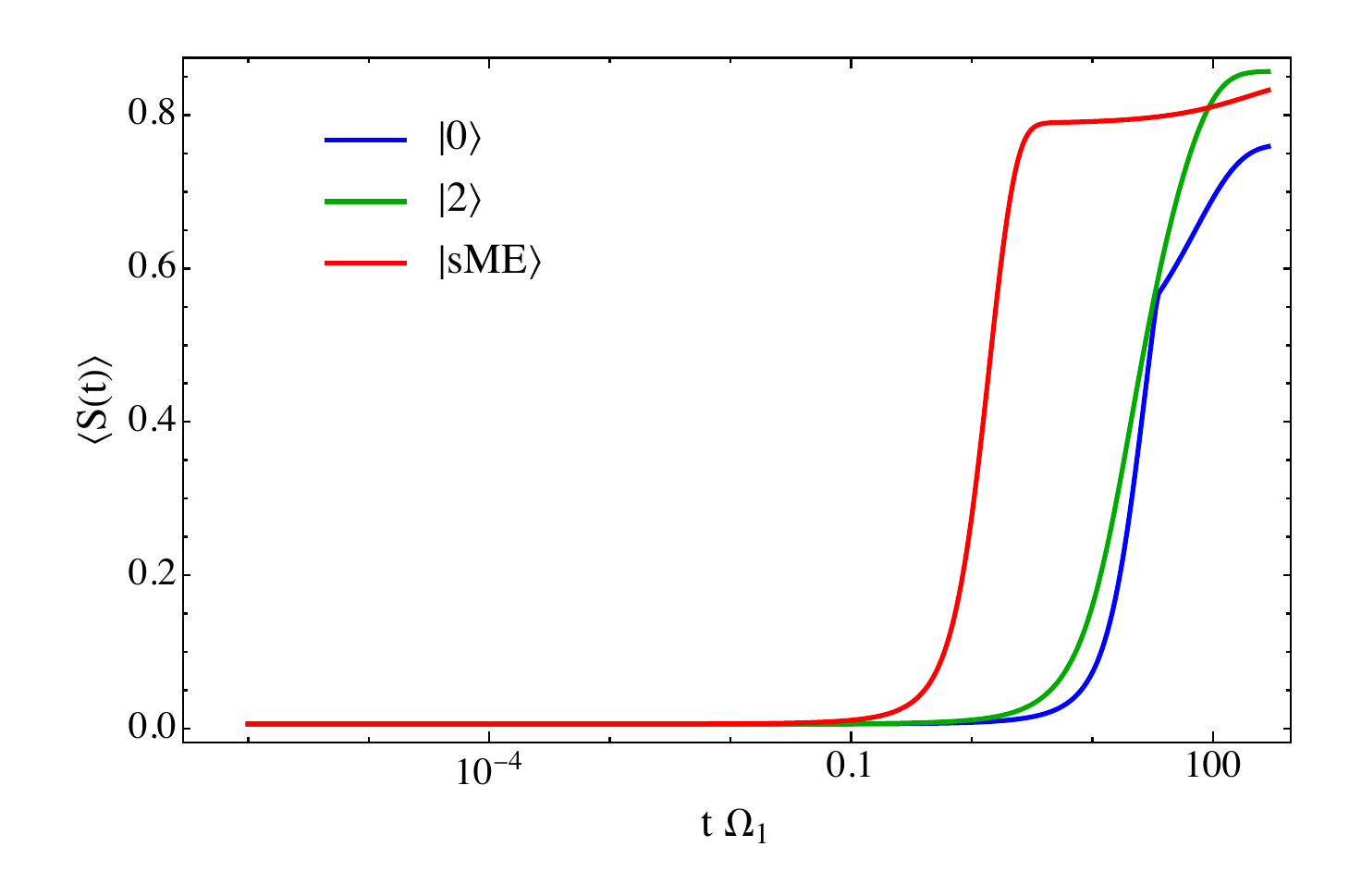}
    \caption{Energy and entropy evolution for the SEAQT model with $\tau_D = \tau_D(t)$: a) state evolution represented on the energy-entropy diagram, b) energy expectation value, and c) entropy expectation value.}\label{fig: E-S dynamics tau(t)}
\end{figure}

\begin{figure}
    \centering
    a)\includegraphics[width=0.7\linewidth]{ 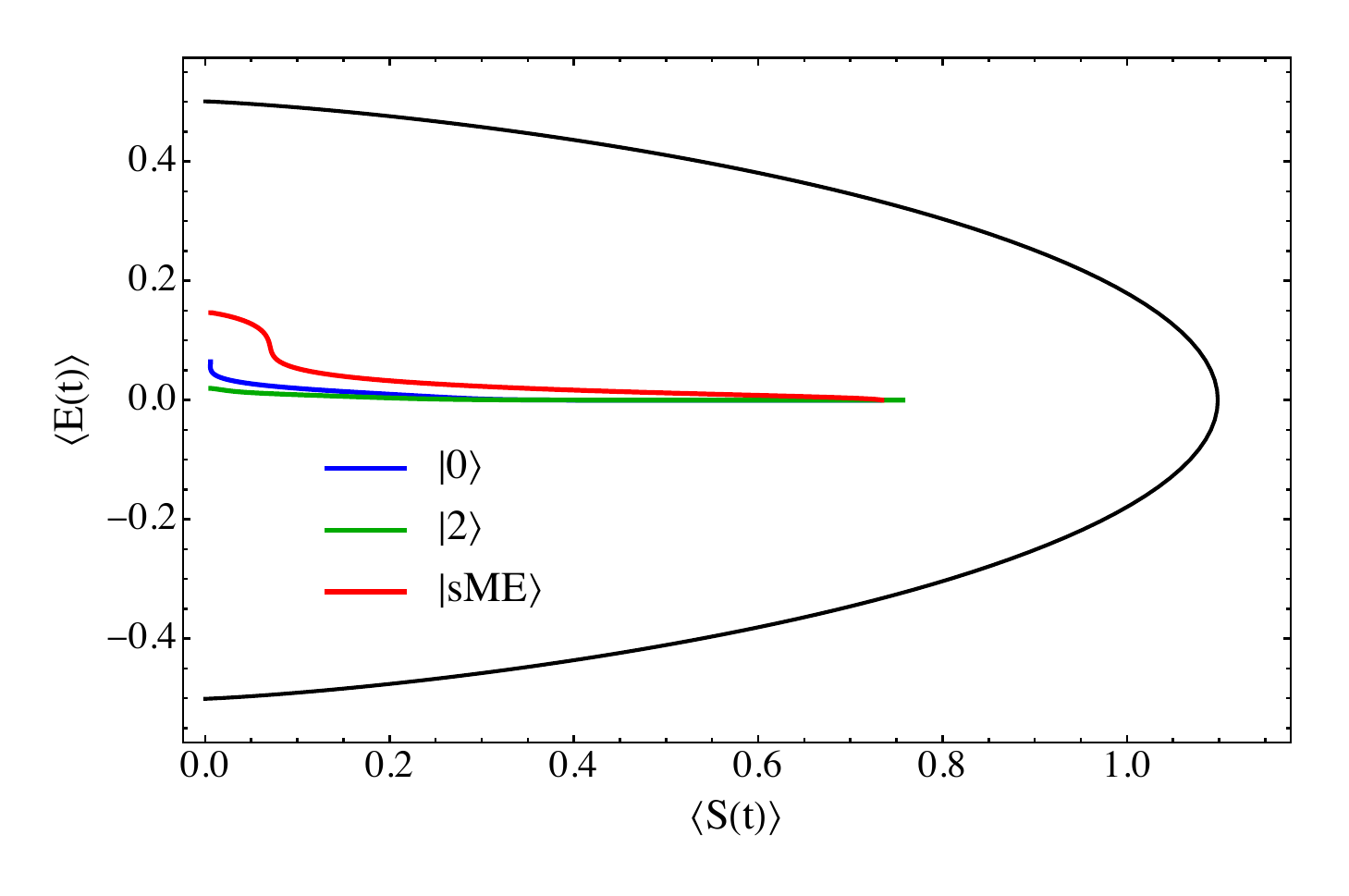}
    b)\includegraphics[width=0.7\linewidth]{ 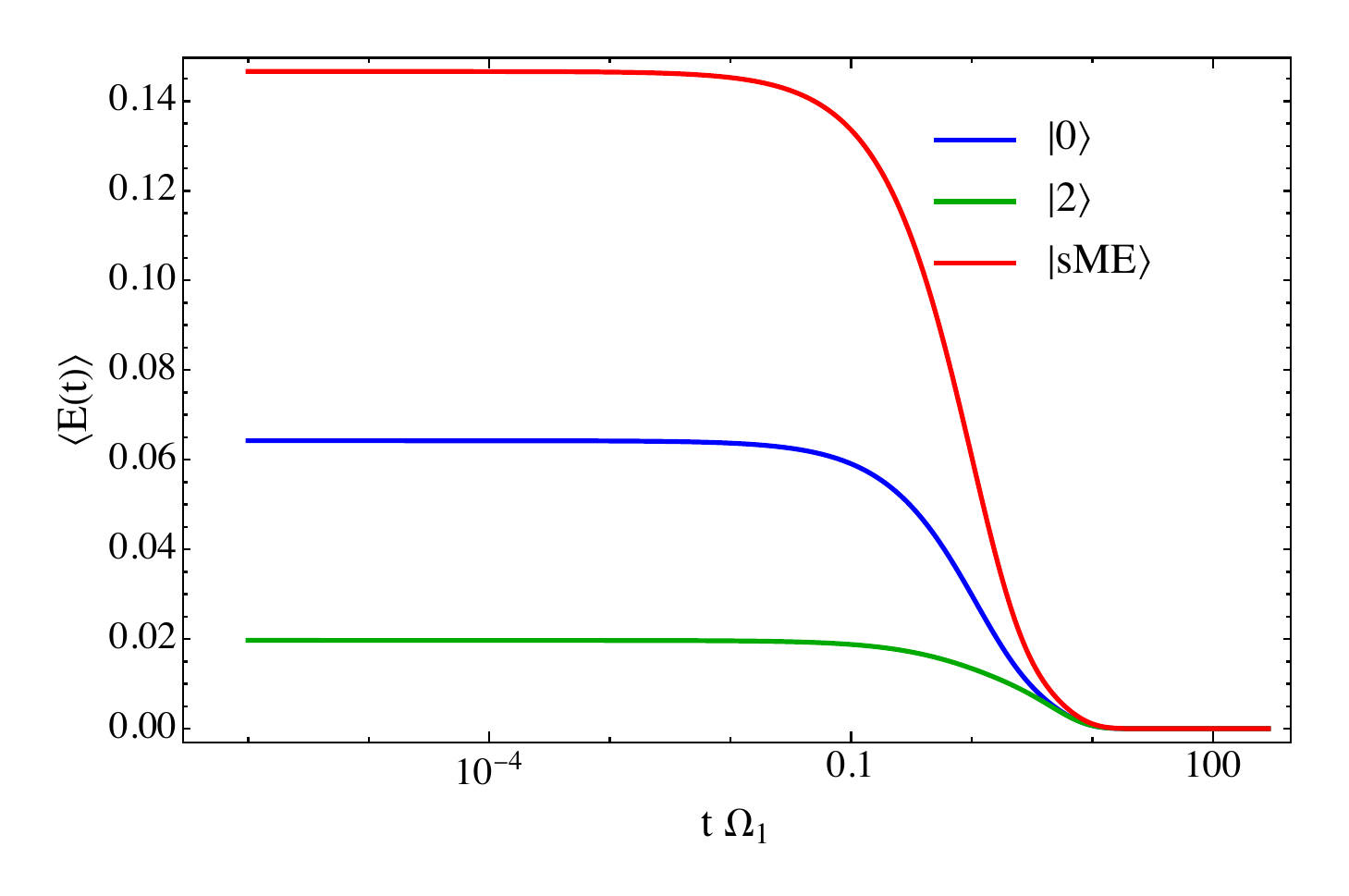}
    c)\includegraphics[width=0.7\linewidth]{ 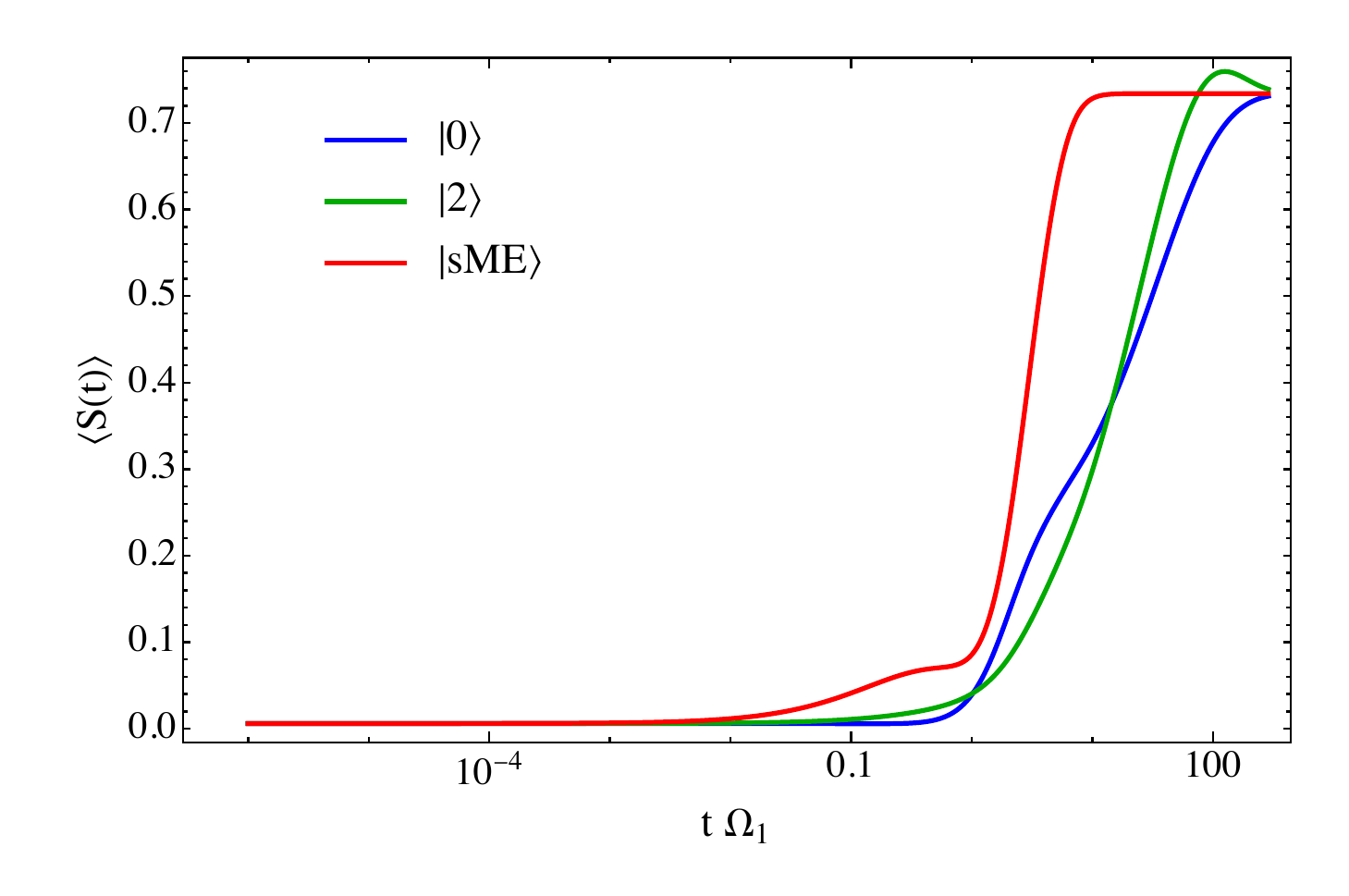}
    \caption{Energy and entropy evolution for the Lindblad model: a) state evolution represented on the energy-entropy diagram, b) energy expectation value, and c) entropy expectation value.}\label{fig: E-S dynamics Lindblad}
\end{figure}

Figure~\ref{fig: E-S dynamics tau(t)} and Figure~\ref{fig: E-S dynamics Lindblad} show the energy and entropy evolution of the system obtained with the SEAQT and Lindblad models, respectively. It is observed that for the Lindblad predictions, the state dynamics evolves towards a final state located inside the nonequilibrium region of the energy-entropy diagram. It is also observed that the dissipative coupling to the environment induces variations on both the expectation values of the energy and entropy of the system during the state evolution. This behavior reflects the open-system nature of the Lindblad formalism, where environmental effects are incorporated via the Born-Markov approximation. It is also observed that for the  SEAQT predictions, the state of the system is driven towards a stable equilibrium state, located on the convex curve of the energy-entropy diagram. It is also observed that the energy remains constant throughout the state evolution, which is a characteristic of the SEAQT model for closed systems. In addition, it is observed that the entropy increases during the state evolution of the system for the three cases presented. The entropy evolution within the SEAQT framework reveals a metastable regime in
the dynamics of the $|\mathrm{sME}\rangle$ initial condition, a characteristic
feature of systems exhibiting multiple relaxation
timescales~\cite{31cp-kq8f,Gerry_2024}. This regime is identified by the plateau
in the corresponding entropy curve of Fig.~\ref{fig: E-S dynamics tau(t)}c,
which extends over roughly $2\lesssim t\,\Omega_1\lesssim 20$ before the entropy
rises to its equilibrium value; the fitted transit time for that state,
$t_{\mathrm{c}}=4.80$, lies inside the plateau. The trajectory issuing from
$\hat{\rho}^{(0)}_{\rm in}$ shows no comparable plateau. The
$|\mathrm{sME}\rangle$ state nevertheless relaxes fastest, in agreement with
experiment, so that the acceleration characterizing the Mpemba effect is
associated here with the suppression of the slow dissipative mode and not with
the avoidance of the metastable stage.

\begin{figure}
    \centering
    a)\includegraphics[width=0.7\linewidth]{ 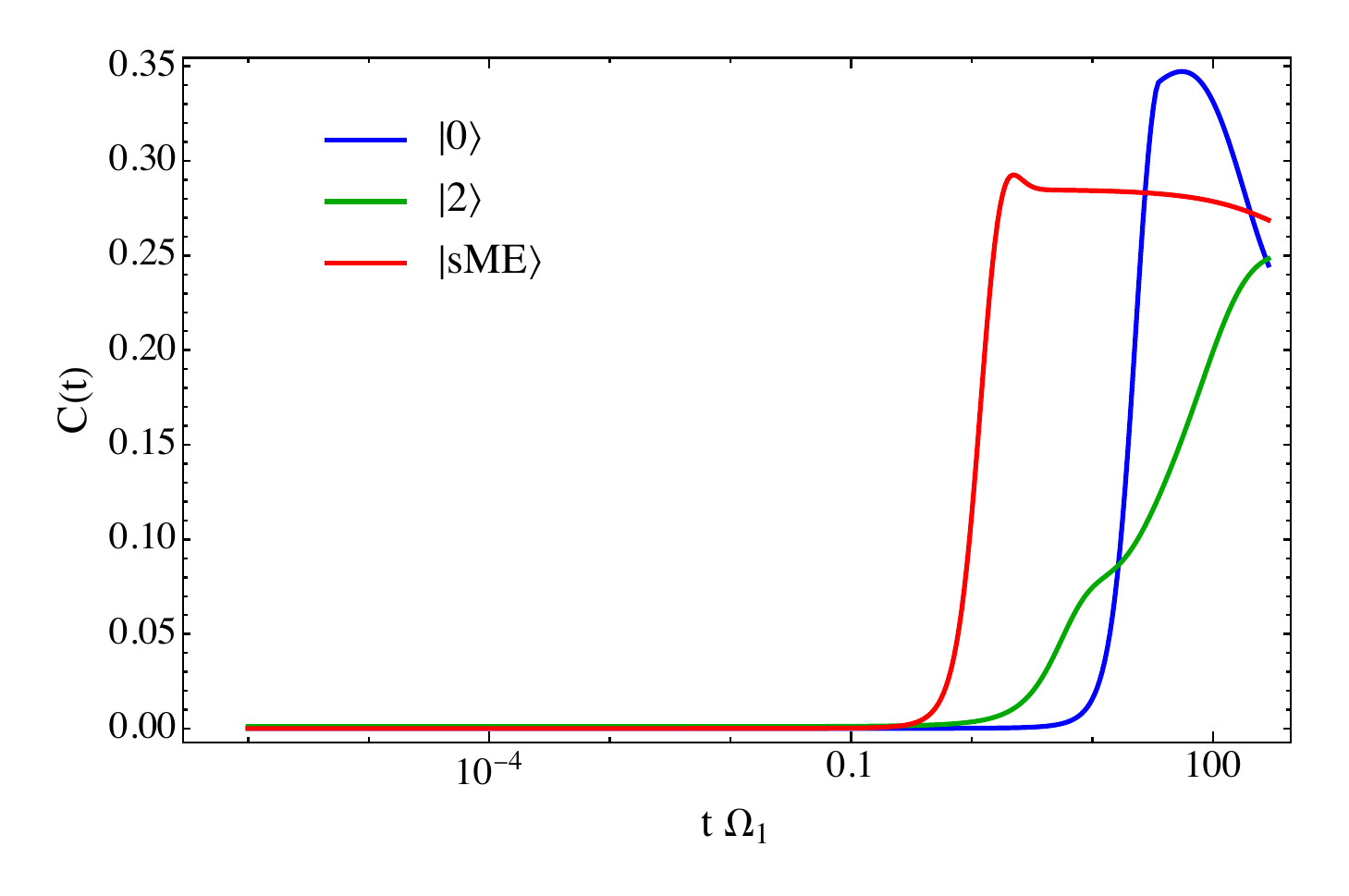}
    b)\includegraphics[width=0.7\linewidth]{ 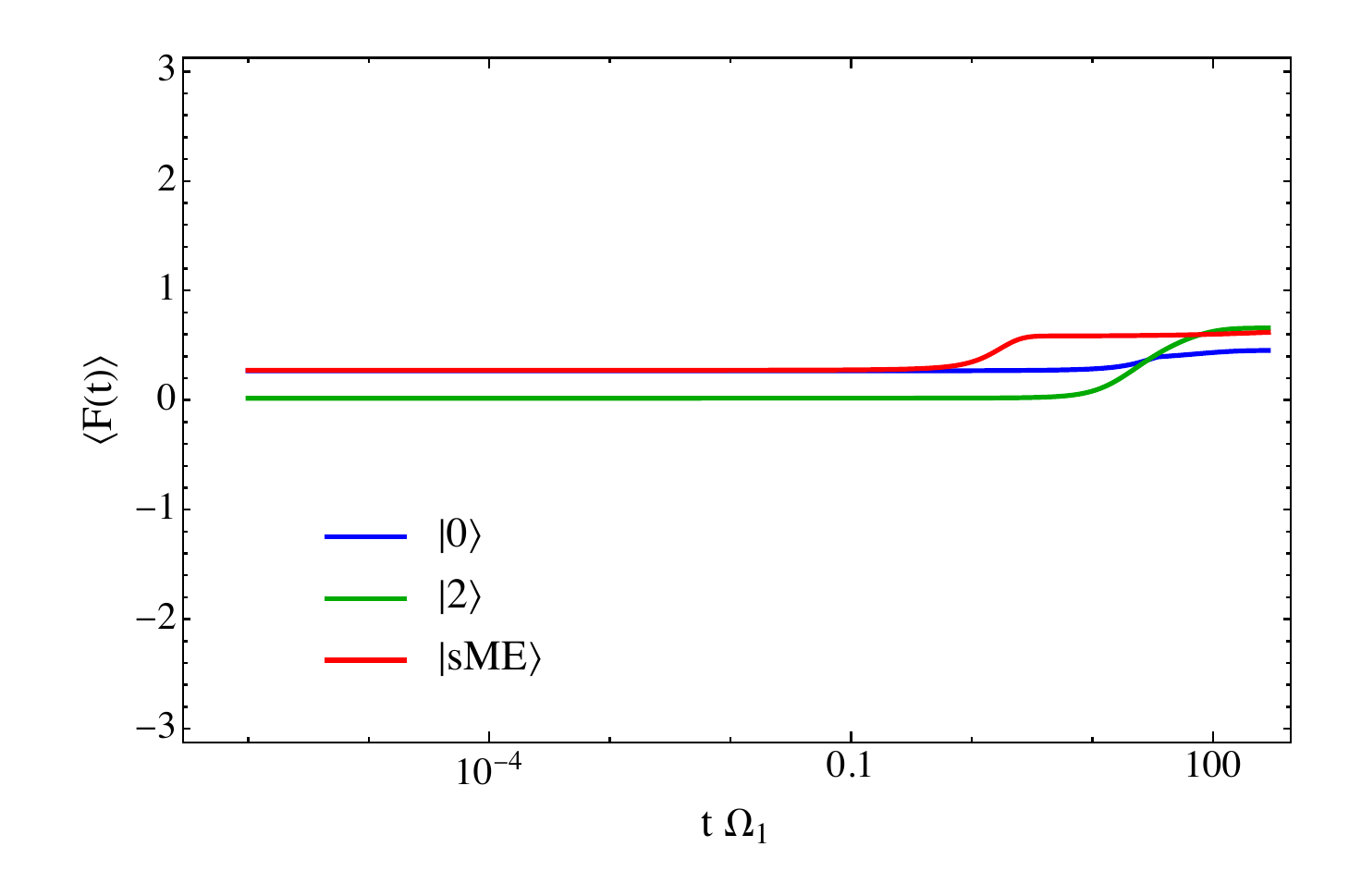}
    \caption{Nonequilibrium a) heat capacity and b) free energy, for the SEAQT framework considering $\tau_D = \tau_D(t)$.}\label{fig: Cv tau(t)}
\end{figure}

\begin{figure}
    \centering
    \includegraphics[width=0.7\linewidth]{ 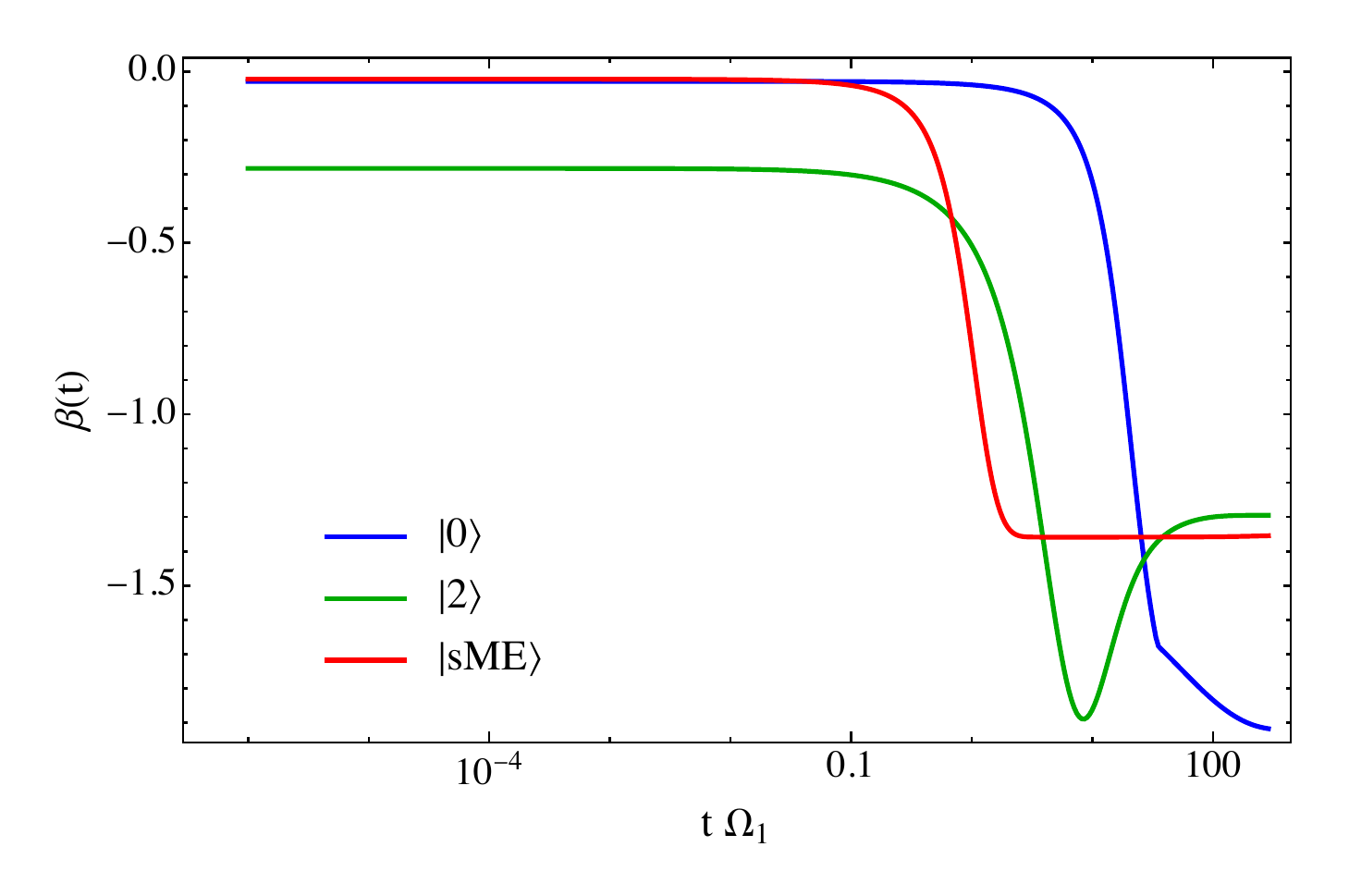}
    \caption{Time evolution of the parameter $\beta$ for the SEAQT framework, considering $\tau_D = \tau_D(t)$.}\label{fig: beta}
\end{figure}

Figure~\ref{fig: Cv tau(t)} shows the time evolution of the nonequilibrium heat
capacity and free energy for the SEAQT framework. Both quantities are smooth along
the whole trajectory. The heat capacity starts near zero, since the prepared states
are nearly pure and $\beta\to 0$ there, and rises together with the entropy of
Fig.~\ref{fig: E-S dynamics tau(t)}c towards $C_{\rm eq}=\sigma^{\rm eq}_{\Sh\Sh}$.
The correlation between the two is not qualitative but exact:
Eq.~(\ref{eq:sigmaFF-closed}) gives
$\sigma_{\Sh\Sh}-C=\beta^{2}\sigma_{\Fh\Fh}=\tau_D\,d\langle\Sh\rangle/dt\ge 0$, so
$C(t)$ is bounded above by the instantaneous entropy variance and the gap between
them is the entropy production. The heat capacity therefore attains its equilibrium
value only when dissipation ceases, and it is a direct measure of how much of the
entropy fluctuation of the state has already been converted into energy
fluctuation. The transient maximum of the $\hat{\rho}^{(0)}_{\rm in}$ curve
reflects the excursion of $\beta$ beyond its equilibrium value seen in
Fig.~\ref{fig: beta}, $C=\beta^{2}\sigma_{\Hh\Hh}$ being quadratic in $\beta$.

For the free energy, $\langle\Fh\rangle=\langle\Hh\rangle-\beta^{-1}\langle\Sh\rangle$
 with $\langle\Hh\rangle$ constant, so all of the structure of
 Fig.~\ref{fig: Cv tau(t)}b comes from $-\langle\Sh\rangle/\beta$. No divergence
occurs even though $\beta$ is small at early times: by Eq.~(\ref{eq: beta}),
$\beta=\sigma_{\Hh\Sh}/\sigma_{\Hh\Hh}$ vanishes together with the entropy scale of
the state as $\rhoh$ approaches purity, so the ratio remains finite and the
apparent singularity of the parametrization of Eq.~(\ref{eq: free energy}) is
removable along the trajectory, consistently with the regularity of
Eq.~(\ref{eq:nonsingular}). Since $\beta<0$ throughout, $-\beta^{-1}>0$ and
$\langle\Fh\rangle$ increases monotonically with $\langle\Sh\rangle$ at fixed
energy; the free-energy curves consequently reproduce the shape of the entropy
curves, metastable plateau of the $|\mathrm{sME}\rangle$ trajectory included. This
is the free-energy--entropy equivalence discussed in Sec.~\ref{sec:Conc}, seen here
directly at the level of the computed trajectories.

Figure~\ref{fig: beta} shows the time evolution of the parameter $\beta$ for the SEAQT framework. It is observed that the transition from an initial nonequilibrium state to a stable equilibrium state occurs fast at the end of the evolution, while at the beginning its value remains almost constant. It is also observed that the values are negative, which is consistent with a state evolution towards a stable equilibrium state located in the upper side of the energy-entropy diagram, see Figure~\ref{fig: E-S dynamics tau(t)}a. In addition, it is observed that $\beta$ coincides with the thermodynamic definition of temperature at the stable equilibrium state.

Because the SEAQT evolution of an isolated system conserves
$\langle\Hh\rangle$, the three initial conditions relax to three distinct stable
equilibrium states, with $\langle\Hh\rangle=0.0571$, $-0.0041$ and $0.0040$
respectively and correspondingly different $\beta_{\mathrm{eq}}$. Comparisons
between initial conditions based on the free energy are therefore meaningful
only at fixed $E_0$; see the discussion in Sec.~\ref{sec:Conc}.


\subsection{\label{subsec:tauD-const}SEAQT case for $\tau_D = \mathrm{constant}$}

For the case when $\tau_D$ is considered as a positive constant, the set $\omega_i$ is reduced to $\omega_i=\{\omega_1,\ \omega_2,\ \omega_3\}$, where $\omega_1$ and $\omega_2$ take the values given in Table~\ref{tab:00}, and $\omega_3=\tau_D$. Thus, the only variable calculated via the optimization procedure is $\tau_D$. The values considered for this case are given in Table \ref{tab:01}.

\begin{table}[!htbp]
    \caption{Optimization results for the case $\tau_D=\mathrm{constant}$, with
    $\omega_1$ and $\omega_2$ held at the values of Table~\ref{tab:00}.}
    \label{tab:01}
    \centering
    \begin{tabular}{c|c|c|c|c}
    \hline\hline
        $\hat{\rho}_{in}$ & $\omega_1$ & $\omega_2$ & $\tau_D$ & $\chi^2_\nu$\\
        \hline
        $\hat{\rho}^{(0)}_{\mathrm{in}}$ & $2.9257$ & $0.01544$ & 39.4103 & 53.1291 \\
        $\hat{\rho}^{(2)}_{\mathrm{in}}$ & $2.9257$ & $0.01544$ & 30.0586 & 30.9077 \\
        $|\mathrm{sME}\rangle\langle\mathrm{sME}|$ & $2.9257$ & $0.01544$ & 1.3462 & 25.992 \\
    \hline\hline
    \end{tabular}
\end{table}

\begin{figure}[]
    \centering
    a)\includegraphics[width=0.7\linewidth]{ 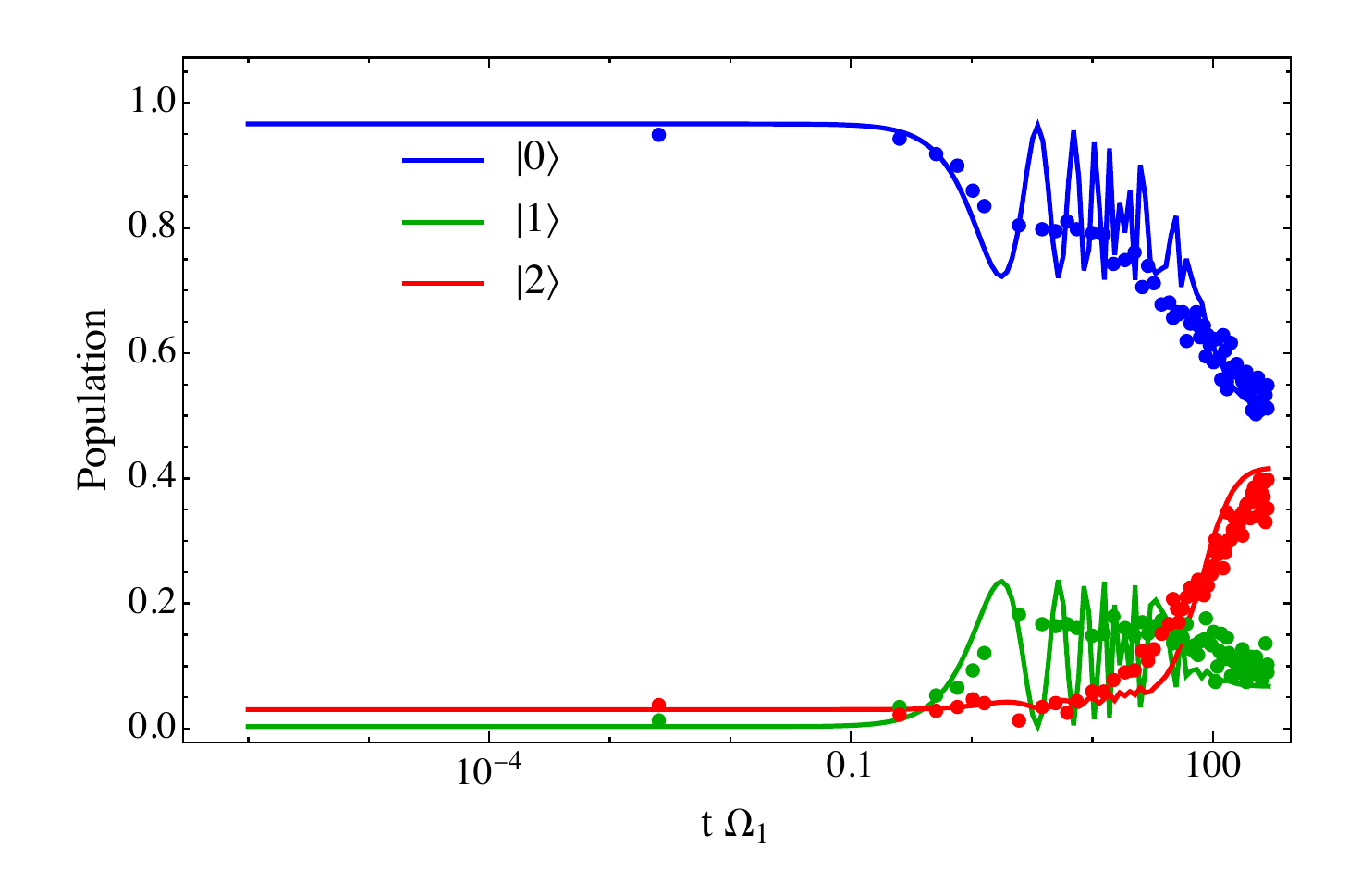}
    b)\includegraphics[width=0.7\linewidth]{ 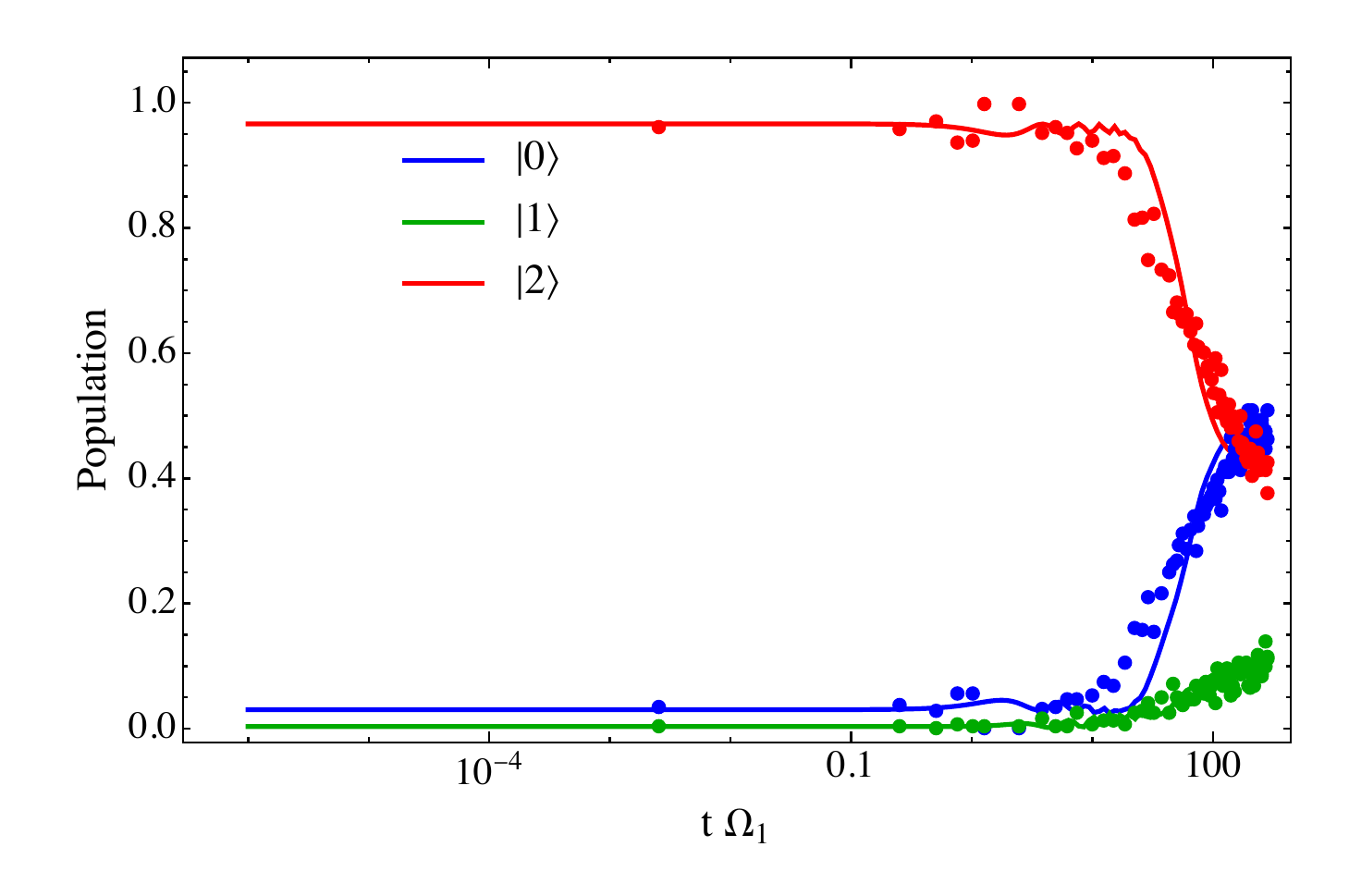}
    c)\includegraphics[width=0.7\linewidth]{ 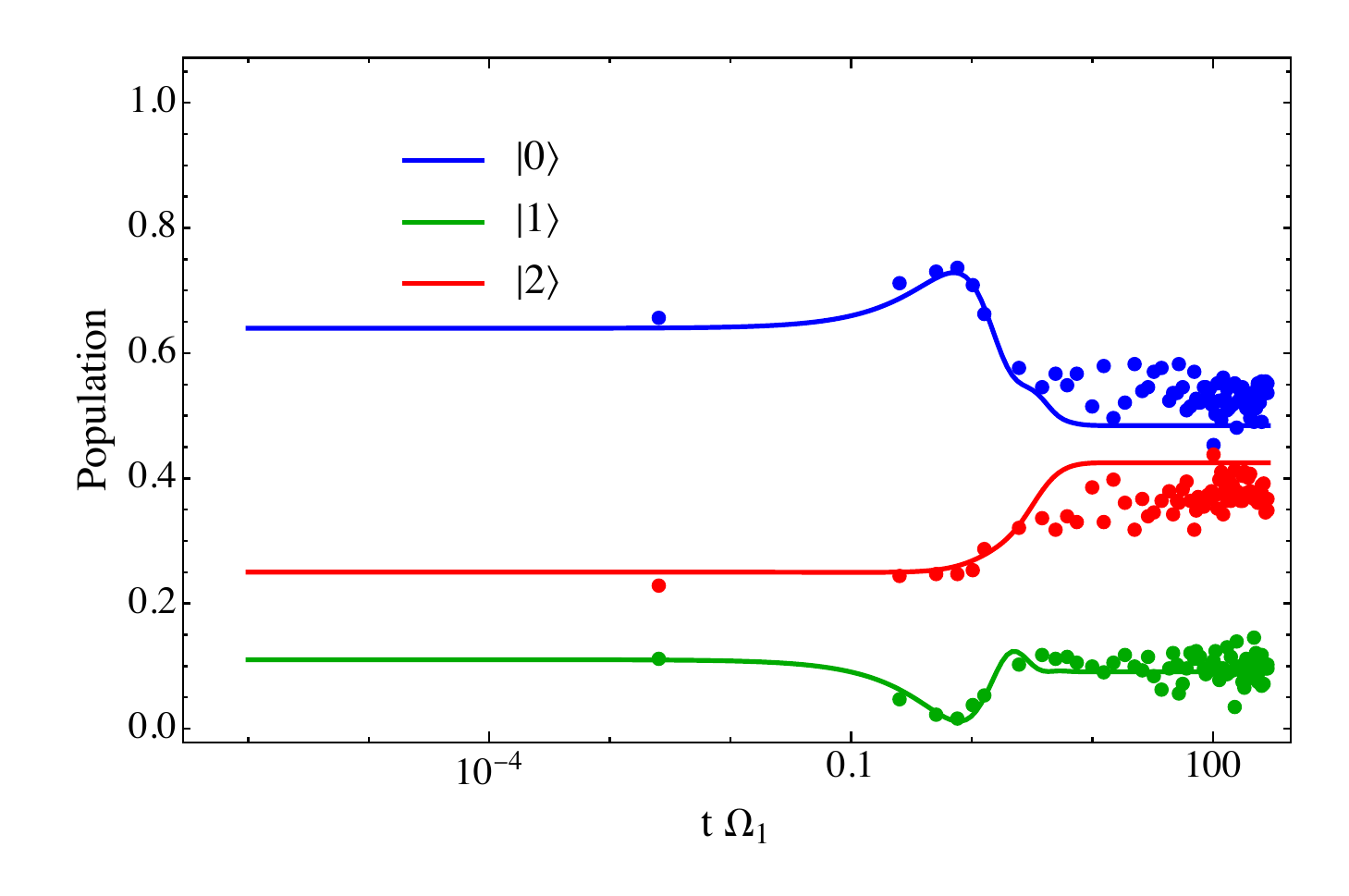}
    \caption{Population dynamics of the SEAQT model with $\tau_D = \mathrm{constant}$ for the three initial conditions: a) $|0\rangle\langle0|$, b)$|2\rangle\langle2|$, and c)$|\mathrm{sME}\rangle\langle\mathrm{sME}|$.}
    \label{fig: Population_Dynamics_sq_tau}
\end{figure}

Figure~\ref{fig: Population_Dynamics_sq_tau} shows the time evolution of the populations for the SEAQT dynamics when a $\tau_D = \mathrm{constant}$ is considered. It is observed that the experimental data is approximated well by the model. In addition, it is also observed that  the dynamics associated with the initial state that leads to accelerated relaxation display a smooth temporal evolution, suggesting that this regime is governed by a single dominant decay channel.

\begin{figure}
    \centering
    a)\includegraphics[width=0.7\linewidth]{ 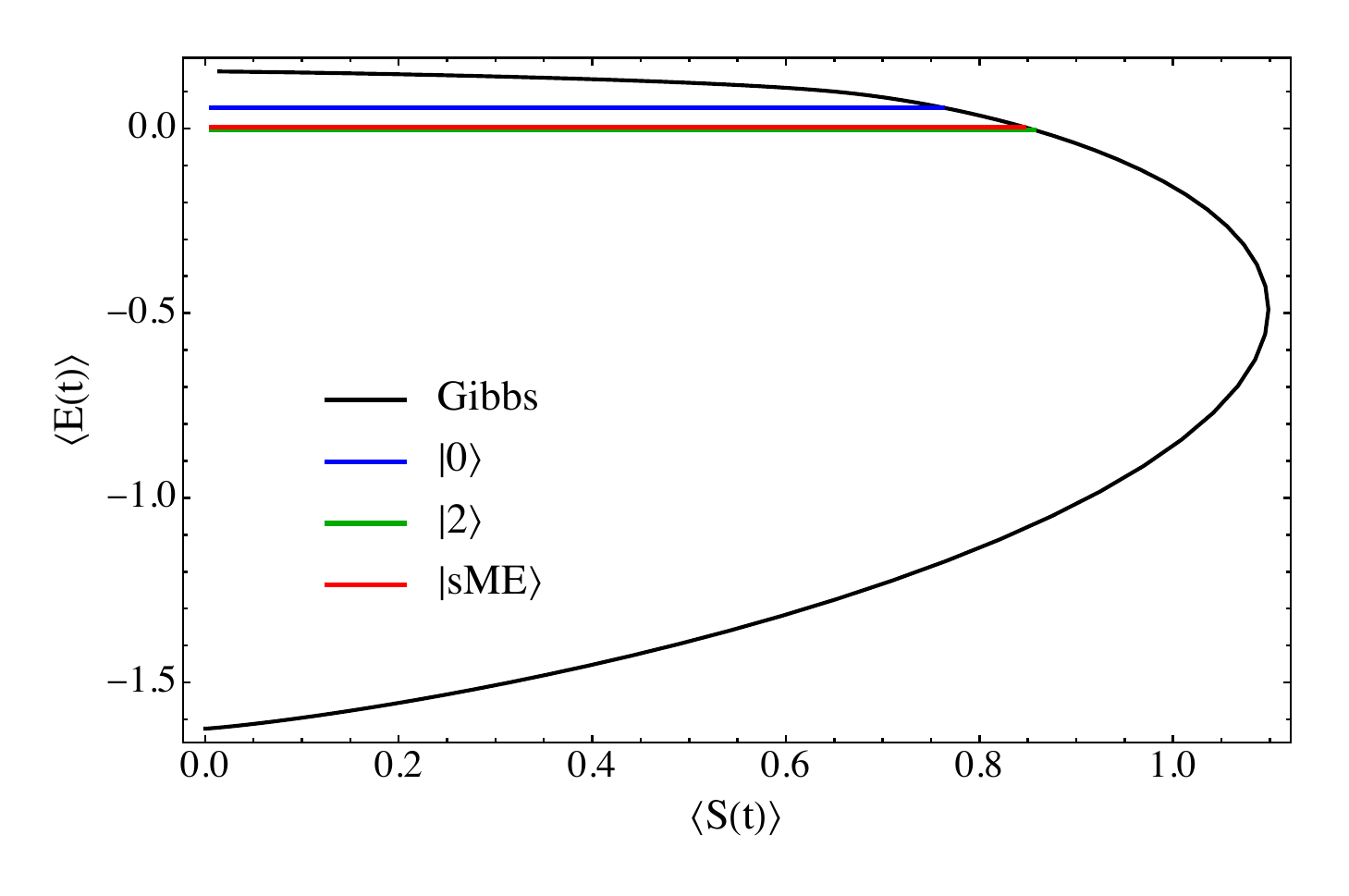}
    b)\includegraphics[width=0.7\linewidth]{ 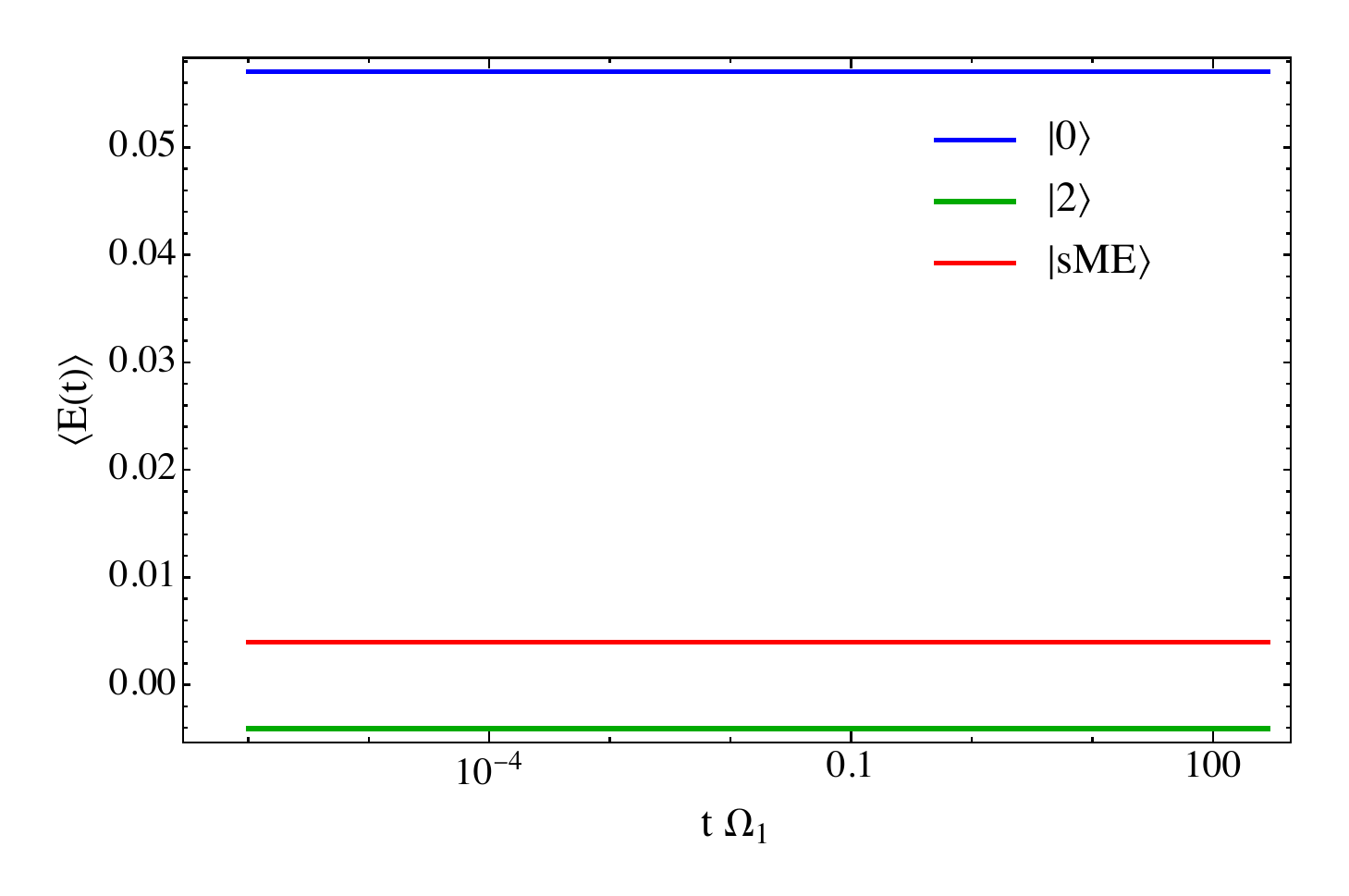}
    c)\includegraphics[width=0.7\linewidth]{ 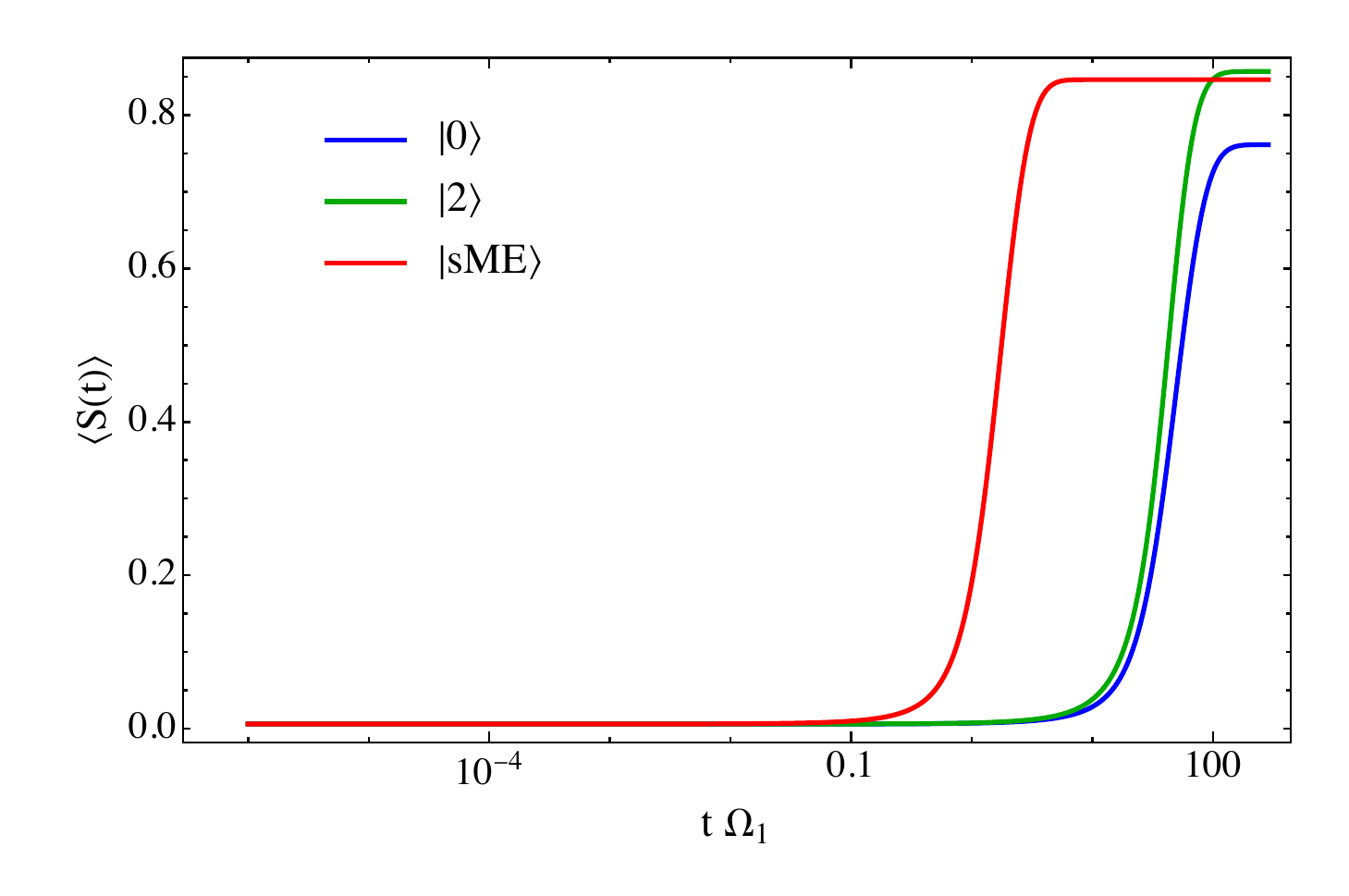}
    \caption{Energy and entropy evolution for the SEAQT model with $\tau_D = \mathrm{constant}$: a) state evolution represented
on the energy-entropy diagram, b) energy expectation value, and c) entropy expectation value.}
    \label{fig: E-S dynamics tau}
\end{figure}

Figure~\ref{fig: E-S dynamics tau} shows the energy and the entropy evolution of the system for the SEAQT model. It is observed that the energy expectation value remains constant at all times, and the expectation value of the entropy increases during the state evolution. For this particular case the entropy curves show no plateau for any of the three
initial conditions, confirming that a state-dependent $\tau_D$ is what carries
the multi-timescale structure of the relaxation.

\begin{figure}
    \centering
    a)\includegraphics[width=0.7\linewidth]{ 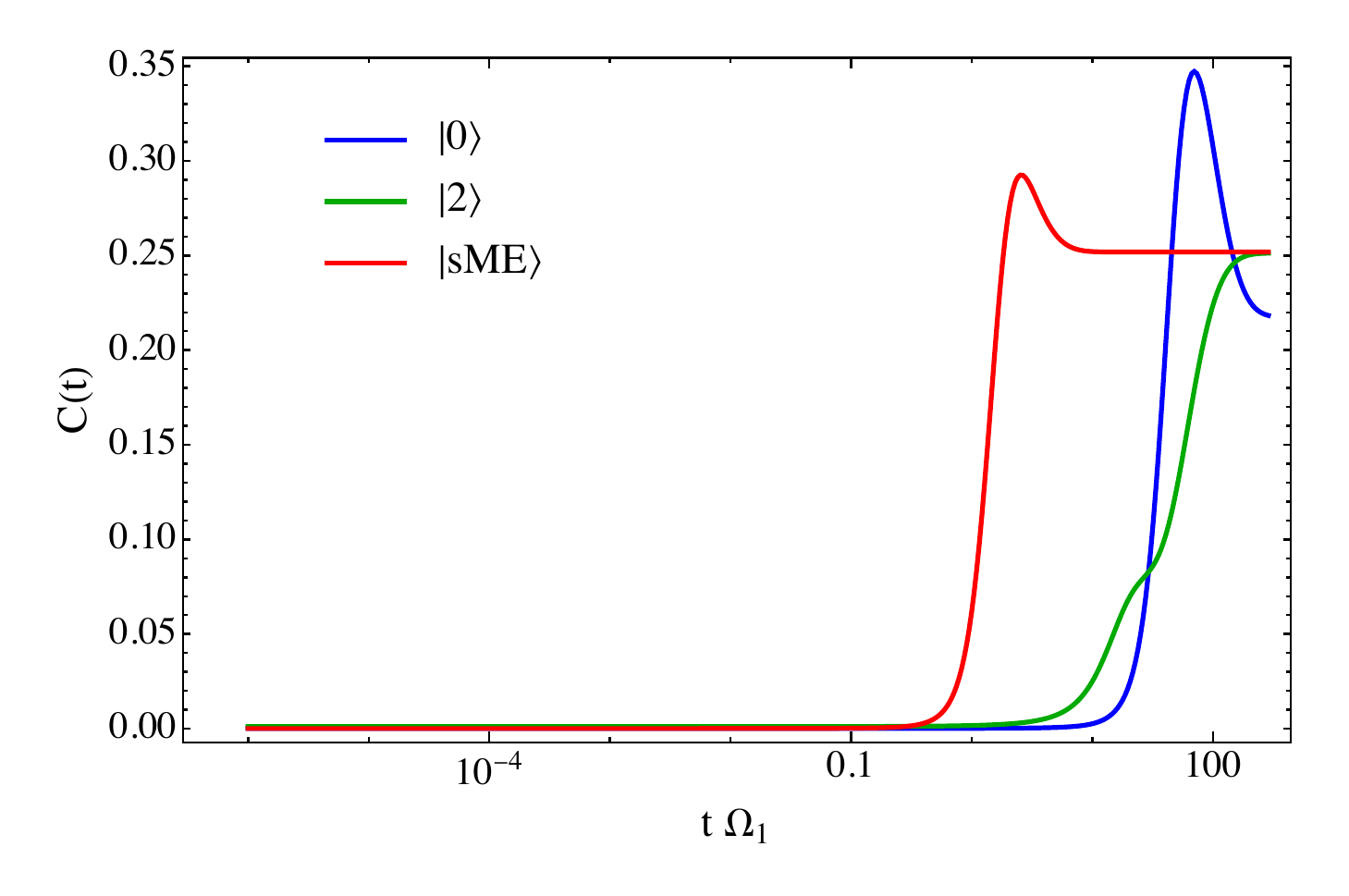}
    b)\includegraphics[width=0.7\linewidth]{ 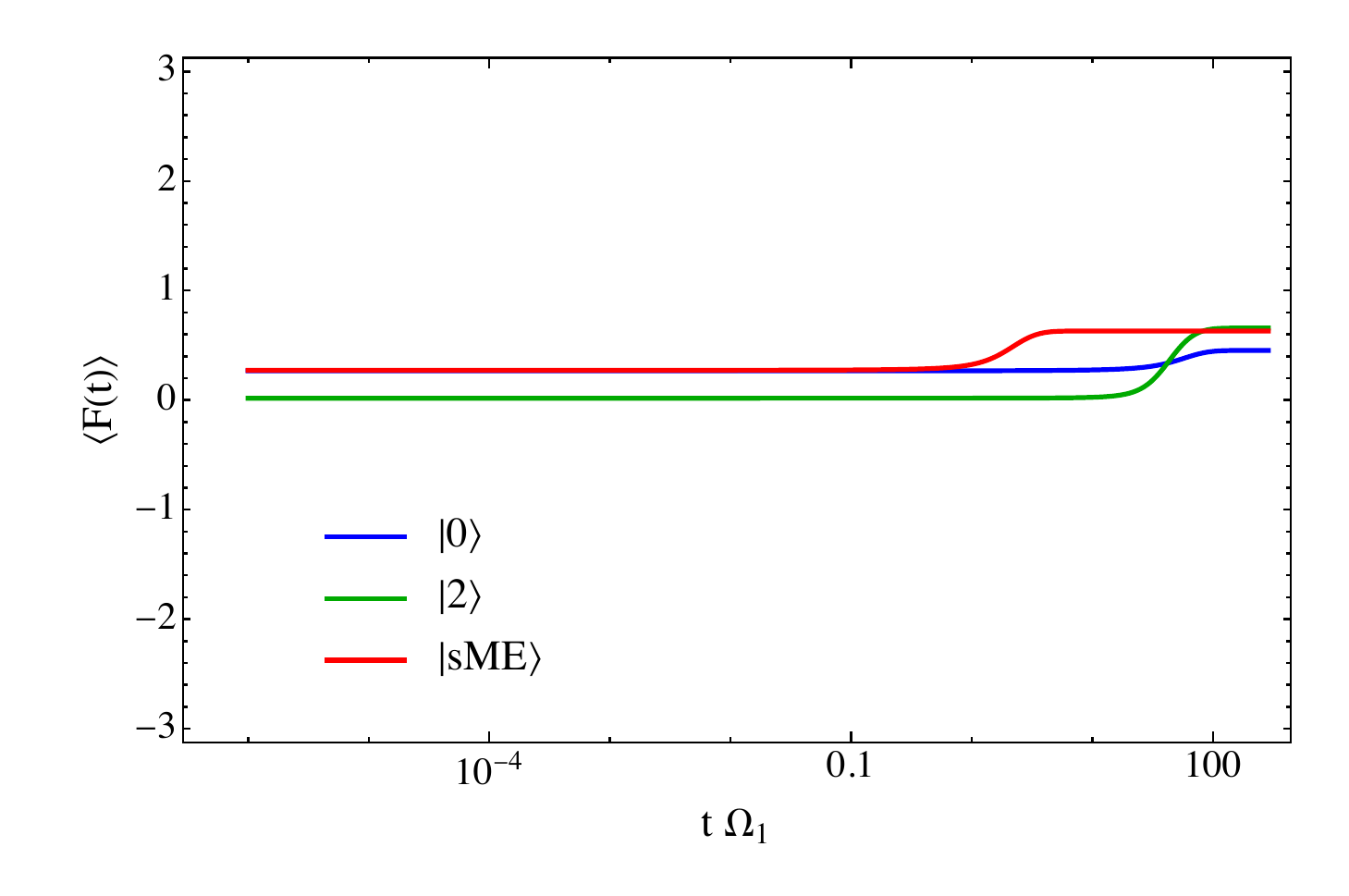}
    \caption{Nonequilibrium a) heat capacity and b) free energy, for the SEAQT framework considering $\tau_D = \mathrm{constant}$.}\label{fig: Cv_cte}
\end{figure}

Figure~\ref{fig: Cv_cte} shows the time evolution of the nonequilibrium heat capacity and free energy, respectively, for the SEAQT framework considering $\tau_D = \mathrm{constant}$. It is observed that for the nonequilibrium heat capacity the system evolves from an initial state where its value is close to zero, towards a final state where the heat capacity stabilizes at a positive constant value. It is also observed that for the nonequilibrium free energy some discontinuities and sharp changes are observed, which are related to transitions between different dynamical regimes.

\begin{figure}
    \centering
    \includegraphics[width=0.7\linewidth]{ 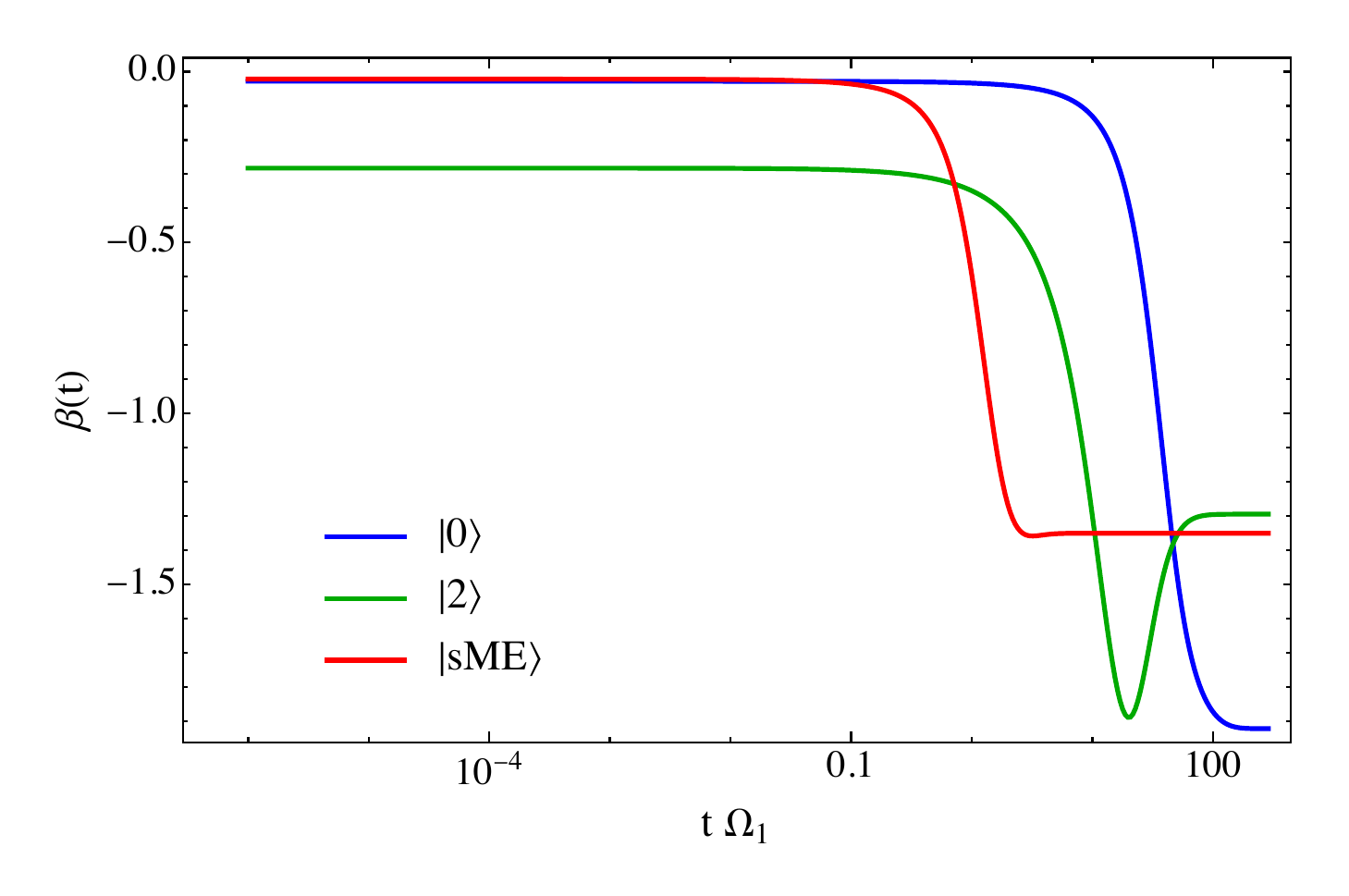}
    \caption{Time evolution of the parameter $\beta$ for the SEAQT framework, considering $\tau_D=\mathrm{constant}$.}
    \label{fig: beta_cte}
\end{figure}

Figure \ref{fig: beta_cte} shows the time evolution of the parameter $\beta$ for the SEAQT framework. It is observed that this parameter evolves from an initial state where its value is close to zero, towards a final state where its value is different from zero. Its values are negative because the state evolution is located in the upper part of the energy-entropy diagram.

\section{\label{sec:Conc}Conclusions}
\vspace*{-5pt}

In this work, the steepest-entropy-ascent quantum thermodynamics framework has been used to model the quantum Mpemba effect observed experimentally in a three-level ion system~\cite{zhang2025observation}. The four-level Hilbert space of the original system was reduced to an effective three-dimensional description via the Feshbach projection, yielding an effective Hermitian Hamiltonian that captures the essential open-system dynamics without explicitly propagating the short-lived auxiliary state. The six free parameters of the model were determined by an alternating
optimization in which the two effective-Hamiltonian coefficients, $\omega_1$ and
$\omega_2$, are shared by all three initial conditions, as required by the fact
that $\Heff$ does not depend on state preparation, while the four coefficients
parametrizing the time-dependent relaxation functional $\tau_D(t)$ are fitted
per initial condition.

The SEAQT and GKSL frameworks reproduced well the experimental population dynamics for all three initial conditions studied. For the case $\tau_D(t)=\mathrm{constant}$, the model still approximates the
data but loses the entropy plateau, which in the time-dependent case appears in
the $|\mathrm{sME}\rangle$ trajectory, confirming that a state-dependent $\tau_D$ is physically necessary for capturing the multi-timescale nature of the relaxation. The SEAQT framework drives the state towards a stable equilibrium state consistent with the first and second laws of thermodynamics, whereas the GKSL framework, based on the Born--Markov approximation, drives the system towards a non-equilibrium steady state. The nonequilibrium inverse temperature parameter, $\beta(t)$, which arises naturally within the SEAQT formalism, converges to the thermodynamic inverse temperature at stable equilibrium, confirming the internal thermodynamic consistency of the model.

Beyond the numerical modelling, the present work establishes three analytical
results for the understanding of dissipative dynamics within the SEAQT
framework.
 
The first is a first-principles determination of the relaxation parameter. By
rearranging the SEAQT entropy production relation, Eq.~\eqref{eq:entprod},
$\tau_D$ is identified in Eq.~\eqref{eq:tauD-exact} as the ratio of the
thermodynamic driving force $\beta^{2}\sigma_{\Fh\Fh}=\sigma_{\Sh\Sh}-C$ to the
actual dissipation rate $d\langle\Sh\rangle/dt$, rather than as a free fitting
parameter. This expression develops a $0/0$ indetermination at the metastable
partially canonical state $\rhoc$, where $\Delta\Fh|_{\rhoc}=0$ forces both
quantities to vanish simultaneously. The indetermination is resolved by the
two-timescale structure of the dissipative dynamics, and yields distinct one-sided limits, Eq.~\eqref{eq:tauD-pm}, whose ratio is the ratio of the
linearized relaxation rates $|\lambda_f|/|\lambda_s|$ at $\rhoc$, not the bare
loss rates $\kappa_1$ and $\kappa_2$, see Sec.~\ref{sec:tauD-indet}, times a
geometric factor $\mathfrak{r}$ fixed by the curvature of the entropy and
free-energy-variance surfaces at $\rhoc$, Eq.~\eqref{eq:step}.
Smoothing this step over the finite crossover duration produces the logistic form of Eq.~\eqref{eq:logistic4} and provides physical readings for each of its coefficients, Eq.~\eqref{eq:ident}: the two one-sided
relaxation times, the transit time through the metastable state, and the inverse
crossover width. The four coefficients are jointly identifiable only when the transit occurs well
inside the measured time window; for two of the three initial conditions studied
here one of the two one-sided limits is not resolved by the data, and no step
height is quoted for them.
 
The second result concerns the dynamical order parameters. The SEAQT order
parameter $\Phi_{\SEAQT}=|\langle\Qh_2\rangle-p^{\mathrm{eq}}_2|$ and the
Lindblad order parameter $\Phi_{\GKSL}=\mathrm{Tr}(\hat L_1\rhoh_{\mathrm{in}})$
characterize the quantum Mpemba effect from complementary perspectives. The
condition $\Phi_{\GKSL}=0$ removes the slowest Liouvillian eigenmode from the
spectral decomposition of the density operator; the condition
$\Phi_{\SEAQT}\simeq 0$ places the initial state on the level set
$\mathcal{M}_0$ of Eq.~\eqref{eq:M0} and thereby bypasses the metastable entropy
plateau geometrically. Both produce the same physical outcome, an accelerated
approach to the final state, and are found to be operationally equivalent for
the three-level system studied here, a correspondence verified numerically
across $100$ randomly generated Mpemba initial states.

The results of this work demonstrate that the SEAQT framework provides a thermodynamically consistent and analytically tractable description of the quantum Mpemba effect, in which the key features of the accelerated relaxation (the step-function form of $\tau_D$, the order parameter separating dynamical regimes, and the free-energy criterion for genuine anomalous relaxation) all emerge naturally from the geometric and thermodynamic structure of the theory without requiring spectral information about a Liouvillian superoperator.


\section*{Acknowledgments}
L. E. Rocha-Soto acknowledges the financial support of the Secretariat of Science, Humanities, Technology and Innovation (SECIHTI), Mexico, under its national scholarship program with Grant No. CVU-1146619. C. E. Damian-Ascencio, and A. Salda\~na-Robles, and S. Cano-Andrade, gratefully acknowledge the financial support of the SECIHTI under its SNII program.


\appendix

\section{Dynamics of multiple randomly generated initial $|\mathrm{sME}\rangle$ states} \label{app:A}

The optimization procedure used to obtain an initial state, where the populations match the experimental data reported in \cite{zhang2025observation}, can be extended to generate multiple random initial states. This enables a comparison of their properties and dynamical behavior within the Lindblad and SEAQT frameworks.

\begin{figure}[]
    \centering
    a)\includegraphics[width=0.25\linewidth]{ 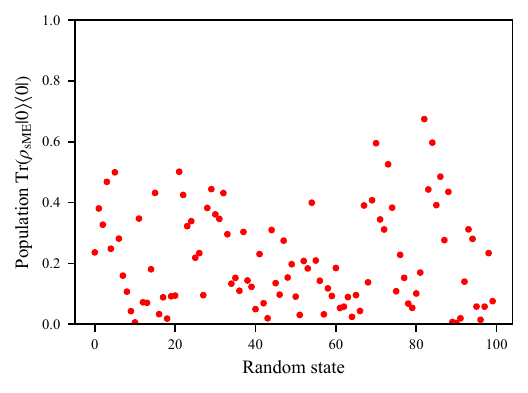}
    b)\includegraphics[width=0.25\linewidth]{ 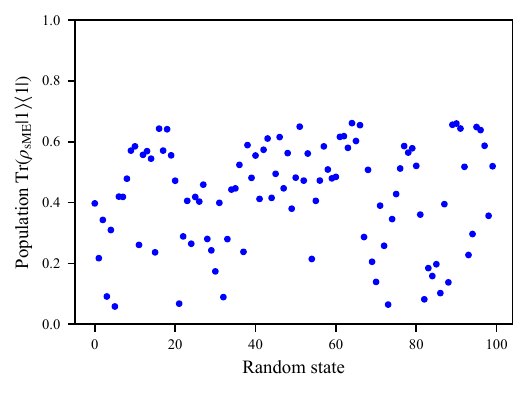}
    c)\includegraphics[width=0.25\linewidth]{ 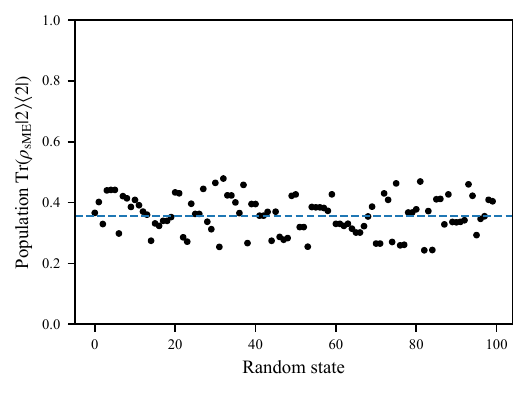}
    d)\includegraphics[width=0.25\linewidth]{ 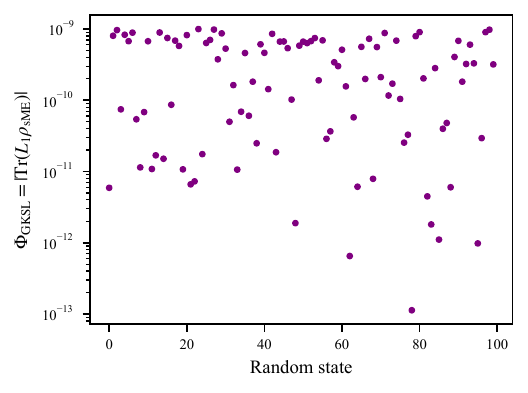}
    e)\includegraphics[width=0.25\linewidth]{ 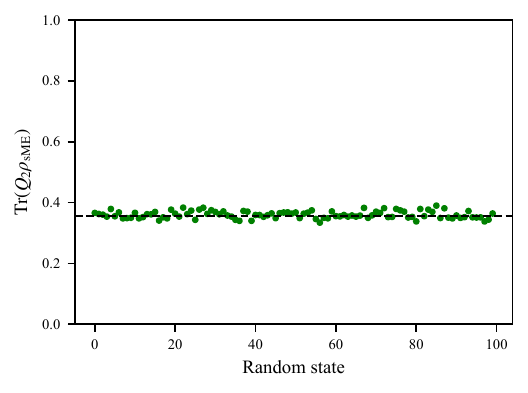}
    f)\includegraphics[width=0.25\linewidth]{ 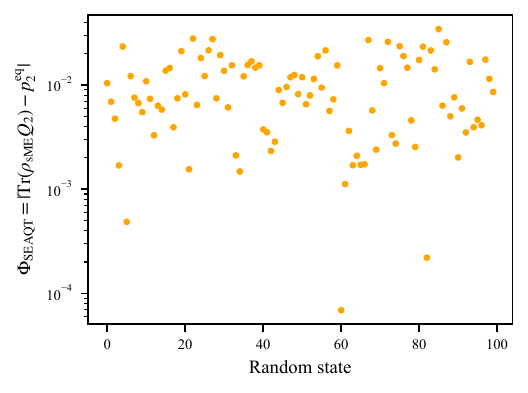}
    \caption{Multiple Mpemba initial $|\mathrm{sME}\rangle$ states randomly generated.}\label{fig: multiple_mpemba}
\end{figure}

A total of 100 different randomly generated initial $|\mathrm{sME}\rangle$ states satisfying the constraints described in Section~\ref{subsec: initial conditions} are generated. Their initial populations are shown in Figure~\ref{fig: multiple_mpemba}. In all cases, it is observed that the faster decay channel dominates the dynamics, resulting in a faster approach to stable equilibrium. It is also observed that the condition for exponential acceleration, $\mathrm{Tr}(\rho_{\mathrm{in}} L_1)=0$, takes values on the order of $10^{-9}$.

\begin{figure}[]
\centering
\includegraphics[width=0.7\linewidth]{ 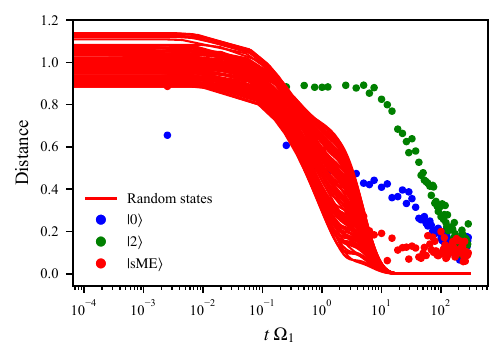}
\caption{Hilbert--Schmidt distance for each $|\mathrm{sME}\rangle$ initial condition.}
\label{fig: multiple_distance}
\end{figure}

The Hilbert--Schmidt distance,
\begin{equation}
D(\rho(t), \rho_{\mathrm{ss}}) = \sqrt{\mathrm{Tr}\left[(\rho(t) - \rho_{\mathrm{ss}})^\dagger (\rho(t) - \rho_{\mathrm{ss}})\right]}
\end{equation}
is calculated for each trajectory that corresponds to the 100 randomly generated $|\mathrm{sME}\rangle$ states. As shown in Figure~\ref{fig: multiple_distance}, all cases show a fast relaxation towards the final stable equilibrium state, which is consistent with the experimental data.

\begin{figure}[]
\centering
a)\includegraphics[width=0.7\linewidth]{ 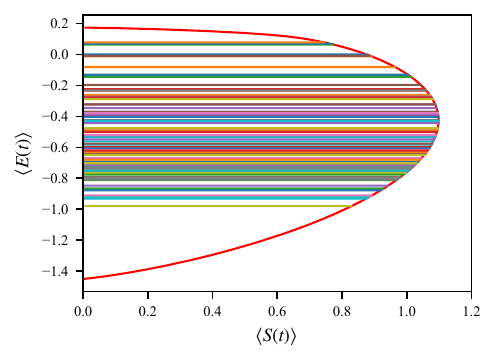}
b)\includegraphics[width=0.7\linewidth]{ 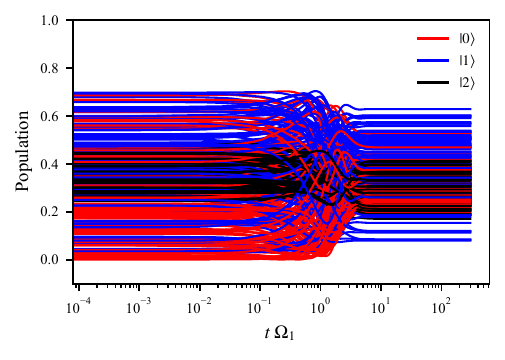}
\caption{State evolution of the different randomly generated initial $|\mathrm{sME}\rangle$ states using the SEAQT framework: a) representation on the energy-entropy diagram, and b) population dynamics.}
\label{fig: multiple-dynamics sq}
\end{figure}

\begin{figure}[]
\centering
a)\includegraphics[width=0.7\linewidth]{ 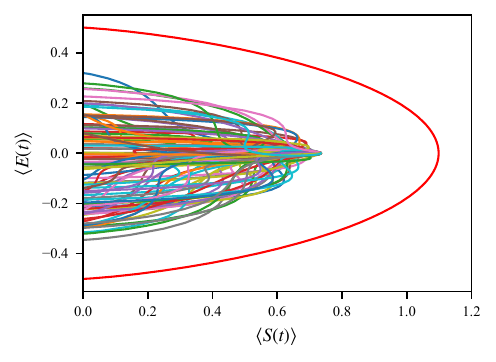}
b)\includegraphics[width=0.7\linewidth]{ 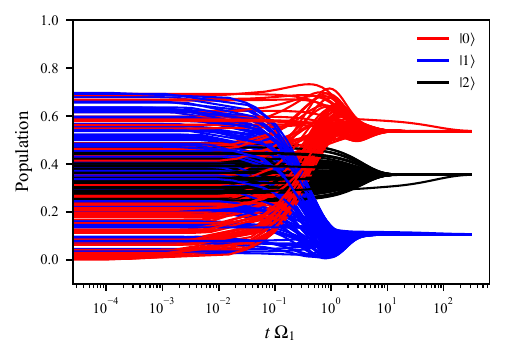}
\caption{State evolution of the different randomly generated initial $|\mathrm{sME}\rangle$ states using the Lindblad framework: a) representation on the energy-entropy diagram, and b) population dynamics.}
\label{fig: multiple-dynamics ld}
\end{figure}

Finally, the population dynamics obtained with the SEAQT, shown in Figure \ref{fig: multiple-dynamics sq}, and Lindblad, shown Figure \ref{fig: multiple-dynamics ld}, frameworks are compared. It is observed that the Lindblad framework drives the state evolution towards the same final state, which is located inside the energy-entropy diagram. For the SEAQT framework, each state evolves towards a unique final stable equilibrium state, keeping the energy of the system constant throughout each state evolution.

\section{Definition of $\beta$ at stable equilibrium}\label{app:B}

At stable equilibrium, the density operator is given as
\begin{equation}
    \hat{\rho}
    = \frac{e^{-\beta \hat H}}{Z(\beta)}
\end{equation}
where $Z(\beta)=\mathrm{Tr}\!\left(e^{-\beta \hat H}\right)$ is the partition function.

Under these conditions, all expectation values given in Section~\ref{subsec:seaqt} can be expressed in terms of derivatives of $\ln Z$ with respect to $\beta$, such as
\begin{equation}
    \langle \hat{H} \rangle = \mathrm{Tr}(\hat{\rho}\hat{H}) = -\frac{\partial \ln Z}{\partial \beta}
\end{equation}
\begin{equation}
    \langle \hat{H}^2 \rangle = \mathrm{Tr}(\hat{\rho}\hat{H}^2) = \frac{\partial^2 \ln Z}{\partial \beta^2} + \left(\frac{\partial \ln Z}{\partial \beta}\right)^2
\end{equation}
\begin{equation}
    \langle \hat{S} \rangle = -k_B \mathrm{Tr}(\hat{\rho}\ln\hat{\rho}) = -k_B \beta\frac{\partial \ln Z}{\partial \beta} + k_B \ln Z 
\end{equation}
\begin{align}  
    \langle \hat{H} \hat{S} \rangle =& -\frac{1}{2}k_B \mathrm{Tr}(\hat{\rho}\{\hat{H},\ln\hat{\rho}\} ) 
    = k_B\beta \left(\frac{\partial^2 \ln Z}{\partial \beta^2}+ \left(\frac{\partial \ln Z}{\partial \beta}\right)^2 \right) \nonumber \\
    & - k_B \ln Z \frac{\partial \ln Z}{\partial \beta}
\end{align}
where from these expressions, the energy variance and the energy-entropy covariance, respectively, can be expressed as
\begin{equation}
    \sigma_{\hat{H}\hat{H}} = \langle \hat{H}^2 \rangle - \langle \hat{H} \rangle^2 = \frac{\partial^2 \ln Z}{\partial \beta^2}
\end{equation}
\begin{equation}
    \sigma_{\hat{H}\hat{S}} = \langle \hat{H}\hat{S} \rangle - \langle \hat{H} \rangle \langle \hat{S} \rangle = k_B\beta \frac{\partial^2 \ln Z}{\partial \beta^2}
\end{equation}
Thus, at stable equilibrium the energy-entropy covariance satisfies
\begin{equation}
    \sigma_{\hat{H}\hat{S}} = k_B \, \beta \, \sigma_{\hat{H}\hat{H}}
\end{equation}
which implies that
\begin{equation}
    \beta = \frac{1}{k_B} \frac{\sigma_{\hat{H}\hat{S}}}{\sigma_{\hat{H}\hat{H}}}
\end{equation}
where it can observed that has dimensions of inverse temperature.

Defining the entropy as
\begin{equation}
\hat{S} = -k_B \hat{B}\ln \hat{\rho}
\end{equation}
and $\hat{H}$ expressed in energy units, ensures that the dimensions of the parameter $\beta$ given by Equation~(\ref{eq: beta}) coincides with the inverse temperature at stable equilibrium state.

\section{Definition of the parameter $\Phi_{\mathrm{SEAQT}}(\hat{\rho})$}
\label{app:slow_population_order_parameter}

The numerical analysis of the multiple initial states satisfying the order-parameter condition $\Phi_{\mathrm{GKSL}}\simeq 0$ is shown in Fig.~\ref{fig: multiple_mpemba}. Panels (a)--(c) display the three bare-state populations of 100 randomly generated initial states selected to satisfy $\Phi_{\mathrm{GKSL}}\simeq 0$. The populations of $|0\rangle$ and $|1\rangle$ span broad intervals, whereas the population of $|2\rangle$ remains confined to a comparatively narrow range around the reference value.

This behavior suggests that the suppression of the slow relaxation stage is primarily associated with a restricted coordinate of the state space that is strongly influenced by the population of the slowly relaxing state $|2\rangle$. However, the population $p_2=\mathrm{Tr}(\hat\rho\hat\Pi_2)$ alone is insufficient to characterize the full set of Mpemba initial states, since coherent contributions may satisfy the fast-relaxation condition even when $p_2\neq p_2^{\mathrm{eq}}$. This observation motivates the construction of an effective slow-population coordinate that incorporates the coherent directions dynamically coupled to $\hat\Pi_2$ through the effective Hamiltonian.

Let

\begin{equation}
    \hat\Pi_2 = |2\rangle\langle2|
\end{equation}

denote the population projector associated with the state having the smallest dissipative decay rate. We introduce the Hermiticity-preserving commutator map

\begin{equation}
    \mathcal{D}_H[\hat A]
    =
    \frac{i}{e}
    [\hat H_{\mathrm{eff}},\hat A],
    \label{eq:DH_slow}
\end{equation}

where $e>0$ is an energy scale introduced to render the map dimensionless.

For any Hermitian operator $\hat A$,

\begin{equation}
    \mathcal{D}_H[\hat A]^\dagger
    =
    \mathcal{D}_H[\hat A],
\end{equation}

and, consequently, all operators
$\mathcal{D}_H^n[\hat\Pi_2]$ are Hermitian.

Starting from $\hat\Pi_2$, successive applications of
$\mathcal{D}_H$ generate the operator Krylov subspace

\begin{equation}
    \mathcal{K}_2
    =
    \mathrm{span}
    \left\{
        \hat\Pi_2,
        \mathcal{D}_H[\hat\Pi_2],
        \mathcal{D}_H^2[\hat\Pi_2],
        \ldots
    \right\}.
    \label{eq:Krylov_slow}
\end{equation}

This subspace contains the population direction associated with $|2\rangle$, together with the coherent operator directions through which the Hamiltonian dynamically couples this population to the remaining states.

For the three-level effective Hamiltonian,

\begin{equation}
    \hat H_{\mathrm{eff}}
    =
    \sum_{n,m}
    H_{nm}|n\rangle\langle m|,
\end{equation}

the first Hamiltonian-generated direction is

\begin{align}
    \mathcal{D}_H[\hat\Pi_2]
    =
    \frac{i}{e}
    \Big[
        &
        H_{02}|0\rangle\langle2|
        +
        H_{12}|1\rangle\langle2|
        \nonumber\\
        &
        -
        H_{20}|2\rangle\langle0|
        -
        H_{21}|2\rangle\langle1|
    \Big].
    \label{eq:first_slow_direction}
\end{align}

Consequently, for the real Hamiltonian considered here,

\begin{equation}
    \left\langle
        \mathcal{D}_H[\hat\Pi_2]
    \right\rangle
    =
    \frac{2}{e}
    \left[
        H_{02}\operatorname{Im}\rho_{02}
        +
        H_{12}\operatorname{Im}\rho_{12}
    \right].
\end{equation}

Thus, the first Hamiltonian-generated correction contains the imaginary coherent quadratures that contribute directly to the unitary population current of the slow state.

The second action,

\begin{equation}
    \mathcal{D}_H^2[\hat\Pi_2]
    =
    -
    \frac{1}{e^2}
    [\hat H_{\mathrm{eff}},
    [\hat H_{\mathrm{eff}},\hat\Pi_2]],
\end{equation}

introduces both population corrections and complementary real coherent quadratures. Higher-order actions generate additional Hamiltonian-coupled operator directions. Therefore, rather than requiring the individual bare coherences $\rho_{02}$ and $\rho_{12}$ to vanish independently, the relevant information may be incorporated into a single Hamiltonian-dressed slow coordinate.

We define

\begin{equation}
    \hat Q_2(\boldsymbol{\alpha})
    =
    \hat\Pi_2
    +
    \sum_{n=1}^{N}
    \alpha_n
    \mathcal{D}_H^n[\hat\Pi_2]
    \label{eq:Qslow}
\end{equation}

where the dimensionless coefficients
$\boldsymbol{\alpha}=(\alpha_1,\ldots,\alpha_N)$ determine the coherent dressing of the slow-population coordinate. Importantly, these coefficients do not modify either the SEAQT metric or the corresponding equation of motion. Instead, they define a state-space coordinate used exclusively to characterize the initial conditions.

Accordingly, $\hat Q_2$ is not introduced ad hoc as an arbitrary linear combination of populations and coherences. Rather, it is constructed from the dynamical orbit of $\hat\Pi_2$ generated by the Hamiltonian flow. From this perspective, $\hat Q_2$ identifies an effective state-space coordinate associated with the slow sector after accounting for the coherent mixing induced by $\hat H_{\mathrm{eff}}$. It therefore provides a dynamically motivated candidate for an order parameter within the SEAQT description.

For the present three-state model, the numerical analysis indicates that the truncated representation

\begin{equation}
    \hat Q_2
    \simeq
    \hat\Pi_2
    +
    \alpha_1\mathcal{D}_H[\hat\Pi_2]
    +
    \alpha_2\mathcal{D}_H^2[\hat\Pi_2]
    +
    \alpha_3\mathcal{D}_H^3[\hat\Pi_2]
    \label{eq:Qslow3}
\end{equation}
\begin{table}[!h]
\centering
\caption{Convergence of the Hamiltonian-dressed slow coordinate.
$\theta_N$ is the misalignment angle between $\hat{L}_1^{\perp}$ and
${\rm span}\{\hat{\Pi}_2^{\perp},\mathcal{D}_H^{1..N}[\hat{\Pi}_2]^{\perp}\}$;
$\varsigma_N$ is the standard deviation of ${\rm Tr}(\hat\rho\hat{Q}_2)$
over $100$ states on the $\Phi_{\rm GKSL}=0$ level set.  $N=0$ is the
undressed projector $\hat{\Pi}_2$.}
\begin{tabular}{cccc}
\hline
$N$ & $\theta_N$ (deg) & $\varsigma_N$ & $\boldsymbol{\alpha}$ \\
\hline
0 & 13.73 & $2.12\times10^{-2}$ & --- \\
1 & 13.34 & $2.07\times10^{-2}$ & $\{0.29\}$ \\
2 & 11.52 & $1.77\times10^{-2}$ & $\{0.29,\,0.38\}$ \\
3 &  2.36 & $3.52\times10^{-3}$ & $\{-41.84,\,0.38,\,-15.97\}$ \\
4 &  2.21 & $3.14\times10^{-3}$ & --- \\
5 &  0.87 & $1.52\times10^{-3}$ & --- \\
6 &  0.08 & $1.41\times10^{-4}$ & --- \\
\hline
\end{tabular}
\end{table}

The expansion is exact at $N=6$, as established above by the
Cayley--Hamilton bound.  The truncation at $N=3$ used in Eq.~\eqref{eq:Qslow3}
reduces the misalignment from $13.73^{\circ}$ to $2.36^{\circ}$ and the
spread of the slow coordinate by a factor of six; this is the sense in
which it is a controlled truncation, and the residual $2.36^{\circ}$ is
the origin of the finite scatter of $\Phi_{\rm SEAQT}$ in Fig.~\ref{fig: multiple_mpemba}(f). Taking $e=\Omega_1$, and using a training ensemble $\{\hat\rho_k\}$ composed of the states shown in Fig.~\ref{fig: multiple_mpemba}, together with the equilibrium population $p_{2}^{\mathrm{eq}}$ as the reference value, the coefficients are obtained from

\begin{equation}
    \boldsymbol{\alpha}
    =
    \underset{\boldsymbol{\alpha}}{\operatorname{arg\,min}}
    \sum_{k}
    \left|
        \operatorname{Tr}
        \left(
            \hat\rho_k
            \hat Q_2(\boldsymbol{\alpha})
        \right)
        -p_{2}^{\mathrm{eq}}
    \right|^2.
    \label{eq:alpha_fit_population}
\end{equation}

Thus, the repeated train--test fits yield approximately
\begin{equation}
    \alpha_1\simeq-4.41,
    \qquad
    \alpha_2\simeq0.26,
    \qquad
    \alpha_3\simeq-1.79.
\end{equation}

These values are model-specific and should therefore be interpreted as parameters of the slow-coordinate representation rather than as universal constants.

The reference value defining the zero of the order parameter follows directly from the final equilibrium state. Let the SEAQT equilibrium state be

\begin{equation}
    \hat\rho_{\mathrm{eq}}
    =
    \frac{
        \exp(-\beta\hat H_{\mathrm{eff}})
    }{
        \mathrm{Tr}\left[
            \exp(-\beta\hat H_{\mathrm{eff}})
        \right]
    }.
\end{equation}

Since

\begin{equation}
    \left[\hat\rho_{\mathrm{eq}},\hat H_{\mathrm{eff}}\right]=0,
\end{equation}

for the projector $\hat{\Pi}_2$ one obtains

\begin{align}
    \mathrm{Tr}
    \left(
        \hat\rho_{\mathrm{eq}}
        \mathcal{D}_H[\hat{\Pi}_2]
    \right)
    &=
    \frac{i}{e}
    \mathrm{Tr}
    \left(
        \hat\rho_{\mathrm{eq}}
        [\hat H_{\mathrm{eff}},\hat{\Pi}_2]
    \right)
    \nonumber\\
    &=
    \frac{i}{e}
    \mathrm{Tr}
    \left(
        [\hat\rho_{\mathrm{eq}},\hat H_{\mathrm{eff}}]
        \hat{\Pi}_2
    \right)
    =0.
    \label{eq:gibbs_commutator_zero}
\end{align}

Because $\hat{Q}_2^{\perp}\propto\hat{L}_1^{\perp}$ to within $\theta_N$,
the expectation value ${\rm Tr}(\hat\rho\hat{Q}_2)$ is invariant along the
level set: every state annihilating the slow Liouvillian mode carries the
same value of the dressed coordinate,
\begin{equation}
q_{\star}\;\equiv\;{\rm Tr}(\hat\rho\,\hat{Q}_2)
\Big|_{{\rm Tr}(\hat{L}_1\hat\rho)=0},
\end{equation}
a constant of the model obtainable in closed form from
$\hat{Q}_2^{\perp}=s\,\hat{L}_1^{\perp}$ and the normalisation
${\rm Tr}\hat\rho=1$.  For the present model $q_{\star}=0.1847$.  The
SEAQT order parameter is then
\begin{equation} \label{eq: phi_seaqt_final}
\Phi_{\rm SEAQT}(\hat\rho)=\big|\,{\rm Tr}(\hat\rho\hat{Q}_2)-q_{\star}\,\big|,
\end{equation}
and the Mpemba condition is $\Phi_{\rm SEAQT}(\hat\rho_{\rm M})\simeq0$.

The same argument applies recursively, yielding

\begin{equation}
    \mathrm{Tr}
    \left(
        \hat\rho_{\mathrm{eq}}
        \mathcal{D}_H^n[\hat\Pi_2]
    \right)
    =0,
    \qquad
    n\geq1.
    \label{eq:higher_zero}
\end{equation}

Using Eq.~(\ref{eq:Qslow}), it follows that

\begin{align}
    \mathrm{Tr}
    \left(
        \hat\rho_{\mathrm{eq}}\hat Q_2
    \right)
    &=
    \mathrm{Tr}
    \left(
        \hat\rho_{\mathrm{eq}}\hat\Pi_2
    \right)
    \nonumber\\
    &=
    p_2^{\mathrm{eq}}.
    \label{eq:Qfinal}
\end{align}

Therefore, the equilibrium population $p_2^{\mathrm{eq}}$ is not introduced as an arbitrary threshold. Rather, it is exactly the equilibrium expectation value of the Hamiltonian-dressed coordinate, independently of the particular values of the coefficients $\alpha_n$.

This result motivates the following dynamical order parameter:

\begin{equation}
    \Phi_{\mathrm{SEAQT}}^{(2)}(\hat\rho)
    =
    \left|
        \mathrm{Tr}[
            \hat\rho\hat Q_2
        ]
        -
        p_2^{\mathrm{eq}}
    \right|
    =
    \left|
        \langle \hat Q_2\rangle-p_2^{\mathrm{eq}}
    \right|.
    \label{eq:new_order_parameter}
\end{equation}

The condition

\begin{equation}
    \Phi_{\mathrm{SEAQT}}^{(2)}
    (\hat\rho_{\mathrm{in}})
    \simeq0
    \label{eq:new_mpemba_condition}
\end{equation}

indicates that the Hamiltonian-dressed slow coordinate is already close to its final equilibrium value, even though
$\hat\rho_{\mathrm{in}}\neq\hat\rho_{\mathrm{eq}}$.
The remaining population and coherent directions may therefore still undergo substantial evolution while the displacement along the effective slow coordinate is strongly suppressed.

When the Hamiltonian-dressing contributions to
$\langle\hat Q_2\rangle$ are negligible for the states under consideration,

\begin{equation}
    \langle\hat Q_2\rangle
    \simeq
    \langle\hat\Pi_2\rangle
    =
    p_2,
\end{equation}

and Eq.~(\ref{eq:new_order_parameter}) approximately reduces to

\begin{equation}
    \Phi_{\mathrm{SEAQT}}^{(2)}
    \simeq
    |p_2-p_2^{\mathrm{eq}}|.
\end{equation}

This limiting behavior provides an explanation for the clustering of the $|2\rangle$ population observed in Fig.~\ref{fig: multiple_mpemba}. For states with non-negligible coherent contributions, however, the higher-order terms in Eq.~(\ref{eq:Qslow}) shift the effective slow coordinate and may allow
$\Phi_{\mathrm{SEAQT}}^{(2)}\simeq0$ even when
$p_2\neq p_2^{\mathrm{eq}}$, as occurs for the $|\mathrm{sME}\rangle$ initial condition.

\begin{figure}[]
    \centering
    a)\includegraphics[width=0.25\linewidth]{ 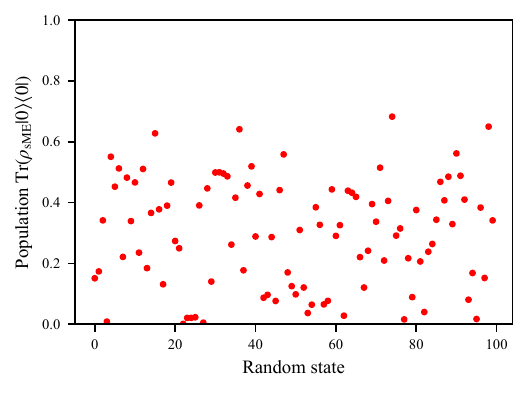}
    b)\includegraphics[width=0.25\linewidth]{ 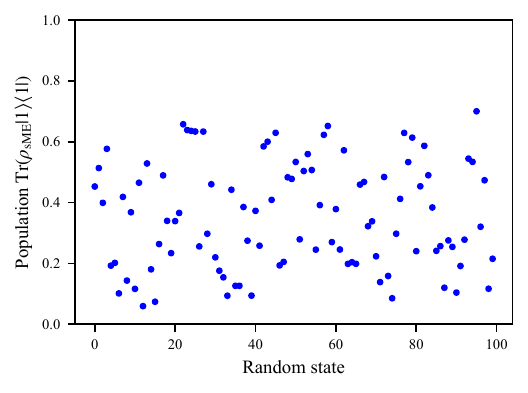}
    c)\includegraphics[width=0.25\linewidth]{ 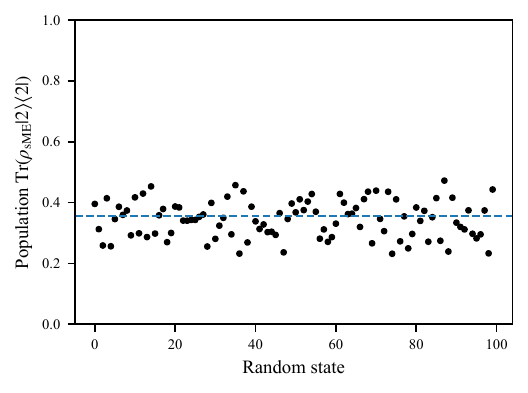}
    d)\includegraphics[width=0.25\linewidth]{ 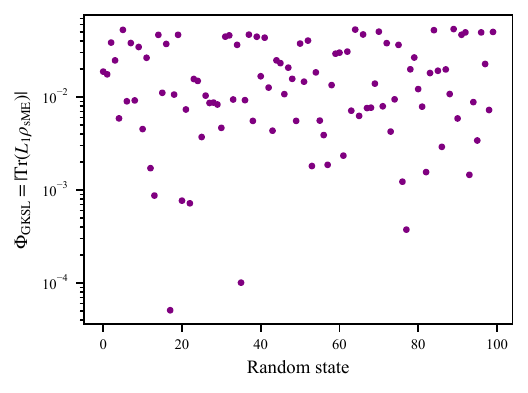}
    e)\includegraphics[width=0.25\linewidth]{ 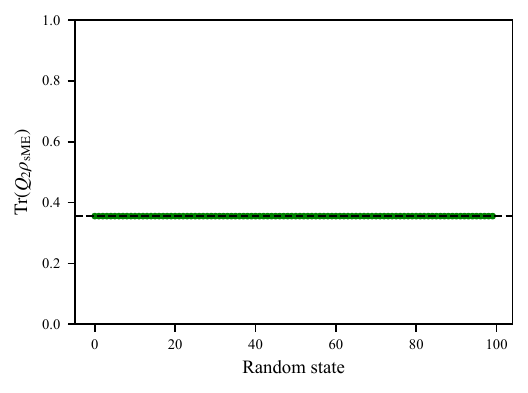}
    f)\includegraphics[width=0.25\linewidth]{ 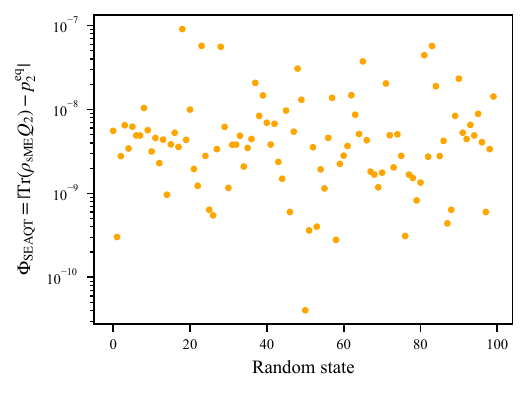}
    \caption{Properties of 100 randomly generated initial states satisfying the SEAQT slow-coordinate condition $\Phi_{\mathrm{SEAQT}}^{(2)}\simeq0$. The corresponding GKSL order parameter is subsequently evaluated to assess the consistency between the two criteria.}
    \label{fig: multiple_mpemba_seaqt}
\end{figure}

To provide an independent test of the proposed criterion and reduce possible selection bias, the analysis is also performed in the reverse direction. Specifically, 100 random initial states satisfying
$\Phi_{\mathrm{SEAQT}}\simeq0$ are generated independently, and their properties are subsequently evaluated using the order parameter $\Phi_{\mathrm{GKSL}}$. The resulting distributions are shown in Fig.~\ref{fig: multiple_mpemba_seaqt}. This complementary analysis tests whether the class of states identified by the SEAQT slow-coordinate criterion is consistent with the slow-mode suppression independently identified within the GKSL description.

Finally, the connection between the proposed slow coordinate and the SEAQT equation of motion can be made explicit. For

\begin{equation}
    \frac{d \hat{\rho}}{dt}
    =
    - \frac{i}{\hbar}
    \left[
        \hat{H}_{\mathrm{eff}},
        \hat{\rho}
    \right]
    -
    \frac{\beta}{2 \tau_D}
    \left\{
        \Delta \hat{F},
        \hat{\rho}
    \right\},
\end{equation}

the instantaneous velocity of the slow coordinate is

\begin{equation}
    \frac{d}{dt}
    \langle\hat Q_2\rangle
    =
    -\frac{i}{\hbar}
    \left\langle
        [\hat Q_2,\hat H_{\mathrm{eff}}]
    \right\rangle
    -
    \frac{\beta}{\tau_D}
    \sigma_{FQ_2},
    \label{eq:Qslow_dynamics}
\end{equation}

where the symmetrized covariance is defined as

\begin{align}
    \sigma_{FQ_2}
    &=
    \frac{1}{2}
    \mathrm{Tr}
    \left[
        \hat\rho
        \left\{
            \Delta\hat F,
            \Delta\hat Q_2
        \right\}
    \right]
    \nonumber\\
    &=
    \frac{1}{2}
    \left\langle
        \left\{
            \hat F,
            \hat Q_2
        \right\}
    \right\rangle
    -
    \langle\hat F\rangle
    \langle\hat Q_2\rangle.
\end{align}

Equation~(\ref{eq:Qslow_dynamics}) shows that satisfying the proposed order-parameter condition does not imply the vanishing of the total thermodynamic driving force. Instead, the condition identifies initial states for which the coordinate associated with the slow sector has only a small remaining displacement from its equilibrium value. Both the unitary and irreversible contributions may remain finite along other directions of the state space, thereby allowing substantial evolution while suppressing the component associated with the slow-relaxation sector. This provides a dynamical interpretation of the accelerated relaxation observed along the Mpemba trajectories.

The expansion of Eq.~(\ref{eq:Qslow}) terminates, and the truncation order can be
bounded exactly. Let $\{|e_a\rangle\}$ be the eigenbasis of $\Heff$ and
$\omega_{ab}=E_a-E_b$ the Bohr frequencies. The map $\mathcal{D}_H$ of
Eq.~(\ref{eq:DH_slow}) acts diagonally in the operator basis
$\{|e_a\rangle\langle e_b|\}$ with eigenvalues $\omega_{ab}/e$; its kernel is the
three-dimensional span of the diagonal projectors and its range the
six-dimensional span of the off-diagonal directions. Consequently
$\mathcal{D}^n_H[\hat\Pi_2]\in\mathrm{ran}\,\mathcal{D}_H$ for all $n\ge 1$, so
that $\dim\mathcal{K}_2\le 7$ and, by the Cayley--Hamilton theorem applied to
$\mathcal{D}_H$ restricted to $\mathcal{K}_2$, the terms with $n\ge 7$ are linear
combinations of the lower-order ones. For the effective Hamiltonian used here the
Bohr frequencies are nondegenerate and the bound is saturated: the family
$\{\mathcal{D}^n_H[\hat\Pi_2]\}_{n=1}^{7}$ has numerical rank $6$. The expansion
is therefore exact at $N=6$, and any $N<6$ is a controlled truncation.
 
The quality of that truncation can be quantified. The slowest nonzero Liouvillian
eigenvalue is real, so $\hat L_1$ is Hermitian and both order parameters are
affine functionals on the trace-one slice of state space. Writing
$\hat X^{\perp}=\hat X-\tfrac{1}{3}\mathrm{Tr}(\hat X)\hat I$, the conditions
$\mathrm{Tr}(\hat L_1\hat\rho)=0$ and
$\mathrm{Tr}(\hat Q_2\hat\rho)=p^{\mathrm{eq}}_2$ define the same hyperplane if
and only if $\hat Q^{\perp}_2\propto\hat L^{\perp}_1$. The coefficients
$\boldsymbol{\alpha}$ may therefore be obtained directly from the orthogonal
projection of $\hat L^{\perp}_1$ onto the truncated Krylov basis, without a
training ensemble, and the residual misalignment angle between the two
hyperplane normals measures how nearly the two criteria coincide. That residual
is the origin of the finite spread of $\Phi_{\mathrm{SEAQT}}$ seen in
Fig.~\ref{fig: multiple_mpemba}(f) for states selected by
$\Phi_{\mathrm{GKSL}}\simeq 0$, and of $\Phi_{\mathrm{GKSL}}$ in
Fig.~\ref{fig: multiple_mpemba_seaqt}(d) for states selected by the SEAQT
criterion.

\section{Correspondence between the Lindblad and SEAQT order parameters} \label{sec:correspondence}
The order parameters $\Phi_{\GKSL}$ and $\Phi_{\SEAQT}$ are constructed within
different mathematical frameworks --- the spectral decomposition of the
Liouvillian superoperator and the Hamiltonian-dressed slow coordinate of the
SEAQT state space, respectively --- yet both enforce the same physical outcome,
the elimination of the $\kappa_2$-dominated slow relaxation stage.
 
Within the Lindblad framework the Mpemba condition
$\Phi_{\GKSL}(\rhoh_{\mathrm{in}})=\mathrm{Tr}(\hat L_1\rhoh_{\mathrm{in}})=0$
removes the slowest-decaying term from the spectral sum in Eq.~(\ref{eq: solution_Lindblad}). The residual dynamics then decays at the faster rate $|\mathrm{Re}(\lambda_2)|\sim\kappa_1$,
producing the exponential acceleration.
 
Within the SEAQT framework the dynamics is nonlinear and no state-independent
spectral decomposition exists. The condition
$\Phi_{\SEAQT}(\rhoh_{\mathrm{in}})\simeq 0$ instead places the initial state on
the level set
\begin{equation}
\mathcal{M}_0=\Big\{\rhoh:\ \langle \Qh_2\rangle_{\rhoh}=p^{\mathrm{eq}}_2\Big\} ,
\label{eq:M0}
\end{equation}
on which the Hamiltonian-dressed slow coordinate already coincides with its
equilibrium value while the remaining population and coherent directions are
still far from equilibrium. By Eq.~(\ref{eq:Qslow_dynamics}) the velocity of $\Qh_2$ is then small throughout
the evolution, so the displacement along the slow coordinate is suppressed while
the remaining directions relax through the fast channel. This suppression is what
produces the accelerated approach to the final state; it does not require the
trajectory to avoid $\rhoc$, and indeed for the model studied here the
$|\mathrm{sME}\rangle$ trajectory transits the metastable region. 

The two conditions therefore act by different mechanisms, the Lindblad
condition spectrally, by removing the slowest eigenmode of $\mathcal{L}$; the
SEAQT condition geometrically, by bypassing the metastable entropy plateau,
and coincide in their physical consequence. We emphasize that this coincidence
is established here numerically rather than analytically: all $100$ randomly
generated initial states satisfying $|\Phi_{\GKSL}|\lesssim 10^{-9}$ exhibit

$\tau_D\simeq\text{const}$ and $\Phi_{\SEAQT}\simeq 0$ (~\ref{app:A}), and the
reverse test of Fig.~(\ref{fig: multiple_mpemba_seaqt}), in which states are generated from the
$\Phi_{\SEAQT}$ criterion and subsequently evaluated with $\Phi_{\GKSL}$,
confirms the correspondence in the opposite direction. Whether the equivalence
holds beyond this three-level model remains open.


\bibliographystyle{elsarticle-harv} 

\end{document}